\documentclass[useAMS,usenatbib,usegraphicx]{mn2e}

\pdfoutput=1
\usepackage{hyperref}
\usepackage{color}
\usepackage{subfigure}
\usepackage{verbatim}

\title[Searching for nova shells]{Searching for nova shells around
  cataclysmic variables} \author[D. I. Sahman et
  al]{D. I. Sahman,$^{1}$\thanks{E-mail: d.sahman@sheffield.ac.uk;
    vik.dhillon@sheffield.ac.uk} V. S. Dhillon,$^{1}$
  C. Knigge,$^2$  T. R. Marsh$^{3}$\\ 
$^{1}$Department of Physics and
  Astronomy, University of Sheffield, Sheffield S3 7RH,
  UK\\
$^{2}$School of Physics \& Astronomy, University of
  Southampton, Southampton SO17 1BJ, UK\\
 $^{3}$Department of Physics, University of Warwick, Coventry
  CV4 7AL, UK\\}

\begin{document}

\date{Accepted 2015 May 16. Received 2015 April 17}

\maketitle

\label{firstpage}

\begin{abstract}

We present the results of a search for nova shells around 101
cataclysmic variables (CVs), using H$\alpha$ images taken with the
4.2-m William Herschel Telescope (WHT) and the 2.5-m Isaac Newton
Telescope Photometric H$\alpha$ Survey of the Northern Galactic Plane
(IPHAS). Both telescopes are located on La Palma. We concentrated our
WHT search on nova-like variables, whilst our IPHAS search covered all
CVs in the IPHAS footprint.  We found one shell out of the 24
nova-like variables we examined. The newly discovered shell is around
V1315 Aql and has a radius of $\sim2.5\arcmin$, indicative of a nova
eruption approximately 120 years ago. This result is consistent with
the idea that the high mass-transfer rate exhibited by nova-like
variables is due to enhanced irradiation of the secondary by the hot
white dwarf following a recent nova eruption. The implications of our
observations for the lifetime of the nova-like variable phase are
discussed.

We also examined 4 asynchronous polars, but found no new shells around
any of them, so we are unable to confirm that a recent nova eruption
is the cause of the asynchronicity in the white dwarf spin.  We find
tentative evidence of a faint shell around the dwarf nova V1363
Cyg. In addition, we find evidence for a light echo around the nova
V2275 Cyg, which erupted in 2001, indicative of an earlier nova
eruption $\sim300$ years ago, making V2275 Cyg a possible recurrent
nova.

\end{abstract}

\begin{keywords} stars: novae, cataclysmic variables.
\end{keywords}

\section{Introduction}

Cataclysmic variables (CVs) are close binary systems in which a white
dwarf (WD) accretes material from a secondary star, via Roche-lobe
overflow (see \citealt{warner95a} for a review).  CVs are classified
observationally into 3 main sub-types -- the novae, the dwarf novae
and the nova-like variables. The {\em novae} are defined as systems in which
only a single nova eruption has been observed. Nova eruptions have
typical amplitudes of 10 magnitudes and are believed to be due to the
thermonuclear runaway of hydrogen-rich material accreted onto the
surface of the white dwarf. The {\em dwarf novae} (DNe) are defined as
systems which undergo quasi-regular (on timescales of weeks-months)
outbursts of much smaller amplitude (typically 6 magnitudes). Dwarf
nova outbursts are believed to be due to instabilities in the
accretion disc causing the sudden collapse of large quantities of
material onto the white dwarf. The {\em nova-like} variables (NLs) are
the non-eruptive CVs, i.e. objects which have never been observed to
show nova or dwarf nova outbursts. The absence of dwarf nova outbursts
in NLs is believed to be due to their high mass-transfer rates,
producing ionised accretion discs in which the disc-instability
mechanism that causes outbursts is suppressed \citep{osaki74}. The
mass transfer rates in NLs are $\dot{M} \sim 10^{-8}-10^{-9}$
M$_{\odot}$ yr$^{-1}$ whereas DNe have rates of $\dot{M} \sim
10^{-10}-10^{-11}$ M$_{\odot}$ yr$^{-1}$. Note that throughout this
paper, when we refer to NLs we mean non-magnetic systems, i.e. we do
not include in our definition systems that accrete via magnetic field
lines, such as polars and intermediate polars.

Our understanding of CV evolution has made great strides in recent
years (e.g. see \citealt{knigge11}, \citealt{knigge11a}).  For example,
there is now strong evidence that the 2--3 hr period gap in the
orbital period distribution of CVs is indeed due to disrupted angular
momentum loss \citep{patterson05}, the first \textquotedblleft
period-bounce\textquotedblright \, CVs, in which the secondary star
has lost so much mass to the WD that it falls below the
hydrogen--burning limit, have now been discovered
\citep{littlefair08}, and the predicted spike of systems at the period
minimum has been revealed \citep{gansicke09}.

One of the remaining unsolved problems in CV evolution is: how can the
different types of CV co-exist at the same orbital period?  Theory
predicts that all CVs evolve from longer to shorter orbital periods on
timescales of gigayears, and as they do so the mass-transfer rate also
declines (e.g. see Fig.~19 of \citealt{knigge11}). At periods longer
than approximately 5\,hrs, all CVs should have high mass-transfer
rates and appear as NLs, whereas below this period the lower
mass-transfer rate allows the disc-instability mechanism to operate
and all CVs should appear as DNe. This theoretical expectation,
however, is in stark contrast to observations, which show that the
fraction of nova-like variables to dwarf novae is actually highest at 3\,hr
periods and then declines to longer periods (e.g. see Fig.~18 of
\citealt{knigge11}).

It is possible that CVs cycle between NL and DN states on timescales
shorter than the gigayear evolutionary timescale of the binary,
thereby explaining the co-existence of NLs and DNe at the same orbital
period.  Two mechanisms for such a cycle have been proposed. Both
mechanisms invoke cyclical variation in the irradiation of the
secondary, which in turn drives cyclical variation of $\dot{M}$ with
timescales of the order of $\sim10^{4}-10^{7}$ yrs.

The first idea is that there is an irradiation feedback mechanism. The
flux from the WD illuminates the inner face of the secondary which
flattens the temperature gradient in the photosphere, leading to an
expansion in the radius of the secondary and an increase in $\dot{M}$
above the secular mean (see \citealt{buning04} and references
therein). The enhanced $\dot{M}$ drives an increase in the radius of
the secondary's Roche lobe. Eventually the expansion of the secondary
star cannot keep pace with the Roche lobe expansion, leading to a
lower $\dot{M}$, and hence a reduction in the irradiating
flux. Consequently, the secondary begins to shrink and the feedback
mechanism operates in reverse as the mass transfer rate
reduces. \citet{buning04} found that this mechanism could produce
limit cycles in $\dot{M}$ of the appropriate timescales (see Fig.~5 of
\citealt{knigge11}), causing CVs to cycle between DN and NL
states. However, their models show that systems just above the period
gap are actually stable and do not undergo cycles.  Hence, although
this model explains why some NLs and DNe may co-exist at the same
orbital period, it does not explain why the irradiation-driven
feedback mechanism would make the NL fraction highest around 3\,hr and
decline towards longer periods.

The second hypothesis for variable $\dot{M}$ is a nova-induced
cycle. Some fraction of the energy released in the nova event will
heat up the WD, leading to increased irradiation and subsequent
bloating of the secondary. Following the nova event, the system would
have a high $\dot{M}$ and appear as a NL. As the WD cools, the radius
of the secondary would return to its secular value, and hence
$\dot{M}$ will reduce and the system changes to a DN. In this model,
therefore, CVs are expected to cycle between nova, NL and DN states,
on timescales of $10^4-10^5$ yrs (see \citealt{shara86}), thereby
explaining the co-existence of these CV sub-types at the same orbital
period. Like the irradiation feedback mechanism, however, this nova
cycle model does not explain why the NL fraction is highest around
3\,hr and declines towards longer periods.

The cyclical evolution of CVs through nova, NL and DN phases recently
received observational support from the discovery that BK Lyn appears
to have evolved through all three phases since its likely nova
outburst in the year AD 101 \citep{patterson13}. A second piece of
evidence has come from the discovery of nova shells around the dwarf
novae Z Cam and AT Cnc (\citealt{shara07}, \citealt{shara13}),
verifying that they must have passed through an earlier nova phase. A
more obvious place than DNe to find nova shells is actually around
NLs, as the nova-induced cycle theory suggests that the high $\dot{M}$
in NLs could be due to a recent nova eruption. Finding shells around the
 highest accretion--rate NLs
would lend further support to the existence of nova-induced cycles and
hence help explain why systems with different $\dot{M}$ are found at the same
orbital period.

In this paper we present the results of an H$\alpha$ imaging survey of
the fields surrounding a sample of 101 cataclysmic variables, 24 of
them NLs, with the aim of identifying nova shells around the central
binary.

\section{Observations and Data Reduction}

\subsection{William Herschel Telescope}

\subsubsection{Search strategy}
\label{search_strategy}

The choices of telescope aperture and field of view for this project
were dictated by the expected brightness and radii of the nova shells
around CVs. Recombination theory tells us that the H$\alpha$
luminosity per unit volume of a nova shell is proportional to density
squared, and hence the total luminosity of the shell is inversely
proportional to volume. If we assume that the shell expands at a
constant velocity, then its volume increases as the cube of time.
Therefore, the luminosity is inversely proportional to time cubed, and
the surface brightness decreases as time to the fifth power.  This
expectation has been empirically confirmed by \citet{downes01} who
found that the H$\alpha$ surface brightness of nova shells diminishes
as $t^{-4.8}$, although novae with strong shock interactions between
the ejecta and any pre-existing circumstellar material, e.g. GK~Per
\citep{shara12} and T~Pyx \citep{shara89}, do not fit this
relationship.

To estimate how bright the shells around NLs might be, we used as a
guide the archetypal old nova DQ Her (Nova Her 1934) which has a
clearly visible shell \citep{slavin95}. Its shell had a brightness of
$ 1.01 \times 10^{-15}$ ergs/s/cm$^{2}$/arcsec$^{2}$ and a radius of
$8\arcsec$ in May 1982, 47.5 years after its nova eruption
\citep{ferland84}. This implies that 100 years after outburst it will
fade to $2.44 \times 10^{-17}$ ergs/s/cm$^{2}$/arcsec$^{2}$, by which
time its radius will have doubled to 16\arcsec, and at 200 years after
outburst it will fade to $7.63 \times 10^{-19}$
ergs/s/cm$^{2}$/arcsec$^{2}$, and its radius will have doubled again
to 32\arcsec. We can use these estimates to determine our observing
strategy and to establish the instrumental setup that is required.

Nova-like variables have apparent magnitudes of $\sim$15, on
average. Novae brighten by $\sim$10 magnitudes during eruption,
implying that most nova-like variables when in outburst would have
been just visible to the naked eye.  Given the much increased rate of
nova detection in the last 100 years, it is very unlikely that such
eruptions would have been missed if they had occurred within the last
$\sim$100 years or so. We thus expect to have to detect shells around
nova-like variables which are at least 100 years old; any which
erupted more recently than this would likely have been detected and
would now be classified as a nova. Assuming the shell of DQ Her is
representative, a 100-year old shell would be at least $16\arcsec$ in
radius and no brighter than $2.44 \times 10^{-17}$
ergs/s/cm$^{2}$/arcsec$^{2}$. Hence to detect such shells we require
deep images with a relatively modest field of view, which led us to
use the Auxiliary Port on the 4.2m William Herschel Telescope (WHT) on
La Palma (see Sec.~2.1.2).

To estimate the age of the faintest shell we might detect with this
setup, we simulated the images we would obtain from a spherical nova
shell in an 1800 second H$\alpha$-exposure. The simulation showed that
a shell with luminosity comparable to the shell of DQ Her would become
too faint to detect in an image at $\sim180$ years after outburst. For
approximately circular, small shells that are centred on the binary,
this detection threshold can be pushed fainter by computing the mean
radial profile of the central object and inspecting the wings for
evidence of a shell (see \citealt{gill98}). The DQ Her shell would be
apparent in the radial profile up to 220 years after
outburst. Assuming DQ Her is representative, this means that we would
be able to detect shells around nova-like variables from nova eruptions up to a
maximum of $\sim$220 years ago using our proposed setup on the WHT.

\subsubsection{Observations}

The observations were taken on the nights of 1997 October 24--26. We
used the 1024$\times$1024 pixels TEK5 CCD chip mounted on the
Auxiliary
Port\footnote{http://www.ing.iac.es/astronomy/observing/manuals/html
  \textunderscore manuals/wht\textunderscore
  instr/pfip/prime3\textunderscore www.html} at the Cassegrain focus
of the WHT to image the fields around our target nova-like
variables. This setup gave a platescale of $0.11\arcsec$ per pixel and
hence a field size of $113\arcsec \times 113\arcsec$. H$\alpha$ is one
of the strongest features in the spectra of nova shells, with typical
velocity widths of up to 2000 km\,s$^{-1}$ (e.g.
\citealt{warner95a}). In order to maximise the detection of light from
the shell and minimise the contribution of sky, we therefore used a
narrow-band (55\AA\ FWHM = 2500 km s$^{-1}$) interference filter
centred on the rest wavelength of H$\alpha$ (ING filter number
61\footnote{http://catserver.ing.iac.es/filter/}). Note that this
filter also includes a contribution from [N{\sc
    ii}]\,6584\AA\ emission, which may dominate the spectra of nova
shells with strong shock interaction of the ejecta with any
pre-existing circumstellar medium, e.g. T~Pyx \citep{shara89}.

As we were planning to compare the radial profiles of the target stars
with field stars, we had to ensure that we did not saturate the target
stars. Hence the CCD chip was used unbinned and in quick readout mode,
in order to decrease dead-time, at the expense of a negligible
decrease in signal-to-noise (thanks to the fact that our observations
were always sky-limited).  The observing conditions were excellent
throughout the run; the sky was always photometric, there was no
evidence of dust and the seeing was usually sub-arcsecond, with an
occasional excursion up to 1.5--2$\arcsec$.

\subsubsection{Target selection}
\label{whttarget}

To ensure we only targeted relatively well-studied systems with
reliable CV classifications, we made our selection from the catalogue of
\citet{ritter03}\footnote{http://www.mpa-garching.mpg.de/RKcat/}
(hereafter RK catalogue). We selected a total of 31 CVs, predominantly
NLs, and searched for nova shells around them. To test our setup we
included three systems with known nova shells, BT Mon, DQ Her and GK
Per. We also took the opportunity to observe two asynchronous polars,
which are CVs with magnetic WDs in which the spin period of the WD is not 
synchronised
with the orbital period \citep{warner83}. The asynchronicity is
believed to be due to a recent nova event, as shown by the system
V1500 Cyg which had a nova eruption in 1975 \citep{stockman88}. The
two asynchronous polars we observed with the WHT were V1432 Aql and BY
Cam. We also included three other non-NL systems, PQ Gem, which is an
intermediate polar, IP Peg which is a DN, and AY Psc which is a Z
Cam-type DN, which were favourably positioned during our observing
run.

\begin{figure}
\centering
  \vspace{10pt}
\includegraphics[width=80mm,angle=0]{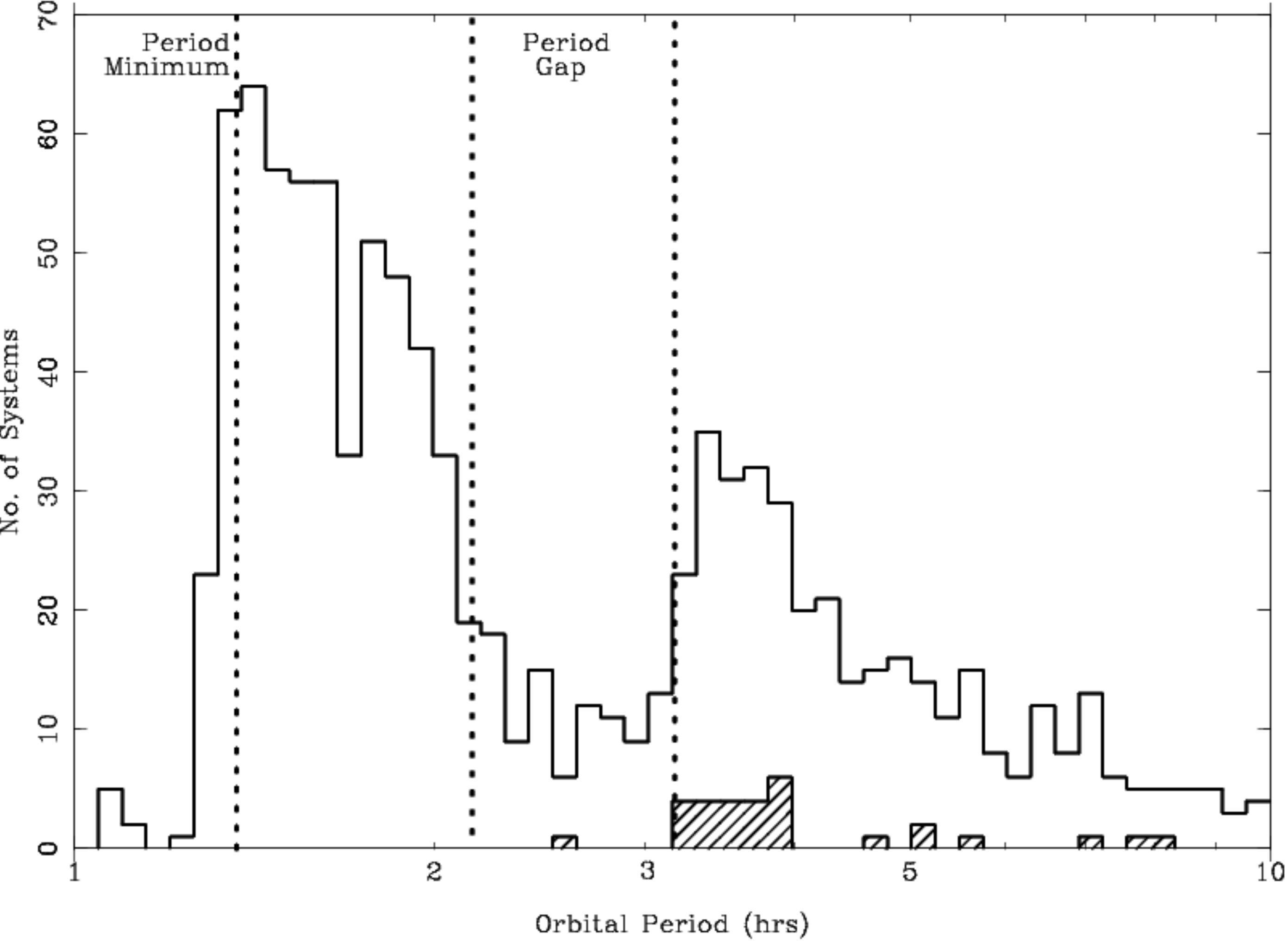}
 \caption{Orbital period distribution of the systems we observed
   with the WHT (hatched) compared to the distribution of all the CVs
   in the RK catalogue. The left-hand dotted line indicates the period
   minimum and the central and right-hand dotted lines show the period
   gap taken from \citet{knigge11}.}
\label{cvhist}
\end{figure}

A full list of the 31 objects observed with the WHT and a journal of
observations is given in Tab.~\ref{tab:journal}. In summary, the
targets comprised 3 old novae with known shells, 1 old nova without a
known shell (V Per), 2 asynchronous polars, 1 intermediate polar, 22
NLs, and 2 DNe. In Fig.~\ref{cvhist} we show the orbital
period distribution of all the systems we observed with the WHT
compared to the total number of systems in the RK catalogue. We
deliberately selected a substantial number of the systems in the 3--4\,hr
 orbital period range, which is where most NLs appear, as shown in
Fig.~18 of \citet{knigge11}.

\subsubsection{Data reduction}

The images were debiased using the median level of the overscan strip
and flat-fielded using normalised twilight sky flats. Where we had
taken multiple images of targets, these were combined to improve the
signal-to-noise ratio. Sky subtraction was performed by subtracting
the median level determined from two blank sky areas of size 100
$\times$ 100 pixels. Each frame suffered from significant vignetting
in the corners due to the circular filter holder, which was not fully
corrected by the flat field. The corners of each image were hence
removed by setting a series of $50 \times 50$ pixel blocks to a fixed
value, so that they appear white in the final images shown in Appendix
A. Pixels affected by cosmic rays were set to the average
value of surrounding pixels. All processing was performed using the
{\sc kappa} and {\sc figaro} packages in the {\sc
  starlink}\footnote{http://starlink.jach.hawaii.edu/starlink} suite
of programs.

\begin{table*}
\caption[]{Journal of WHT observations. The classifications of the CVs
  have been taken from the RK catalogue. The date refers to the start
  time of the first exposure. All shells detected in our WHT
  observations are shown in bold and discussed in Sect.~3.1. Note that
  the RK catalogue classifications for novae are N, Na, Nb.  $^*$We
  detected a shell around V1315 Aql with the INT, not the WHT -- see
  section \ref{v1315}. $^\dagger$This system is a NL.}
\begin{center}
\begin{tabular}{llccccccc}
\hline\hline
& & & & & & & \\
\multicolumn{1}{l}{Object} &
\multicolumn{1}{l}{Classification} &
\multicolumn{1}{c}{Orbital} &
\multicolumn{1}{c}{Date} &
\multicolumn{1}{c}{UTC} &
\multicolumn{1}{c}{UTC} &
\multicolumn{1}{c}{Number of} &
\multicolumn{1}{c}{Total exposure} &
\multicolumn{1}{c}{Visible} \\
& & \multicolumn{1}{c}{period (hrs)} & & \multicolumn{1}{c}{start} &
\multicolumn{1}{c}{end} & \multicolumn{1}{c}{exposures}
& \multicolumn{1}{c}{time (secs)} & \multicolumn{1}{c}{shell?} \\
& & & & & & & & \\
\hline
PX And        & NL SW NS SH & 3.51 & 25/10/97 & 00:06 & 01:11 & 2 & 3600 & N \\
UU Aqr        & NL UX SW SH & 3.93 & 25/10/97 & 22:42 & 23:46 & 3 & 3300 & N \\
HL Aqr        & NL UX SW    & 3.25 & 27/10/97 & 00:13 & 00:54 & 2 & 2400 & N \\
V794 Aql      & NL VY       & 3.68 & 26/10/97 & 21:17 & 21:58 & 2 & 2400 & N \\
V1315 Aql     & NL UX SW    & 3.35 & 26/10/97 & 19:32 & 20:15 & 4 & 2400 & N$^*$ \\
V1432 Aql     & NL AM AS    & 3.37 & 25/10/97 & 19:59 & 21:01 & 2 & 3600 & N \\
WX Ari        & NL UX SW    & 3.34 & 25/10/97 & 02:42 & 03:44 & 2 & 3600 & N \\
KR Aur        & NL VY NS    & 3.91 & 26/10/97 & 03:50 & 04:10 & 1 & 1200 & N \\
V363 Aur      & NL UX SW    & 7.71 & 25/10/97 & 05:09 & 06:10 & 2 & 3600 & N \\
BY Cam        & NL AM AS    & 3.36 & 25/10/97 & 04:00 & 05:01 & 2 & 3600 & N \\
AC Cnc        & NL UX SW    & 7.21 & 27/10/97 & 03:31 & 04:12 & 2 & 2400 & N \\
V425 Cas      & NL VY       & 3.59 & 24/10/97 & 22:29 & 22:37 & 2 &\ 200 & N \\
V751 Cyg      & $^\dagger$VY SW? NS SS & 3.47 & 24/10/97 & 22:07 & 22:27 & 1 & 1200 & N \\
V1776 Cyg     & NL UX SW    & 3.95 & 25/10/97 & 21:31 & 22:32 & 2 & 3600 & N \\
CM Del        & NL UX VY?   & 3.89 & 26/10/97 & 20:22 & 21:05 & 4 & 2400 & N \\
PQ Gem        & NL IP       & 5.19 & 27/10/97 & 02:41 & 03:22 & 2 & 2400 & N \\
\textbf{DQ Her}        & \textbf{Na DQ }      & \textbf{4.65} & \textbf{25/10/97} & 
\textbf{19:31} & \textbf{19:51} & \textbf{1} & \textbf{1200} & \textbf{Y} \\
BH Lyn        & NL SW SH NS & 3.74 & 26/10/97 & 04:24 & 05:05 & 2 & 2400 & N \\
BP Lyn        & NL UX SW    & 3.67 & 27/10/97 & 04:21 & 05:02 & 2 & 2400 & N \\
\textbf{BT Mon}        & \textbf{Na SW}       & \textbf{8.01} & \textbf{26/10/97} & 
\textbf{06:12} & \textbf{06:43} & \textbf{2} & \textbf{1800} & \textbf{Y} \\
\textbf{BT Mon}        & \textbf{Na SW}       & \textbf{8.01} & \textbf{27/10/97} & 
\textbf{06:02} & \textbf{06:32} & \textbf{1} & \textbf{1800} & \textbf{Y} \\
V1193 Ori     & NL UX SW?   & 3.96 & 26/10/97 & 01:05 & 02:38 & 3 & 5400 & N \\
IP Peg        & DN UG       & 3.80 & 25/10/97 & 21:08 & 21:19 & 1 &\ 621 & N \\
LQ Peg        & NL VY SH NS & 3.22 & 26/10/97 & 22:37 & 23:18 & 2 & 2400 & N \\
V Per         & Na NL SW?   & 2.57 & 27/10/97 & 01:06 & 01:48 & 2 & 2400 & N \\
\textbf{GK Per}        & \textbf{Na DN IP}    & \textbf{47.92} & \textbf{25/10/97} & 
\textbf{06:14} & \textbf{06:34} & \textbf{1} & \textbf{1200} & \textbf{Y} \\
AY Psc        & DN ZC NS    & 5.21 & 25/10/97 & 23:51 & 00:53 & 2 & 3600 & N \\
VY Scl        & NL VY       & 3.98 & 24/10/97 & 23:22 & 00:03 & 2 & 2400 & N \\
VZ Scl        & NL VY SW    & 3.47 & 24/10/97 & 22:47 & 23:39 & 2 & 3000 & N \\
SW Sex        & NL UX SW    & 3.24 & 26/10/97 & 05:15 & 05:56 & 2 & 2400 & N \\
RW Tri        & NL UX SW    & 5.57 & 25/10/97 & 01:29 & 02:30 & 2 & 3600 & N \\
DW UMa        & NL SW SH NS & 3.28 & 27/10/97 & 05:09 & 05:50 & 2 & 2400 & N \\
\hline\hline
\end{tabular}
\end{center}
\label{tab:journal}
\end{table*}

\subsection{Isaac Newton Telescope}

\subsubsection{Observations}

In support of our WHT observations, we also examined known CVs in the
2.5m Isaac Newton Telescope (INT) Photometric H$\alpha$ Survey of the
Northern Galactic Plane (IPHAS). IPHAS is a 1800 deg$^2$ survey of the
northern Milky Way spanning the galactic latitude range $-5^{\circ} <
b < +5^{\circ}$ and galactic longitude range $29^{\circ} < l <
215^{\circ}$. Three filters were used, H$\alpha$, Sloan $r'$ and Sloan
$i'$, reaching down to $r'\approx 20\,\,(10\sigma)$. The survey took
place between 2003 and 2008. The survey used the INT Wide Field Camera
(WFC) which offers a pixel scale of $0.33\arcsec$ per pixel and a
field of view of $\sim34\arcmin \times 34\arcmin$. Exposure times were
initially set at 120\,s (H$\alpha$) and 10\,s ($r'$ and $i'$) but
evaluation of the early data led to an increase in the $r'$--band
exposure time to 30\,s -- for full details of the observations and
data reduction see \citet{drew05} and \citet{barentsen14}.

\subsubsection{Target selection}

We cross-matched the RK catalogue to the IPHAS footprint. There were
74 matches of CVs with the classification N, NL or DN (indicating
nova, nova-like variable and dwarf nova, respectively). Each matching IPHAS
field was reviewed visually to determine whether any nebulosity was
apparent around the target CVs. Due to the significant H$\alpha$
nebulosity in the Galactic plane, we did not attempt to compute
radial profiles for the IPHAS targets.

The 74 systems we examined in IPHAS are listed in
Tab.~\ref{tab:iphastab1}. The targets comprised 2 asynchronous polars,
10 polars \& intermediate polars, 5 NLs, 34 DNe, 3 old novae with
known shells and 20 old novae without known shells. Three of the NLs,
V1315 Aql, V363 Aur and V751 Cyg, were also part of our WHT sample, as
was BT Mon, an old nova with a known shell.

\begin{table*}
\caption[]{List of CVs examined in the IPHAS database. The
  classifications of the CVs have been taken from the RK
  catalogue. Note that the RK catalogue classifications for novae are
  N, Na, Nb. All shells detected in IPHAS are shown in bold and are
  discussed in Sect.~3.2.  $^*$We confirmed the detection of a shell
  around V1315 Aql with additional INT observations -- see section
  \ref{v1315}. $^\dagger$This system is a NL. $^\ddagger$Whilst these
  objects are shown as possible Z Cam systems in the RK catalogue,
  \citet{simonsen14} found that they do not exhibit the standstills
  necessary for this classification and that they are actually normal
  DNe.}
\begin{center}
\begin{tabular}{lllcclllc}
\hline\hline
& & & & & & & &  \\
\multicolumn{1}{l}{Object} &
\multicolumn{1}{l}{Classification} &
\multicolumn{1}{l}{Orbital} &
\multicolumn{1}{c|}{Visible} &
\multicolumn{1}{c|}{\hspace*{1.0cm}} &
\multicolumn{1}{l}{Object} &
\multicolumn{1}{l}{Classification} &
\multicolumn{1}{l}{Orbital} &
\multicolumn{1}{c}{Visible} \\
& & \multicolumn{1}{c}{period (hrs)} & \multicolumn{1}{c}{shell?} & & & & 
\multicolumn{1}{c}{period (hrs)} & \multicolumn{1}{c}{shell?} \\

& & & & & & & & \\
\hline
CI Aql      & Nr          & 14.83 & N & & V2468 Cyg   & Na           & 3.49 & N \\   
KX Aql      & DN SU       & 1.45 & N & & V2491 Cyg   & Na            & 2.56 & N \\ 
V368 Aql    & Na          & 16.57 & N & & V446 Her    & Na DN        & 4.97 & N \\
V603 Aql    & Na SH NS    & 3.32 & N & & CP Lac      & Na SW?        & 3.48 & N \\
\textbf{V1315 Aql}   & \textbf{NL UX SW}   & \textbf{3.35} & \textbf{Y$^*$} & & DI Lac      
& Na           & 13.05 & N \\
V1425 Aql   & Na NL? IP?  & 6.14 & N & & \textbf{BT Mon}      & \textbf{Na SW }        
& \textbf{8.01} & \textbf{Y}\\
V1493 Aql   & Na          & 3.74 & N & & CW Mon      & DN UP IP?     & 4.24 & N \\
V1494 Aql   & Na          & 3.23 & N & & V902 Mon    & NL IP         & 8.16 & N \\ 
FS Aur      & DN UG IP PW?& 1.43 & N & & V959 Mon    & N             & 7.10 & N \\
HV Aur      & DN SU       & 1.98 & N & & CZ Ori      & DN UG         & 5.25 & N \\
\textbf{T Aur}       & \textbf{Nb}          & \textbf{4.91} & \textbf{Y} & & V344 Ori    & 
DN ZC$^\ddagger$         & 5.62 & N \\
QZ Aur      & Na          & 8.58 & N & & FO Per      & DN UG? ZC?$^\ddagger$    & 4.13 & N \\ 
V363 Aur    & NL UX SW    & 7.71 & N & & FY Per      & NL VY         & 6.20 & N \\
AF Cam      & DN UG       & 7.78 & N & & TZ Per      & DN ZC         & 6.31 & N \\
FT Cam      & DN SU?      & 1.80 & N & & UV Per      & DN SU         & 1.56 & N \\
V705 Cas    & Na          & 5.47 & N & & V Per       & N NL SW?      & 2.57 & N \\
V709 Cas    & NL IP       & 5.33 & N & & WY Sge      & N DN? SW?     & 3.69 & N \\
V1033 Cas   & NL IP       & 4.03 & N & & DO Vul      & DN SU         & 1.38 & N \\
HT Cas      & DN SU       & 1.77 & N & & QQ Vul      & NL AM         & 3.71 & N \\
KP Cas      & DN SU       & 1.95 & N & & V405 Vul    & DN SU         & 2.71 & N \\
EM Cyg      & DN ZC       & 6.98 & N & & \textbf{V458 Vul}    & \textbf{Na }           & 
\textbf{1.64} & \textbf{Y}\\
EY Cyg      & DN UG SH?   & 1.10 & N & & V498 Vul    & DN SU WZ      & 1.41 & 
N \\
V337 Cyg    & DN SU       & 1.64 & N & & GD 552      & DN? WZ?       & 1.71 & N \\
V503 Cyg    & DN SU NS    & 1.87 & N & & Lanning 420 & DN SU         & 1.45 & 
N \\
V516 Cyg    & DN UG       & 4.11 & N & & J0130+6221  & DN?           & 3.12 & N \\
V550 Cyg    & DN SU       & 1.62 & N & & J0345+5335  & CV DN?        & 7.53 & N 
\\
V751 Cyg    & $^\dagger$VY SW? NS SS & 3.47 & N & & J0506+3547  & DN SU         & 
1.62 & N \\
V1251 Cyg   & DN SU WZ    & 1.77 & N & & J0518+2941  & NL?           & 5.72 & N 
\\
V1316 Cyg   & DN SU       & 1.78 & N & & J0524+4244  & NL AM AS      & 2.62 & 
N \\
\textbf{V1363 Cyg}   & \textbf{DN  ZC?$^\ddagger$}    & \textbf{2.42} & \textbf{Y}? & & 
J0619+1926  & DN SU WZ     & 1.34 & N \\
V1454 Cyg   & DN SU       & 1.36 & N & & J1853-0128  & NL IP         & n/a & N \\
\textbf{V1500 Cyg}   & \textbf{Na NL AM AS} & \textbf{3.35} & \textbf{Y} & & 
J1915+0719  & DN SU  WZ     & 1.37 & N \\
V2274 Cyg   & Na          & 7.20 & N & & J1926+1322  & NL IP         & 4.58 & N \\
\textbf{V2275 Cyg}   & \textbf{Na IP?}      & \textbf{7.55} & \textbf{Y}? & & 
J1953+1859  & DN SU?       & 1.44 & N \\
V2306 Cyg   & NL IP       & 4.35 & N & & J2133+5107  & NL IP         & 7.14 & N \\
V2362 Cyg   & Na          & 1.58 & N & & J2138+5544  & NL IP         & n/a  & N \\
V2467 Cyg   & Na NL IP?   & 3.83 & N & & J2250+5731  & NL AM         & 2.90 & 
N \\
\hline\hline
\end{tabular}
\end{center}
\label{tab:iphastab1}
\end{table*}

\section{Results}
\label{sec:res}

\subsection{WHT Images and Radial Profiles}
\label{im}

\begin{figure}
\centering
  \vspace{10pt}
\includegraphics[width=80mm,angle=0]{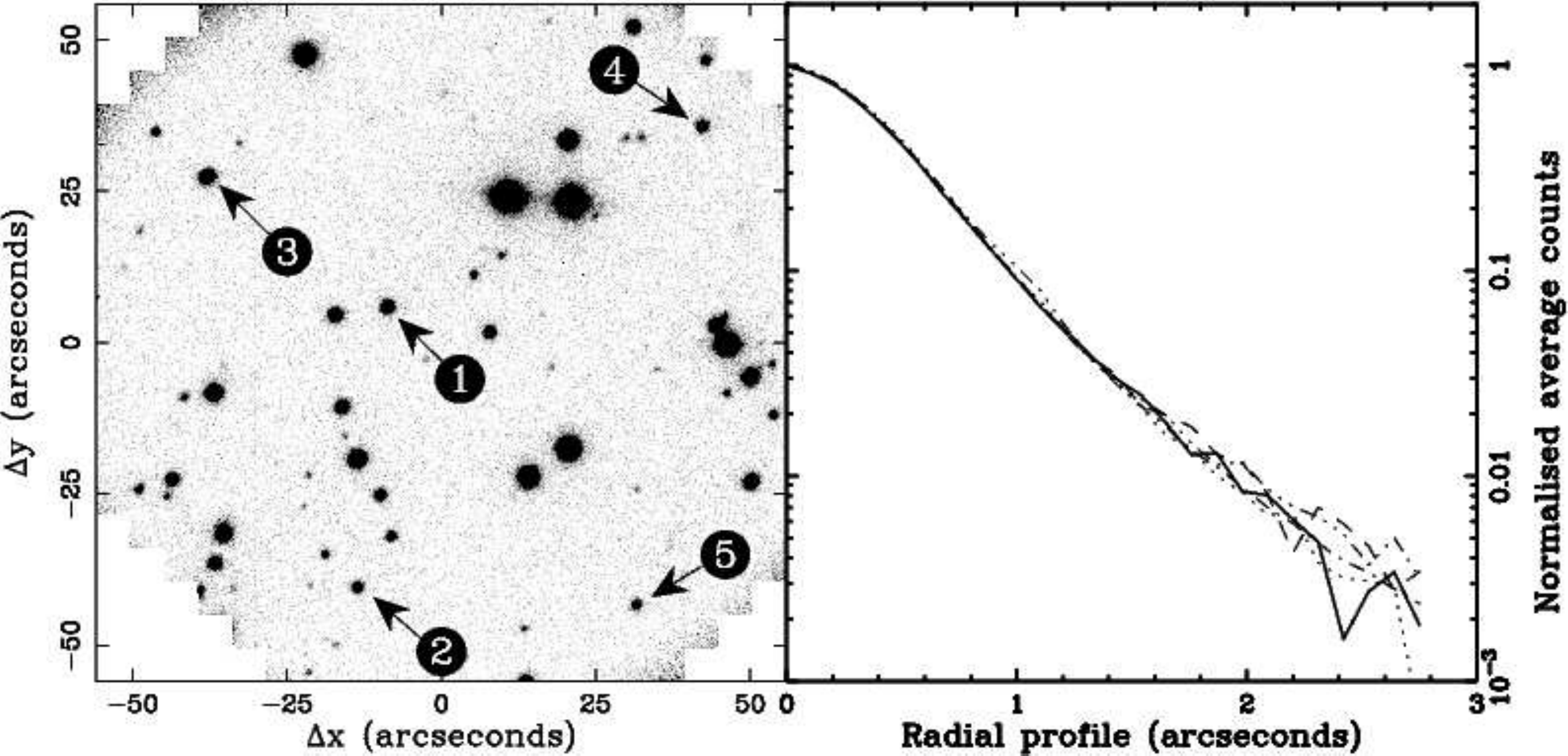}
\label{psf}
 \caption{Left: WHT image of V Per (Star number 1). Right: PSFs for
   the five arrowed stars showing the uniformity across the WHT
   chip. Each numbered star has been plotted as follows 1 - solid
   line, 2 - dashed, 3 - dot-dashed-dot-dashed, 4 - dotted, 5 -
   dashed-dot-dot-dot-dashed. The orientation of the image is the same
   as that shown in Fig.~\ref{junk4}.}
\label{psftest}
\end{figure}

In order to detect shells in the WHT images, we adopted two
strategies. First, we visually examined each image to determine if a
shell is visible. This technique would reveal wide shells with
diameters of more than a few arcseconds. Second, we calculated the
radial profile of each CV and compared it to a number of field stars
in the same image. Any nebulosity around the CV due to a nova shell
would cause the radial profile of the CV to lie above the average
profile of the field stars (for example, see the radial profile of BT
Mon in Fig.~\ref{fig4}). This technique can reveal shells with
diameters of less than a few arcseconds, and was successfully used by
\citet{gill98} to discover four new nova shells. A key assumption in
this technique is that the Point Spread Function (PSF) is uniform
across the WHT chip. Fig.~\ref{psftest} shows the PSFs for five
field stars (arrowed) in the image of V Per. The PSFs show identical
radial profiles irrespective of field position, giving confidence that
the PSFs are uniform across the field of view of the CCD, as expected.

The centroids of the stars were first measured by fitting a
two-dimensional Gaussian. The radial profiles were then generated by
calculating the radial distance of each pixel from the centroid, and
then averaging the fluxes of the pixels falling into bins of
increasing radial distance from the centroid. The radial profiles were
then normalised to unity, and plotted from the centre of the star
until the flux reached $1\sigma$ above the mean background flux.

In Appendix \ref{app1} we show the images and radial profiles for all of
the objects that were observed with the WHT. As expected, the images
for the three old novae with previously known shells (BT Mon, DQ
Her, GK Per) clearly show a shell and each is discussed briefly
below. There are no visible shells in the images of the remaining
objects, nor do any of the radial profiles differ significantly from
the field stars.

\subsubsection{BT Mon}
\label{btmon}

The shell around BT Mon (Nova Mon 1939) was discovered
spectroscopically by \citet{marsh83}.  BT Mon is a high-inclination
system and the system parameters were derived by \citet{smith98}.  The
first image of the shell was reported by \citet{duerbeck87b}, who
found it to be an incomplete clumpy, slightly elliptical ring
with approximate dimensions of $11\arcsec \times 9\arcsec$ and the
major axis pointing in the NW--SE direction. 

Our image and radial profile of BT Mon is shown in
Fig.~\ref{fig4}. The lower right quadrant was not used to calculate
the radial profile in order to remove the flux from the nearby
star. The radial profile of BT Mon clearly deviates from the profile
of the field stars, from approximately 4$\arcsec$ outwards. This is
due to the presence of the shell and gives assurance that our
technique for identifying shells is valid.

\begin{figure}
\centering
  \vspace{10pt}
\includegraphics[width=80mm,angle=0]{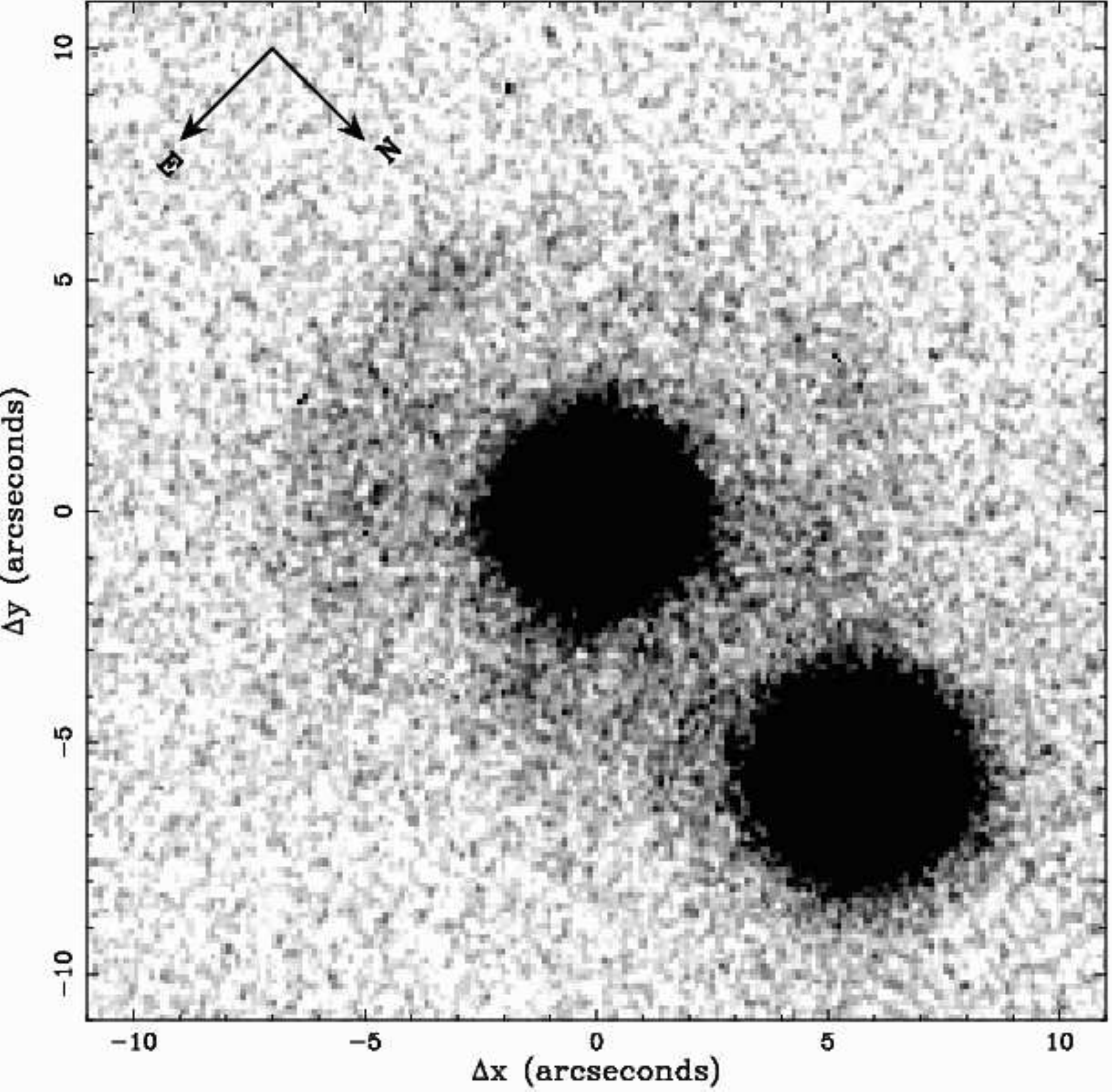}
 \caption{Enlarged image of the BT Mon nova shell. Note that the
   second star to the lower right is an unassociated
   foreground/background star.}
\label{junk4}
\end{figure}

In Fig.~\ref{junk4} we show an enlarged version of our image of the
BT Mon shell. We estimate the shell diameter to be $13\arcsec \pm
1\arcsec$. Assuming a constant shell expansion velocity of $1800 \pm
300$ km\,s$^{-1}$ and a distance of $1.8 \pm 0.1$ kpc as derived by
\citet{marsh83}, together with the date of the nova as $1939.7$, gives
an expected diameter of $12 \pm 3\arcsec$ at the time of our
observations, in agreement with our measured value.

\subsubsection{DQ Her}
\label{dqh}

DQ Her (Nova 1934) is an intermediate polar with system parameters
derived by \citet{horne93}. The nova shell (see Fig.~\ref{fig4}) is a
prolate ellipsoid with a slightly pinched central
ring. \citet{vaytet07} used our WHT image of DQ Her to estimate the
angular size and hence distance of the system. They measured the
angular size of the major and minor axes to be $a=25.31 \pm
0.44\arcsec$ and $b= 18.70 \pm 0.44\arcsec$ . Assuming a constant
expansion velocity of $370 \pm 14$ km\,s$^{-1}$, they derived a
distance of $d=525 \pm28$ pc.

\subsubsection{GK Per}
\label{gkper}

The nova GK Per (1901) is the archetypal nova remnant and has been
extensively studied (see \citet{shara12} for a review). The shell is
boxy in shape, of size approximately $100\arcsec \times 90\arcsec$ and
exhibits clumpy knots (see Fig.~\ref{fig5}). Recently,
\citet{liimets12} derived a three dimensional model of the nova shell
in GK Per, and determined the proper motion and radial velocities of
more than 200 knots in the ejecta. The knots have a wide range of
velocities (600--1000 km\,s$^{-1})$ and have suffered only modest
deceleration.  \citet{shara12} used HST images from 1995 and 1997 to
resolve over 1000 filamentary structures in the ejecta. They also
investigated a jet-like feature, first discovered by
\citet{anupama93}, which they suggest could be the shock interaction
of a collimated flow with the ISM, probably originating from the
accretion disc. The jet extends some 2.7$\arcmin$ to the NW, which is
larger than the field of view of our image. We examined our image of
GK Per but could not find any evidence of the jet-like feature on
smaller spatial scales, most probably due to the lower signal-to-noise
ratio of our image.

\subsection{IPHAS images}

We visually examined the IPHAS images for evidence of nova
shells. Tab.~\ref{tab:iphastab1} lists all of the objects we examined,
and indicates whether a shell is visible. The short exposure times of
the H$\alpha$ images (120\,s) means that only bright, nearby shells
are likely to be visible. We found three old novae with shells that
are visible in the IPHAS footprint: T Aur (Nova Aurigae 1891), V458
Vul (Nova Vul 2007 No. 1) and V1500 Cyg (Nova Cygni 1975), all of
which are well studied systems.  We briefly review these objects
below. We found no definite detections of shells around any other
IPHAS targets with the exception of two systems, V1363 Cyg and V1315
Aql, as discussed below. We did discover a nebula around V2275 Cyg,
which is too large to be associated with its nova event in 2001. This
nebula may be a light echo due either to scattering off, or
flash ionisation of, a pre-existing nebula. We also discuss this
object further below.


\begin{figure*}
\centering 
\subfigure[T Aur, H$\alpha$]{
  \includegraphics[width=55mm,angle=0]{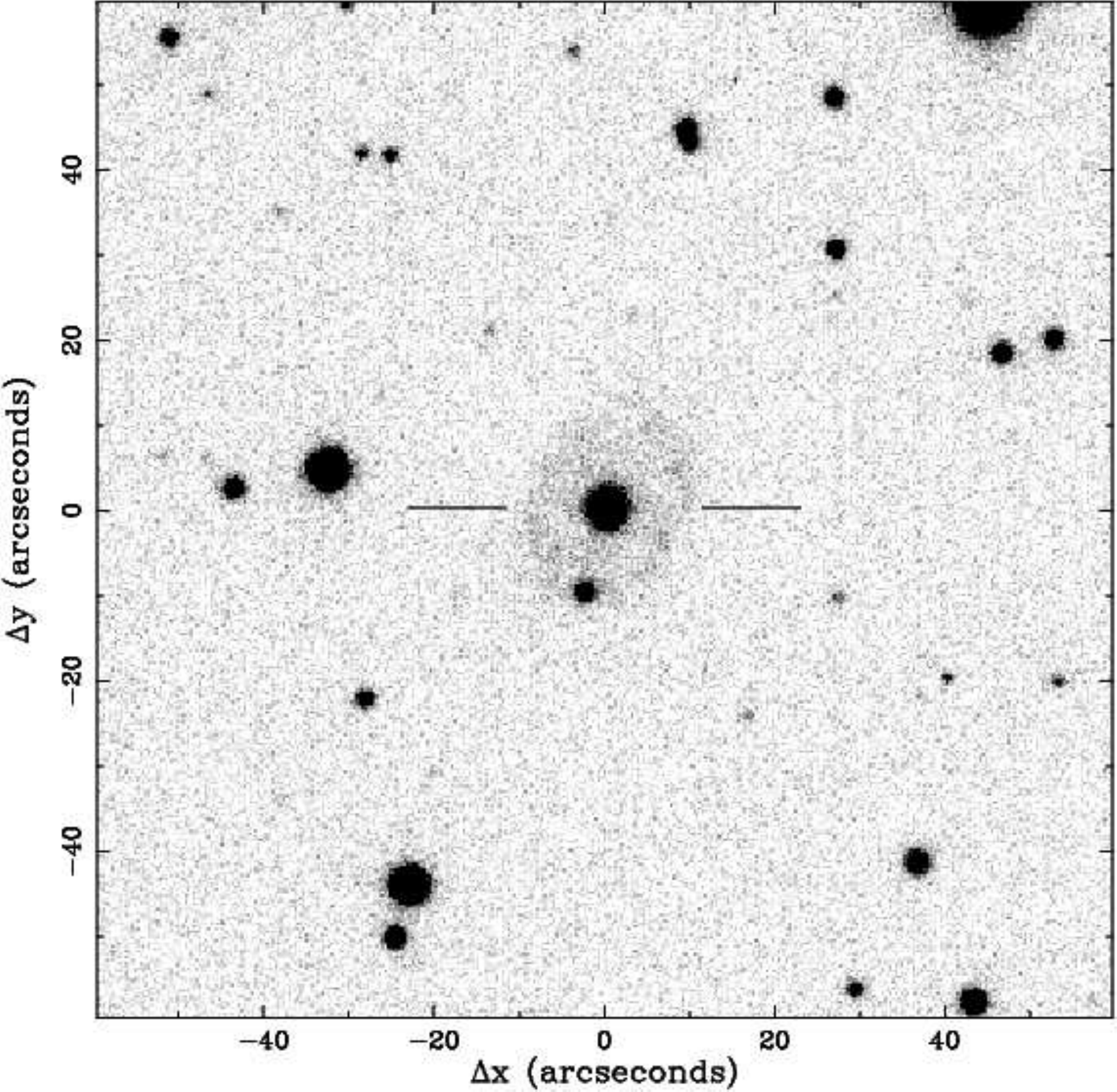}
\label{fig:taur}}
\subfigure[V458 Vul,  H$\alpha$]{
  \includegraphics[width=55mm,angle=0]{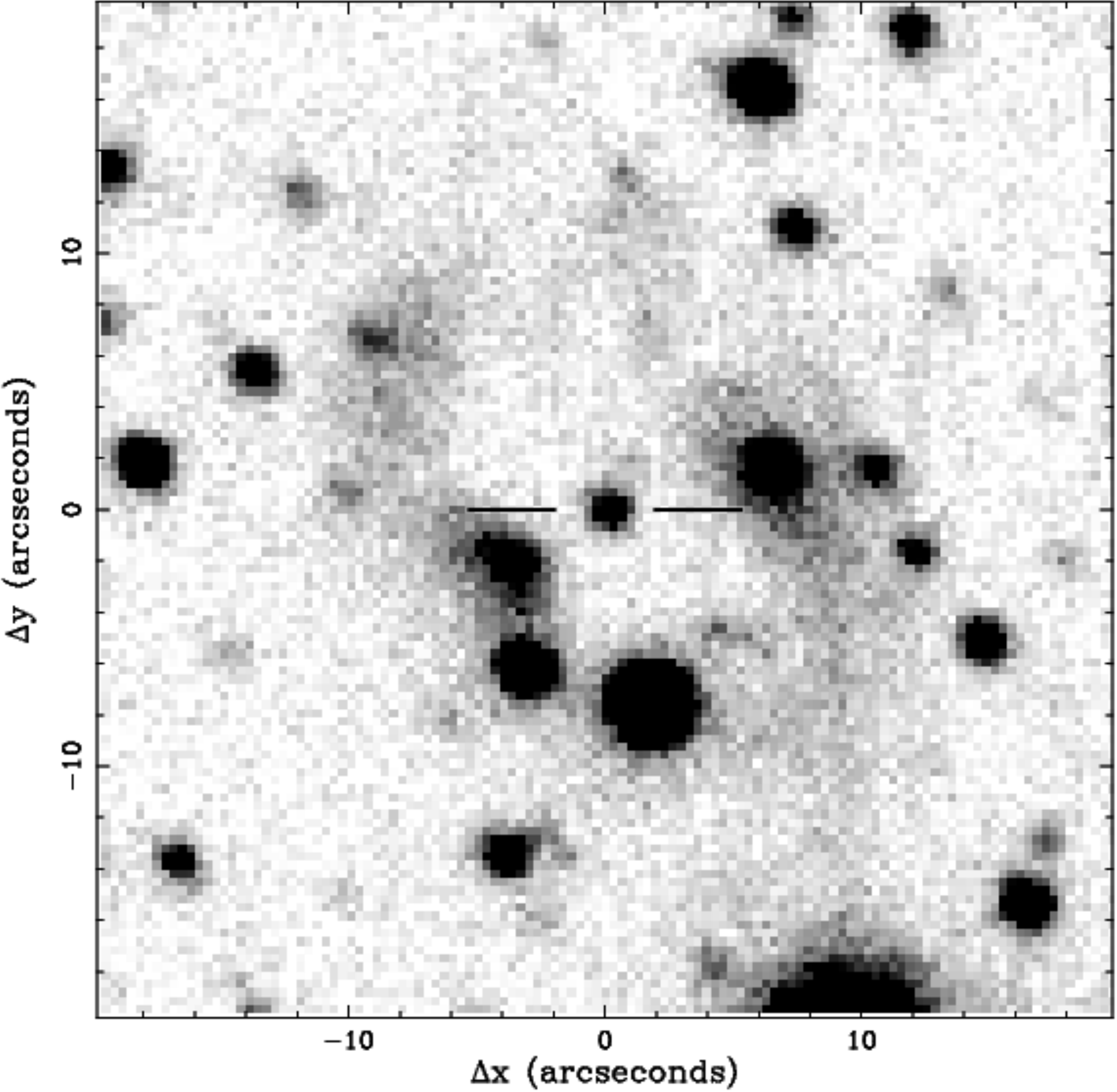}
\label{fig:v458vul}}
\subfigure[V1500 Cyg,  H$\alpha$]{
  \includegraphics[width=55mm,angle=0]{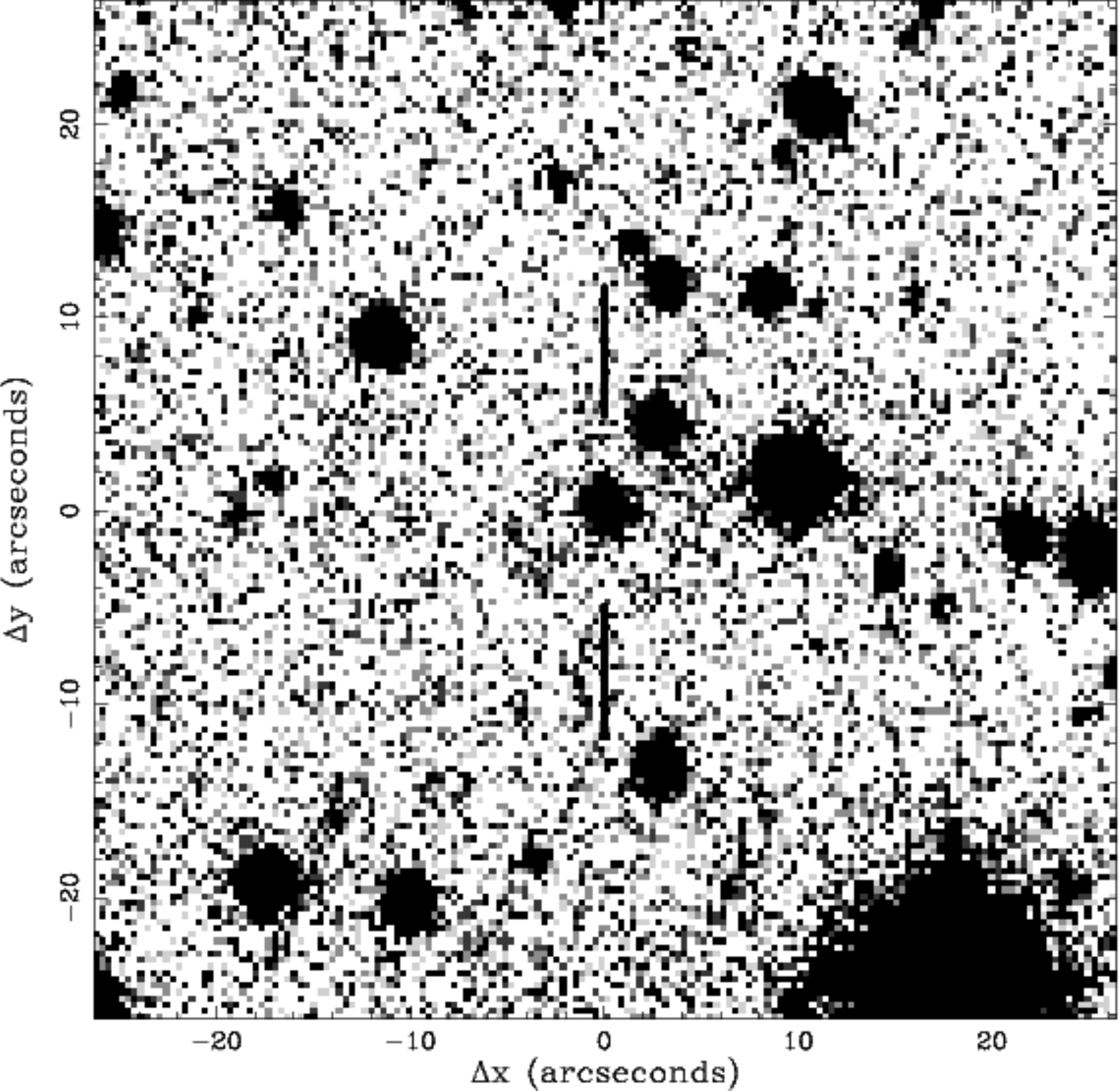}
\label{fig:v1500cygha2}}

\centering
\subfigure[V1363 Cyg, H$\alpha$]{
  \includegraphics[width=55mm,angle=0]{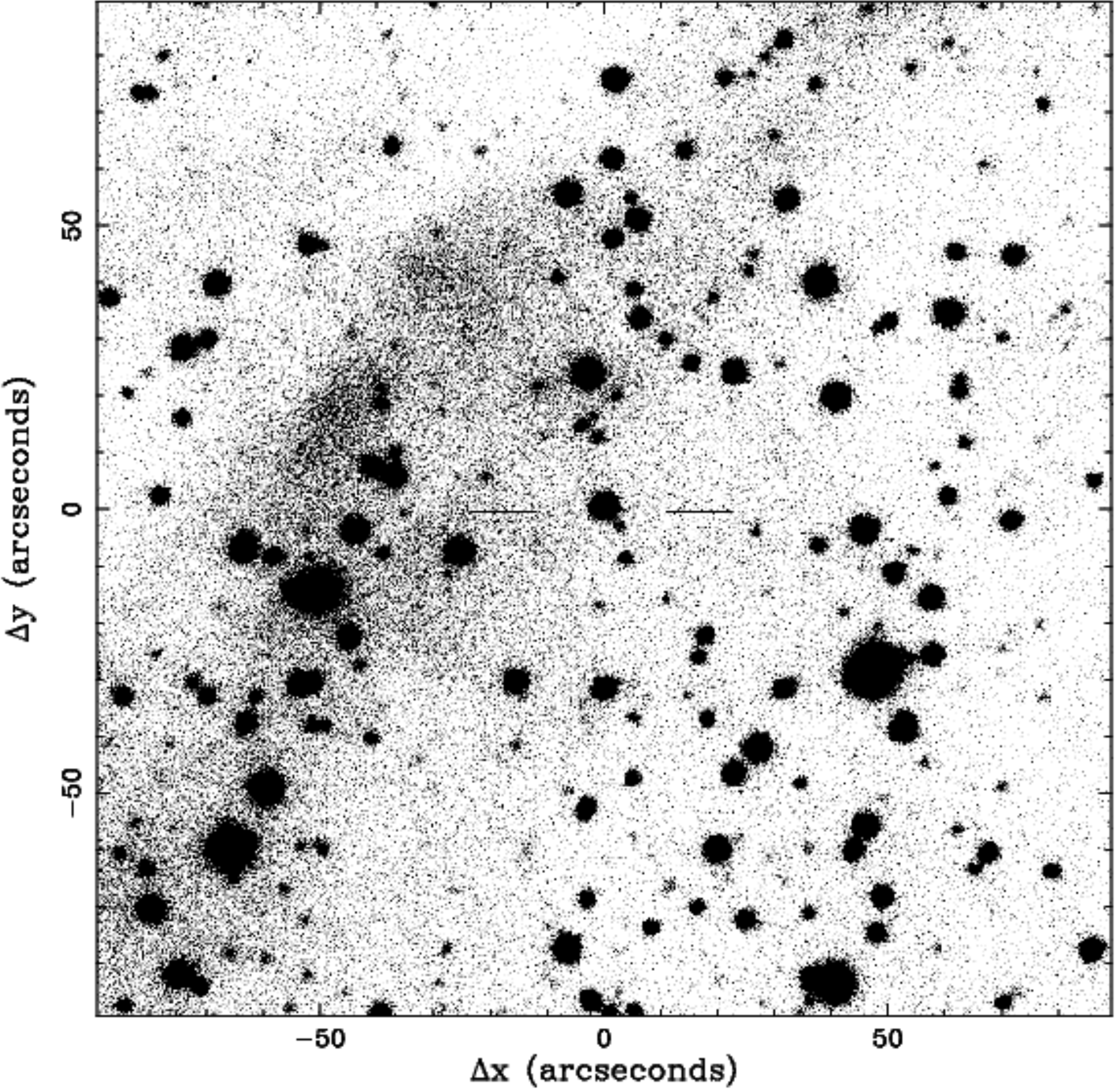}
\label{fig:v1363}}
\subfigure[V1363 Cyg, H$\alpha-r$ wide field]{
  \includegraphics[width=55mm,angle=0]{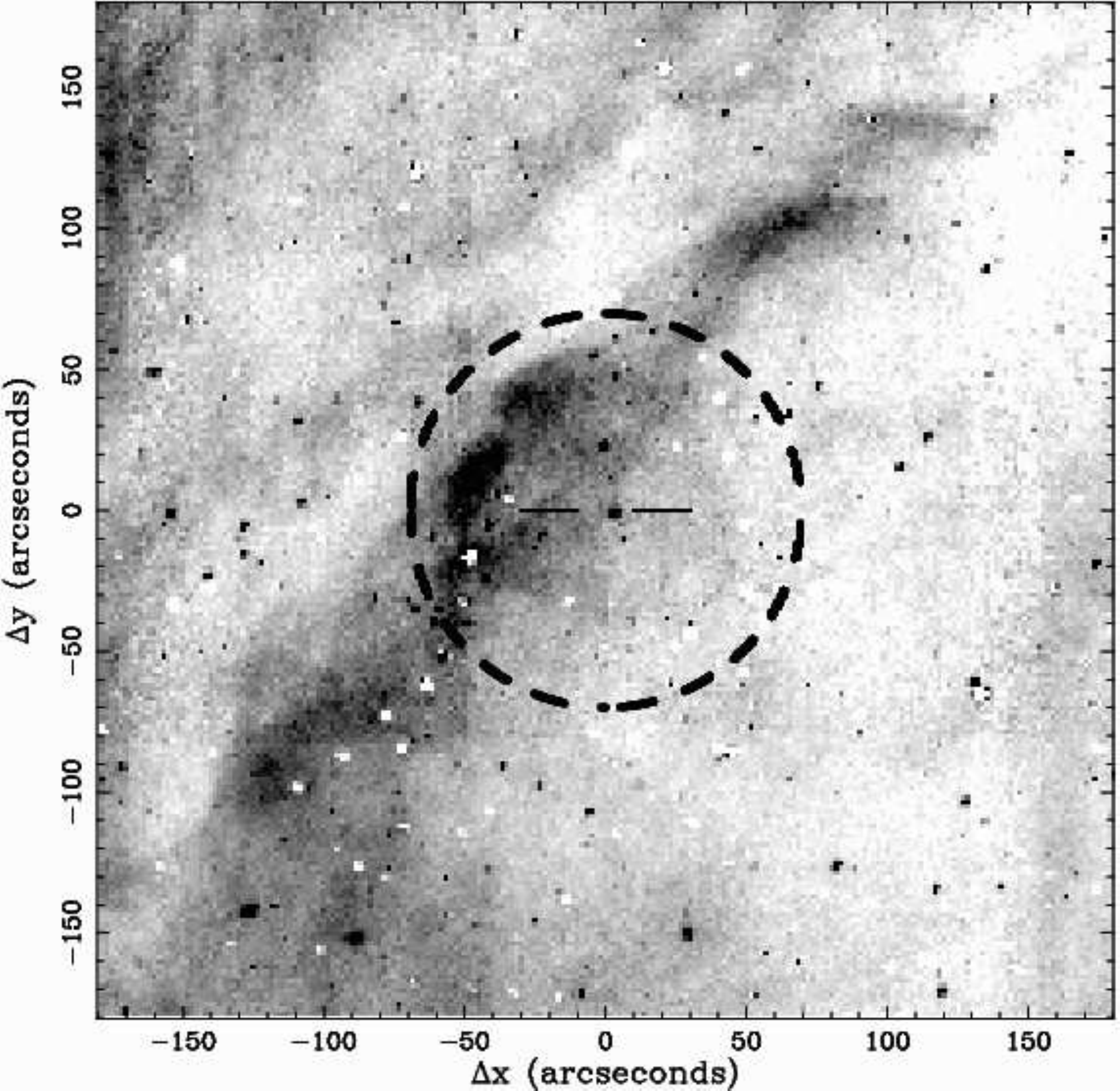}
\label{fig:v1363ha-r3}}
\subfigure[V1363 Cyg, H$\alpha-r'$ zoom]{
  \includegraphics[width=55mm,angle=0]{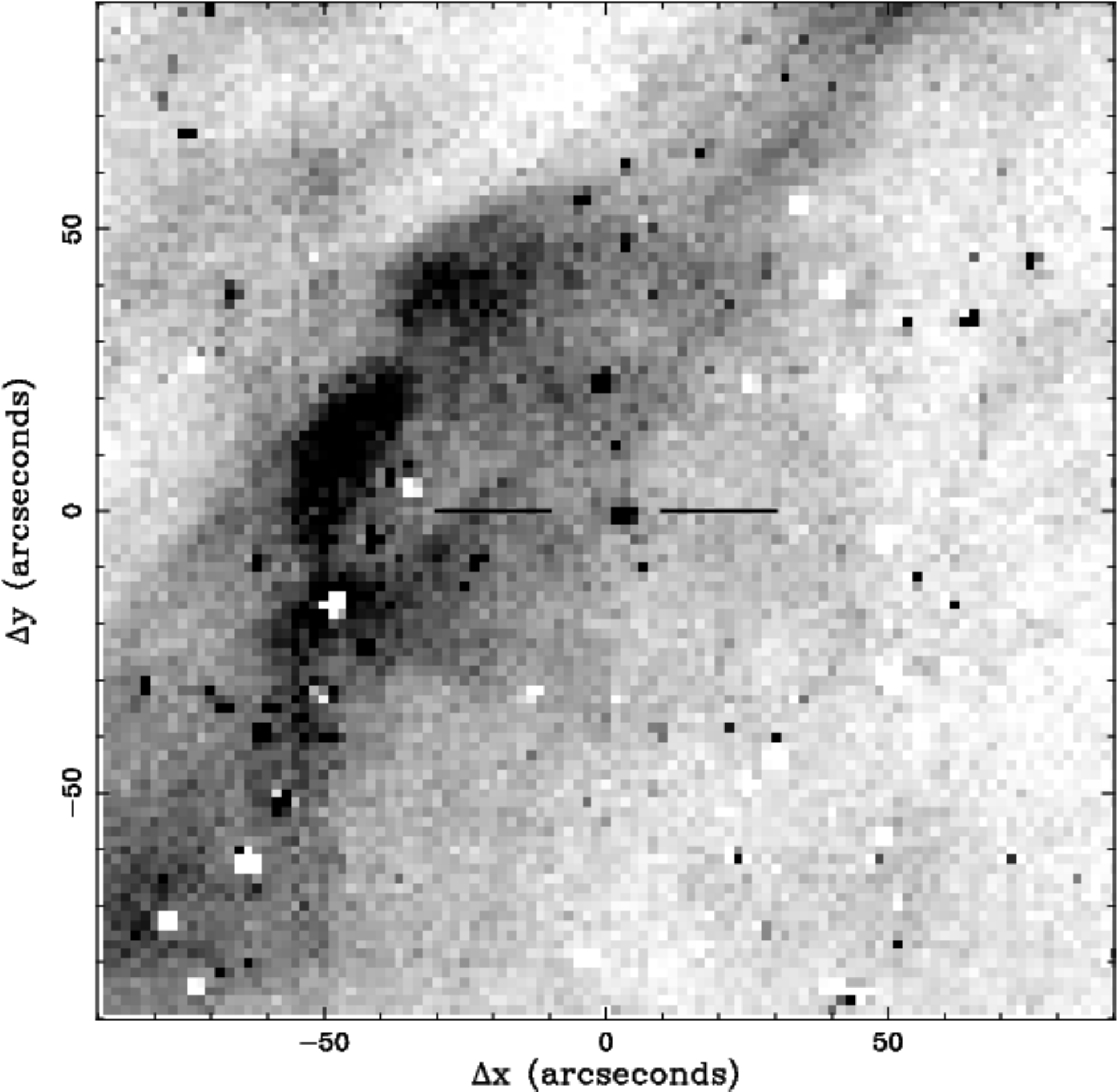}
\label{fig:v1363ha-r2}}

\centering
\subfigure[V1315 Aql, H$\alpha$]{
  \includegraphics[width=55mm,angle=0]{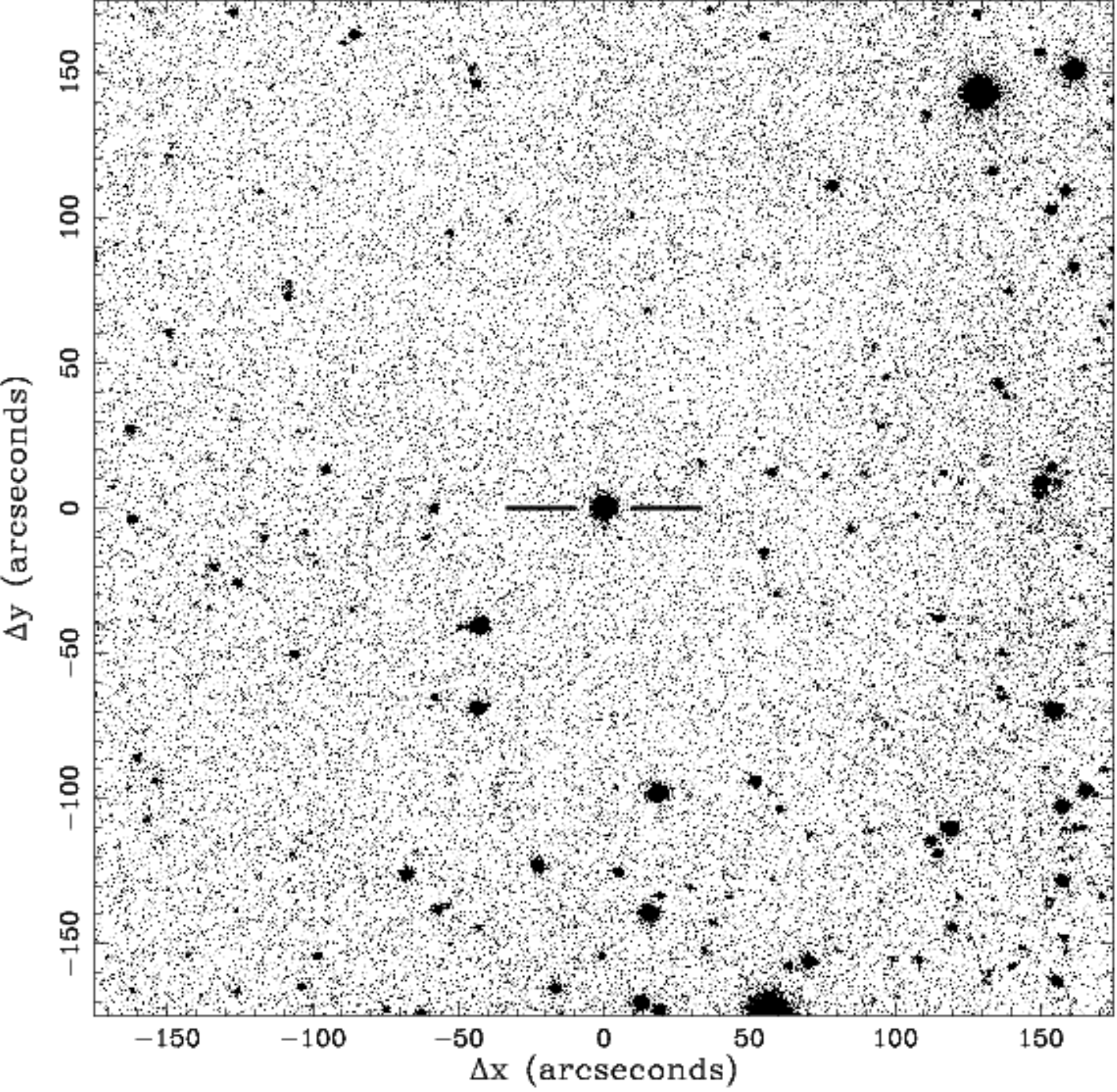}
\label{fig:v1315}}
\subfigure[V1315 Aql, H$\alpha-r'$]{
  \includegraphics[width=55mm,angle=0]{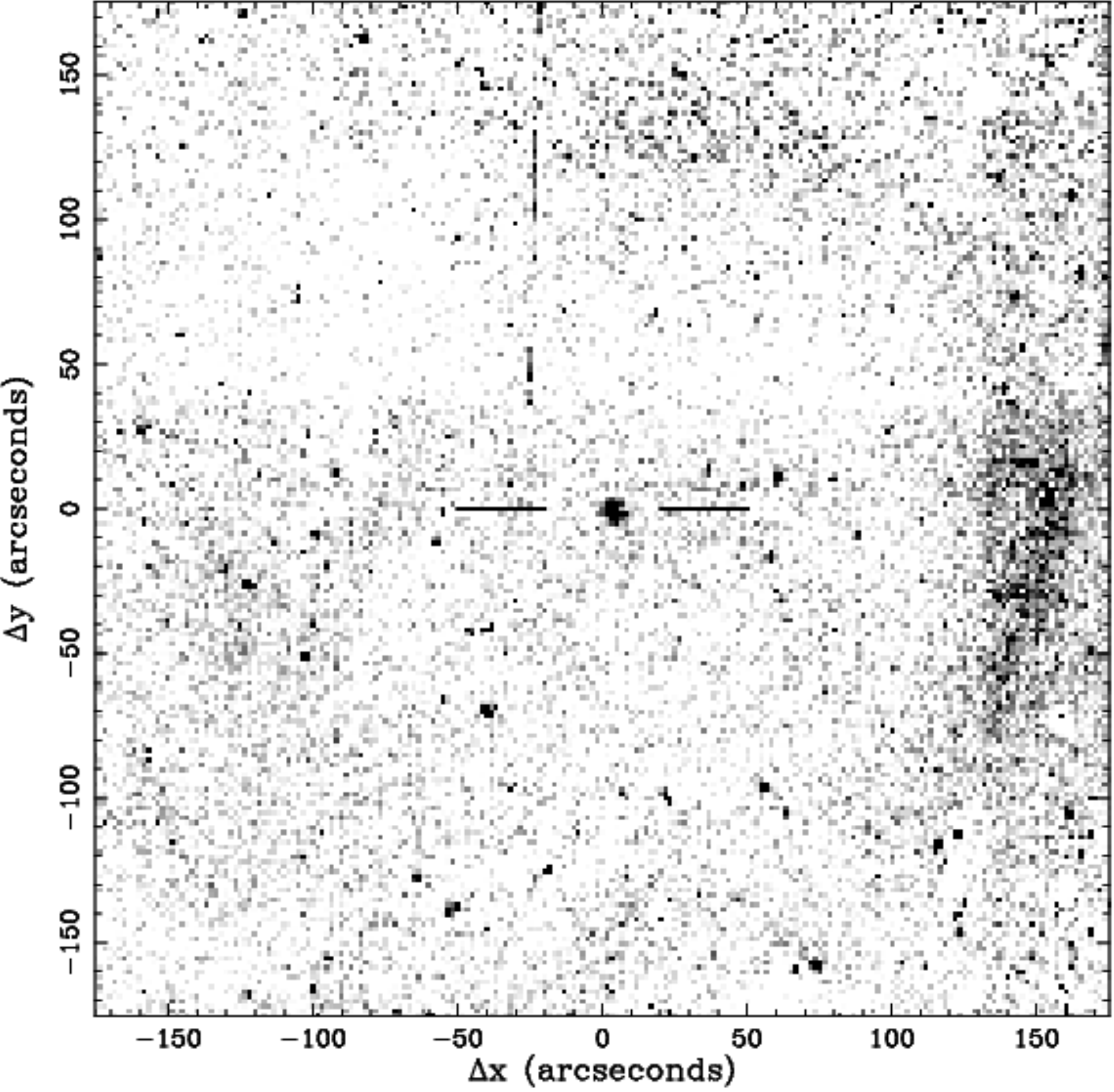}
\label{fig:v1315ha-r}}
\caption{IPHAS H$\alpha$ and H$\alpha-r'$ images. In all images,
  North is up, East is left.}
\label{fig:iphasim}
\end{figure*}

\subsubsection{T Aur}

The IPHAS H$\alpha$ image of T Aur is shown in
Fig.~\ref{fig:iphasim}a. The shell is clearly discernible in the image
giving us confidence that it is possible to see nova shells in the
IPHAS images. The shell structure has been likened to that of DQ Her,
although T Aur is some 43 years older (\citealt{slavin95}). The shell
is elliptical in shape, with major and minor axes of length $\sim
30\arcsec \times 20\arcsec$ respectively.

\subsubsection{V458 Vul}

The H$\alpha$ image of V458 Vul is shown in
Fig.~\ref{fig:iphasim}b. The shell has major and minor axes of
$\approx 30\arcsec \times 20\arcsec$. The image was taken in June
2007, two months before the system underwent a nova explosion in
August 2007. The shell is actually a pre-existing planetary nebula
ejected some 14,000 years ago \citep{wesson08}. The central binary is
most likely a post-double common-envelope binary comprised of a WD of
mass $\sim$ 1.0 M$_\odot$ and a post-AGB secondary of mass $\sim$ 0.6
M$_\odot$ \citep{rodriguez10}.

\subsubsection{V1500 Cyg}
\label{v1500}

V1500 Cyg (Nova Cygni 1975) is a well-studied nova and is the
archetypal asynchronous polar \citep{wade91}. The nova shell was first
imaged four years after outburst by \citet{becker80}, who measured the
radius at $\sim 1.0\arcsec$. Subsequently, \citet{wade91} presented an
image taken in 1987 by which time the shell had expanded to
$\sim1.9\arcsec$, giving an expansion rate of $0.16\arcsec$ per annum,
and \citet{slavin95} presented an image taken in 1993 showing a
nebular radius of $\sim3\arcsec$. The IPHAS H$\alpha$ image taken in
2004 is shown in Fig.~\ref{fig:iphasim}c. The nova shell is
extremely faint and has a radius of $\sim 5\arcsec$, still consistent
with the nebular expansion rate of $\sim 0.16\arcsec$ per annum given
by \citet{wade91}.

\subsubsection{V1363 Cyg}
\label{v1363}

In Fig.~\ref{fig:iphasim}d we show the IPHAS H$\alpha$ image of the
dwarf nova V1363~Cyg. The H$\alpha$ image is heavily populated with
field stars making the surrounding nebula difficult to discern. Hence
we show the H$\alpha-r'$ image in Fig.~\ref{fig:iphasim}e, which
effectively removes most of the flux from the field stars. The field
is extremely crowded and the object lies close to a ribbon of gas,
making the unambiguous detection of a nova shell extremely
difficult. However, there is a faint egg-shaped shell of emission of
$\approx2\arcmin$ diameter, approximately centred on the CV.

\subsubsection{V1315 Aql}
\label{v1315}

Figs.~\ref{fig:iphasim}f \& \ref{fig:iphasim}g show the IPHAS
H$\alpha$ and H$\alpha-r'$ images of V1315 Aql. There is a faint shell
of $\approx2.5\arcmin$ radius approximately centred on the CV, with
more pronounced emission towards the West. We also imaged this object
with the WHT.  However, the small field of view of our WHT image (see
Fig.~\ref{fig1}) is not large enough to confirm the possible
detection of this shell.

\begin{figure}
\centering
  \vspace{10pt}
\includegraphics[width=80mm,angle=0]{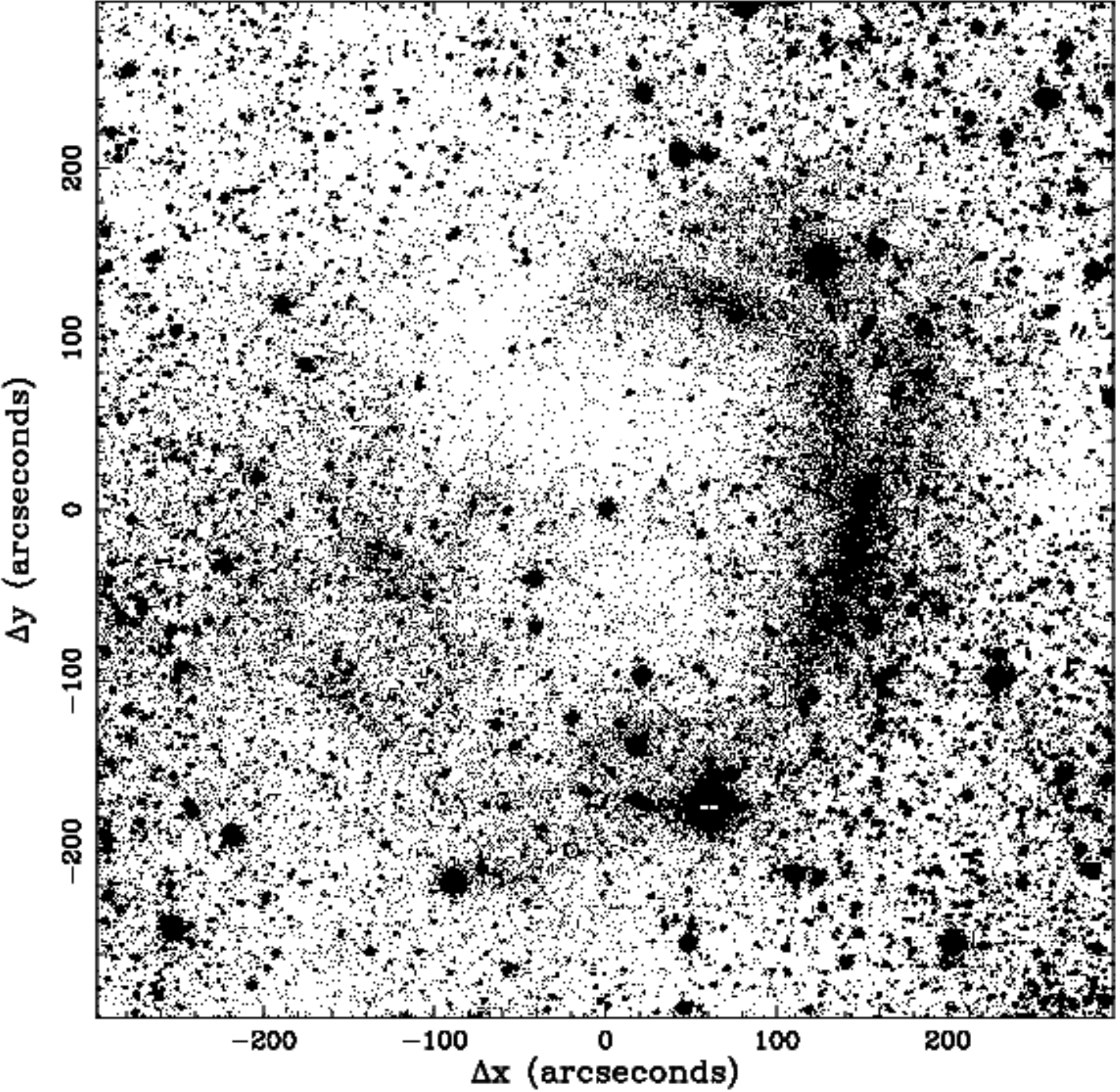}
 \caption{H$\alpha$ image of the nebula surrounding V1315 Aql. The
   image is 10$\arcmin \times 10\arcmin$, with North up and East to
   the left. V1315 Aql is the bright star located at the centre of the
   image.}
\label{v1315cut}
\end{figure}

In order to confirm the detection of the shell around V1315 Aql, we
took a further 13 exposures of V1315 Aql on 2014 August 2 with the WFC
on the INT with a total exposure time of 7200\,s in H$\alpha$.  The
stacked H$\alpha$ image is shown in Fig.~\ref{v1315cut}. The image
clearly shows a shell surrounding the central system, with a radius of
$\sim 3\arcmin$, confirming the proposed detection in
Fig.~\ref{fig:iphasim}. Assuming a shell expansion rate of 2000\,km
s$^{-1}$ and a distance of $356^{+65}_{-80}$ pc (\citealt{ak07}), this
means that V1315 Aql experienced a nova eruption $\sim120$ years ago.

We examined the historic records of nova sightings compiled by
\citet{ho62} and \citet{stephenson76}, but found nothing that
coincides with the position of V1315 Aql.

\subsubsection{V2275 Cyg}
\label{v2275}

\begin{figure*}
\centering \subfigure[November 2003]{
  \includegraphics[width=55mm,angle=0]{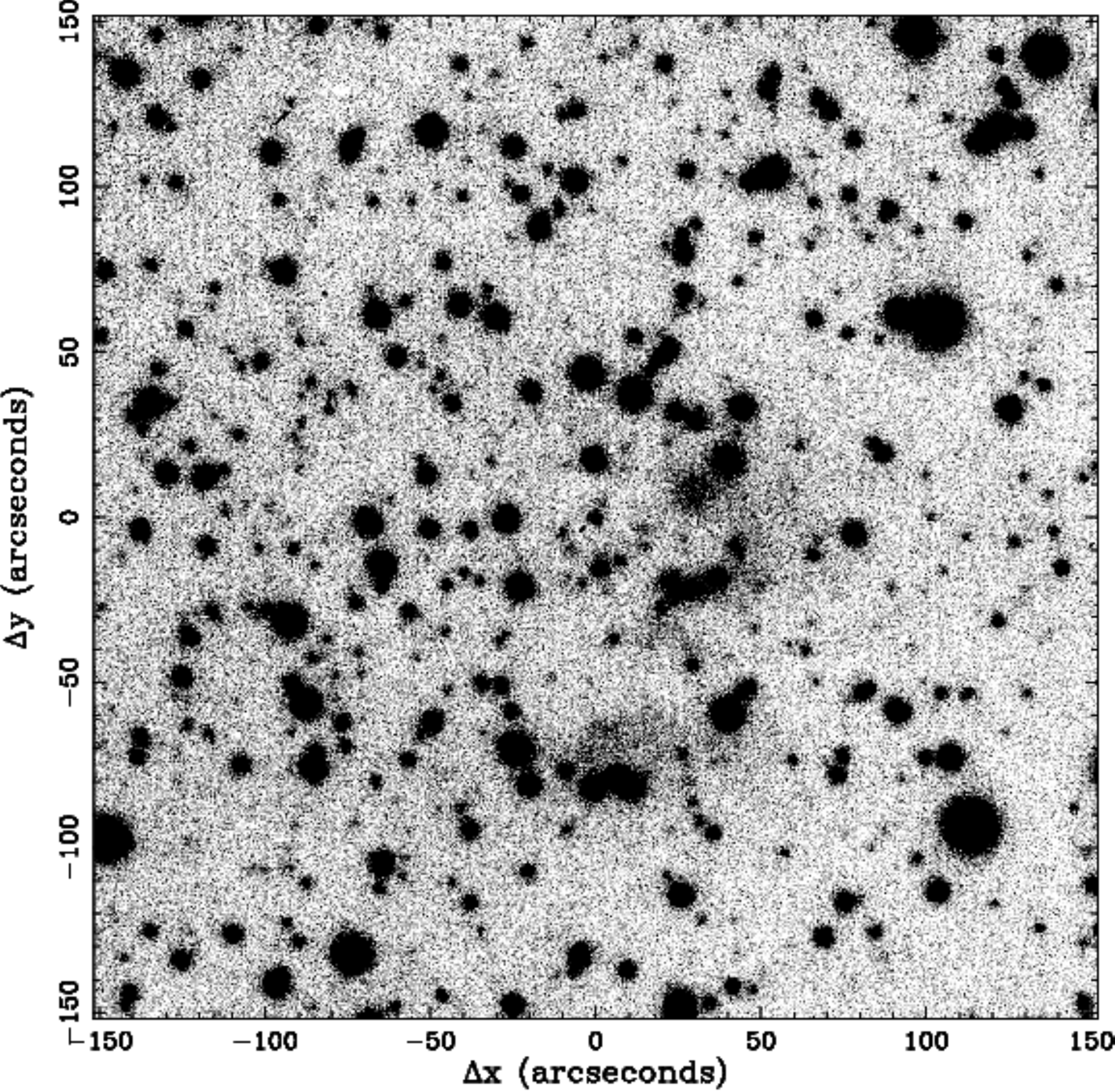}
\label{fig:r375}}
\subfigure[November 2005]{
  \includegraphics[width=55mm,angle=0]{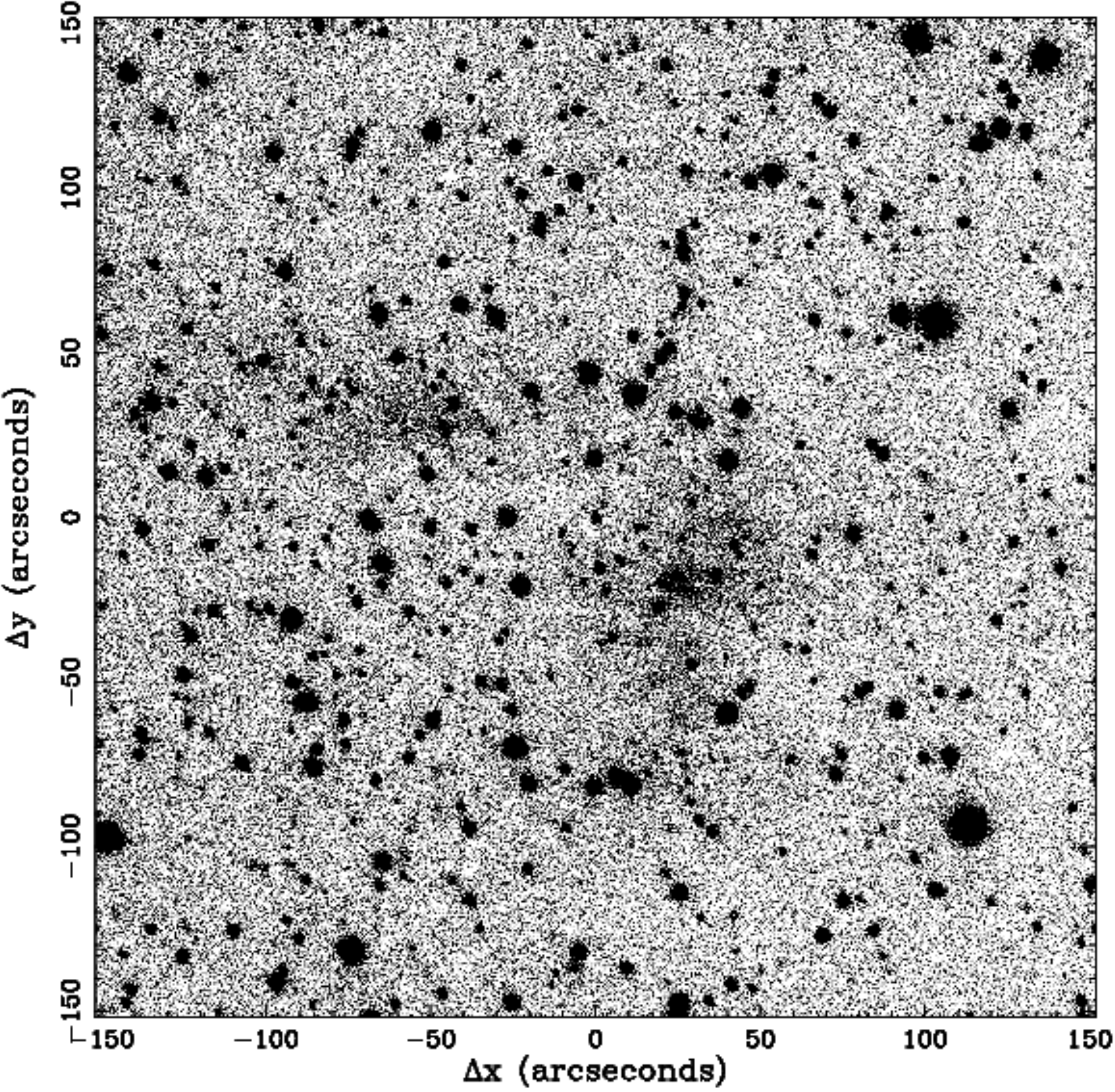}
\label{fig:r482}}
\subfigure[November 2006]{
  \includegraphics[width=55mm,angle=0]{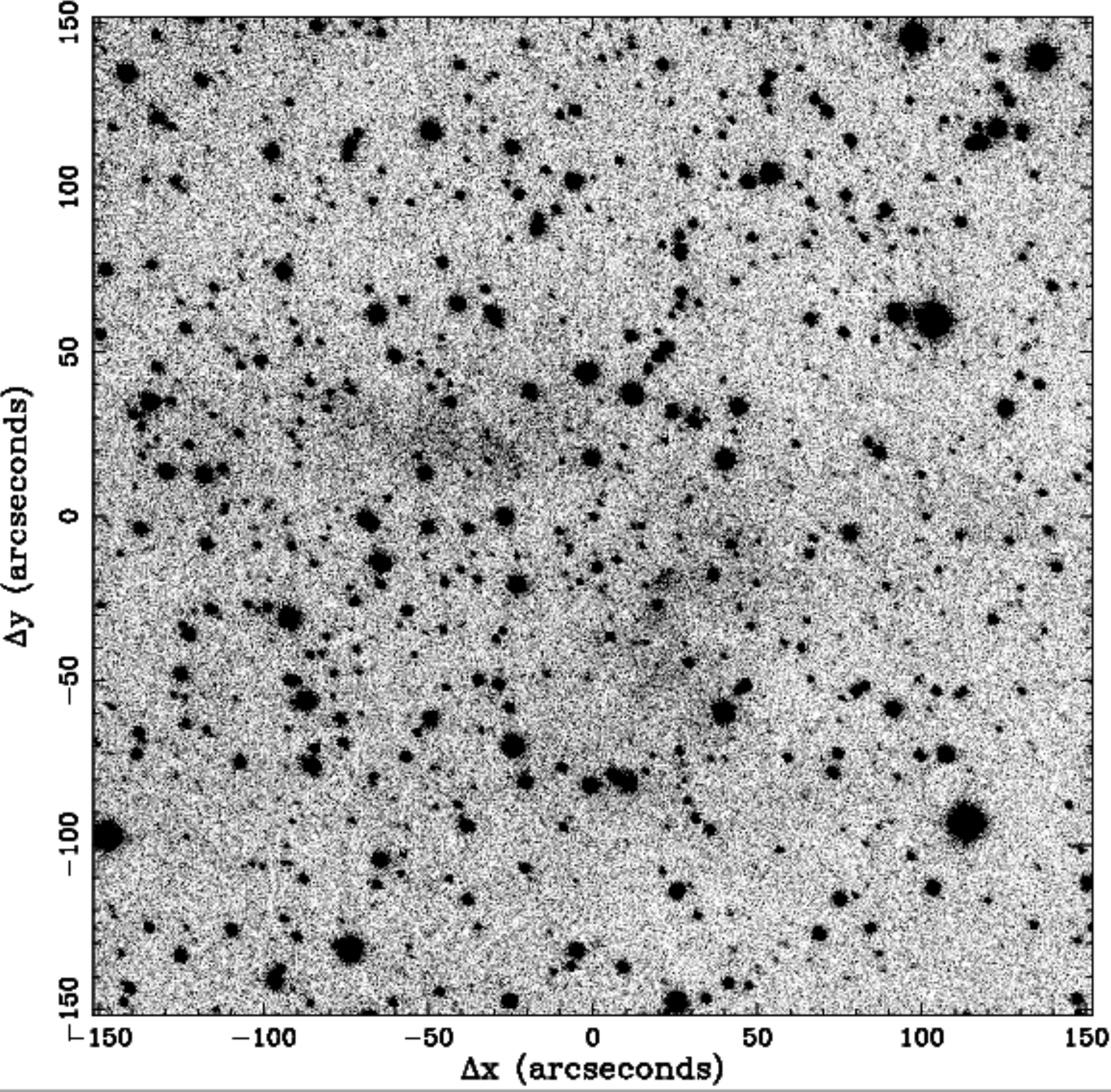}
\label{fig:r536}}

\centering
\subfigure[December 2008]{
  \includegraphics[width=55mm,angle=0]{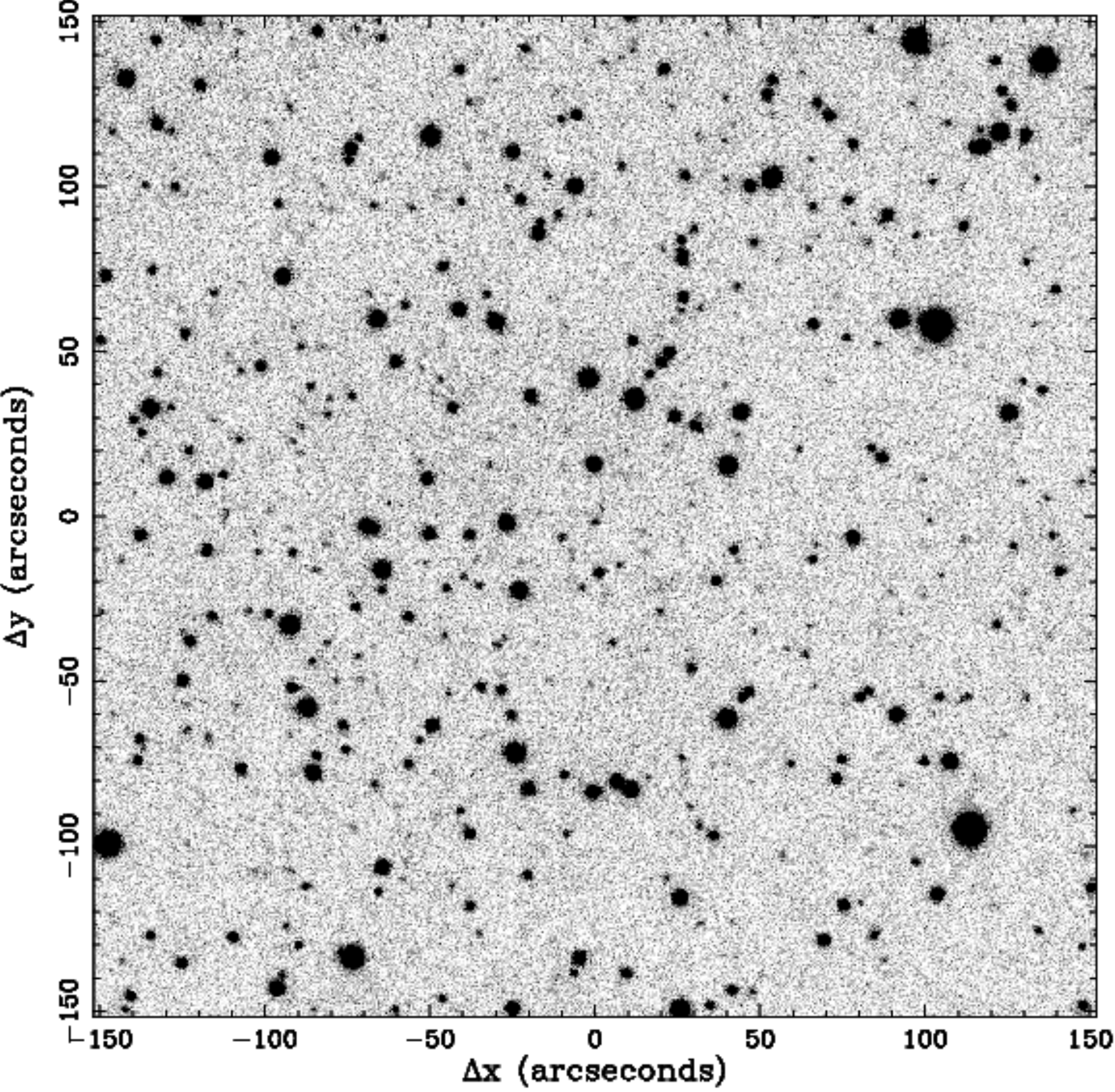}
\label{fig:r622}}
\subfigure[August 2009]{
  \includegraphics[width=55mm,angle=0]{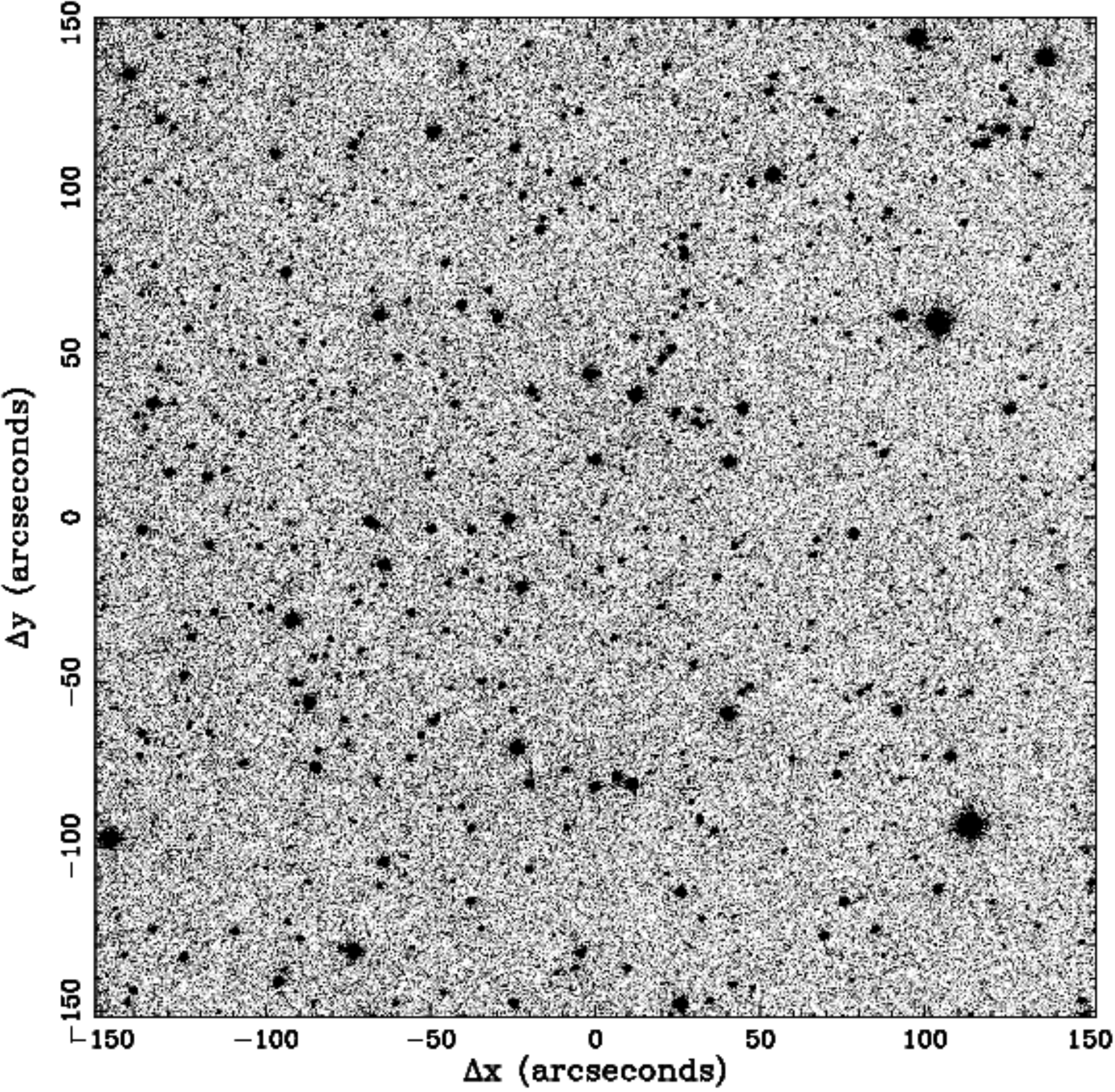}
\label{fig:r686}}
\caption{H$\alpha$ images of V2275 Cyg from IPHAS. A faint nebula is
  apparent in images (a)--(c) but is not present in images (d) and
  (e). In all images, North is up, East is left.}
\label{fig:v2275}
\end{figure*}

The nova eruption of V2275 Cyg (Nova Cygni 2001 No. 2) occurred on
2001 August 19 \citep{nakamura01}. The field around V2275 Cyg was
observed on five epochs during the IPHAS survey. The five images are
shown in Figure \ref{fig:v2275}. 

In the first three images, taken between November 2003 and November
2006, a nebula of $\approx2.7\arcmin$ diameter is clearly apparent,
but it has disappeared in the fourth and fifth images taken in
December 2008 and August 2009. Using the expansion velocity and
minimum distance derived by \citet{kiss02} of approximately 2000
km\,s$^{-1}$ and 3 kpc respectively, the shell from the 2001 nova
event should have been no larger than $0.2\arcmin$ by November 2006,
which is the date of the last IPHAS image the nebula was visible
in. Hence the nebula in the image can not be from the 2001 nova
event. The most obvious explanation is that it is a light echo from
material ejected from the system by a previous event, such as a nova
shell or a planetary nebula. The angular diameter of the shell is
$2.7\arcmin \pm 0.5\arcmin$. Adopting the distance of 3--8 kpc derived
by \citet{kiss02} using maximum magnitude versus rate of decline
relationships, the radius of the shell is 3--12 $\times 10^{16}$m. The
time from the nova in 2001 to the date of the first IPHAS image is 819
days giving a light radius of $2.1 \times 10^{16}$m.  These two radii
are broadly comparable, as expected for a light echo. Assuming a
typical nova shell expansion velocity of 2000 km s$^{-1}$, the age of
the shell can be estimated to be $\sim300$ years. We have reviewed the
literature and can find no previous discussion of a nebula around
V2275 Cyg. Indeed, the presence of light echoes around novae are
relatively rare, and only GK Per, V732 Sgr, V458 Vul and T Pyx have
recorded echoes (\citealt{kapteyn01}, \citealt{swope30},
\citealt{wesson08}, \citealt{sokoloski13}, respectively).

\begin{figure*}
\centering \subfigure[November 2003]{
  \includegraphics[width=50mm,angle=0]{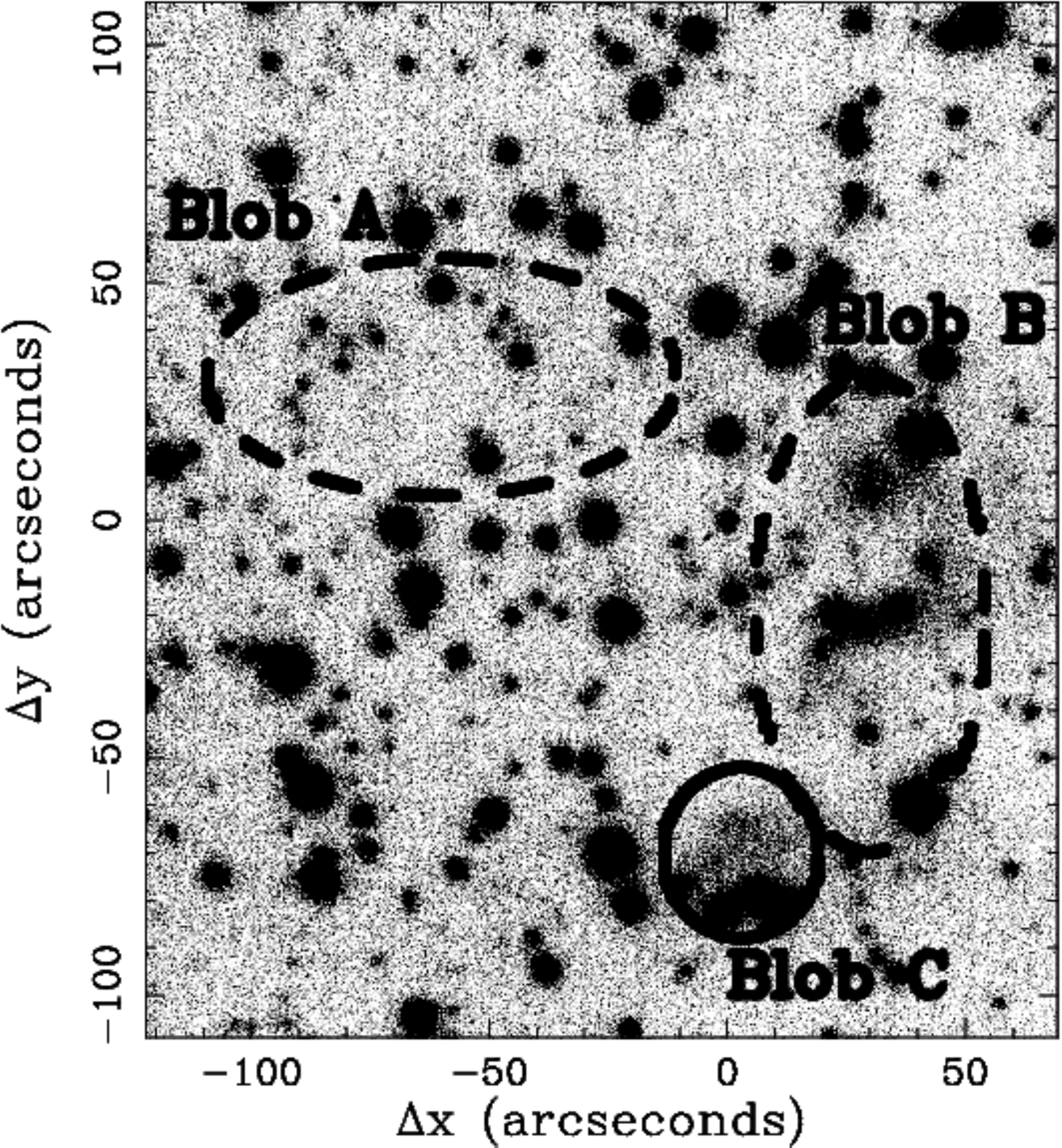}
\label{fig:1}}
\subfigure[November 2005]{
  \includegraphics[width=50mm,angle=0]{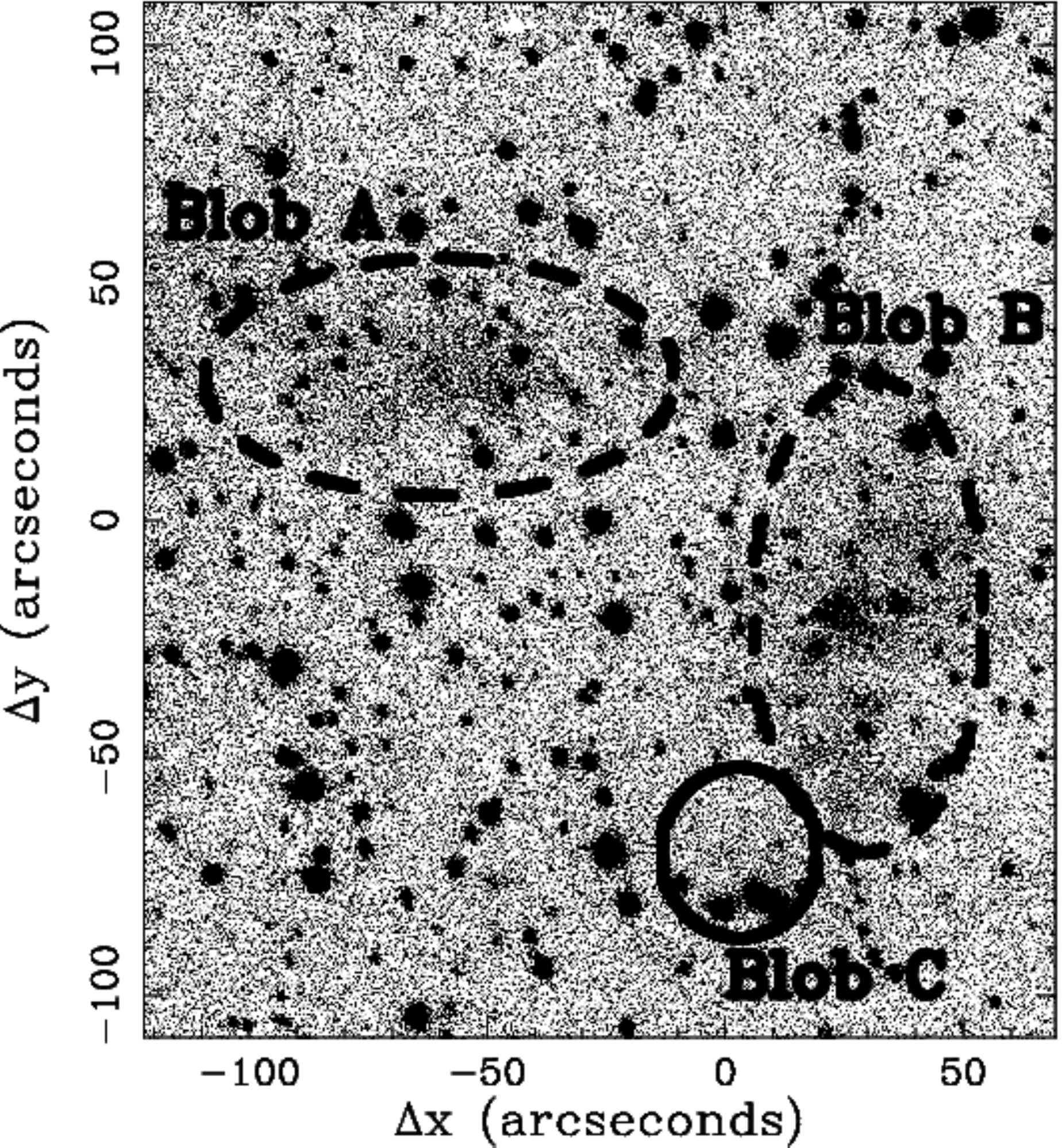}
\label{fig:2}}
\subfigure[November 2006]{
  \includegraphics[width=50mm,angle=0]{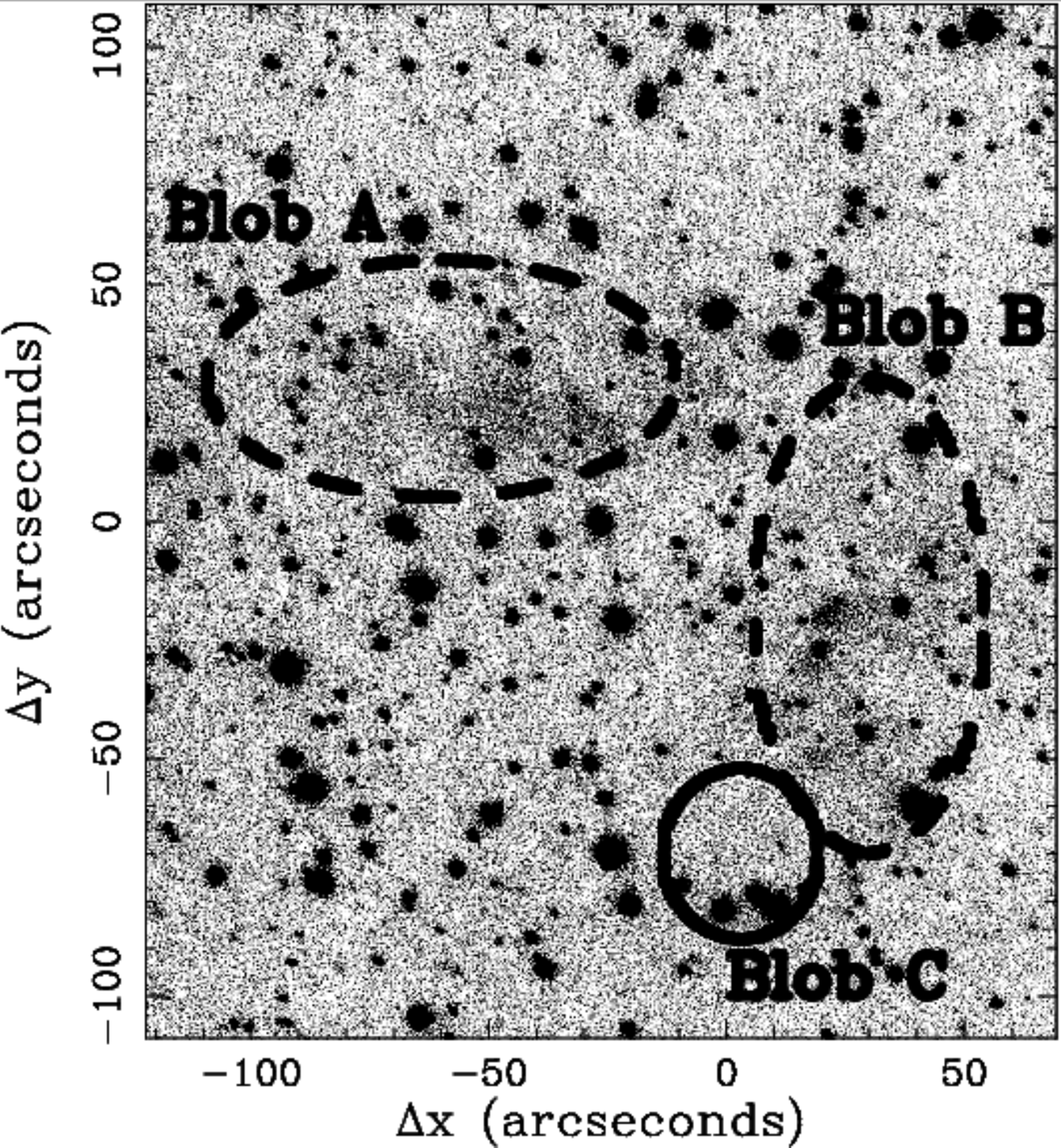}
\label{fig:3}}

\caption{H$\alpha$ images of V2275 Cyg from IPHAS. See text for a
  discussion about the three blobs of material. In all images, North
  is up, East is left.}
\label{fig:blob}
\end{figure*}

There are three principal blobs of material that are apparent in the
images, as highlighted in Figs \ref{fig:blob}\,a--c. Blob A does not
appear in Fig.~\ref{fig:1} but appears in Figs. \ref{fig:2} \&
\ref{fig:3}. It appears to move southwards (towards the right in the
images) by approximately $20\arcsec$. At a distance of 3 kpc, with an
interval of approximately one year between the two images, this
equates to a transverse speed of $2.7\times10^{8}$\,ms$^{-1}$. Clearly
this cannot be bulk motion of material. It is better explained as the
passage of a light pulse through a pre-existing bi-polar nebula, with
the axis of symmetry of the nebula pointing approximately
perpendicular to the plane of the sky. This orientation is suggested
by the lack of eclipses in the light curve of V2275 Cyg
(\citealt{balman05}). Blob B appears in all three images and whilst
different parts change intensity, there is no consistent motion
shown. This blob of material is diagonally opposite blob A and could
be the opposite pole of a bi-polar nebula. Blob C only appears in Fig
\ref{fig:blob}(a).

If the shell is due to a previous nova event this would mean that
V2275 Cyg should be reclassified as a recurrent nova (RN), in
agreement with \citet{pagnotta14} who identified V2275 Cyg as a
likely RN on the basis of its outburst light curve and spectrum.

\section{Discussion}
\label{sec:disc}

Our goal was to search for previously undetected nova shells around
CVs, primarily nova-like variables. The results of our search are shown
in Tab.~\ref{tab:tottab}.

\begin{table*}
\caption[]{Summary of our search for nova shells.}
\begin{center}
\begin{tabular}{lrrrrrr}
\hline\hline
\multicolumn{1}{l}{ } &
\multicolumn{1}{r}{Nova-like} &
\multicolumn{1}{r}{Polars \&} &
\multicolumn{1}{r}{Asynchronous} &
\multicolumn{1}{r}{Dwarf} &
\multicolumn{1}{r}{Old} &
\multicolumn{1}{r}{Total} \\
\multicolumn{1}{l}{ } &
\multicolumn{1}{r}{Variables} &
\multicolumn{1}{r}{Intermediate} &
\multicolumn{1}{r}{Polars} &
\multicolumn{1}{r}{Novae} &
\multicolumn{1}{r}{Novae} &
\multicolumn{1}{r}{} \\
\multicolumn{1}{l}{ } &
\multicolumn{1}{r}{} &
\multicolumn{1}{r}{Polars} &
\multicolumn{1}{r}{} &
\multicolumn{1}{r}{} &
\multicolumn{1}{r}{} &
\multicolumn{1}{r}{} \\
\hline
WHT Targets              & 22 & 1 & 2 & 2 & 4 & 31 \\
IPHAS Targets          & 5 & 10 & 2 & 34 & 23 & 74 \\
\multicolumn{1}{l}{less: Duplicated objects}& --3 & 0 & 0 &  0 & --1 & --4 \\
\hline
\multicolumn{1}{l}{Grand total of systems} & 24 & 11 & 4 & 36 & 26 & 101 \\
\hline\hline
\end{tabular}
\end{center}
\label{tab:tottab}
\end{table*}

\subsection{Nova-like variables}
\label{lifetime}

We surveyed 22 NLs with the WHT and 5 NLs with IPHAS (three NLs were
surveyed in both giving a total of 24 unique NLs), and found no
shells with the WHT and evidence for only one shell in IPHAS, V1315
Aql, which we subsequently confirmed with additional INT observations
(see Sect.~\ref{v1315}).

What can we deduce from our discovery of a shell around one NL? Let us
assume that all novae that occurred in the last $\sim100$ years would
have been observed. These would now be classified as old novae in the
RK catalogue and hence would not appear in our sample of NLs. We also
know that our observations are not sensitive to shells older than
$\sim200$\,yrs (see Sect.~\ref{search_strategy}). Hence our search for
nova shells around NLs is only likely to find shells between 100 and
200 years old. We found one shell in this 100-year window, out of 24
NLs surveyed, indicating that the lifetime of the NL phase lasts
approximately 2400\,yrs. This is consistent with the
order-of-magnitude estimate of 1,000 years derived by
\citet{patterson13} for the NL phase for long-period CVs. Hence our
results lend some support to the nova-induced cycle theory, although
we are dealing with small number statistics; our survey of 24 NLs
represents only 31\% of the 78 NLs in the RK catalogue. We also
acknowledge the incompleteness of our survey due to the small field of
view of the WHT images, and the shallow IPHAS images (see
Sect. \ref{future} for a discussion). We also note that our IPHAS
search included 7 novae with previously discovered shells but we only
found 4, and searches for shells around known novae tend to recover
shells around only half of the targets \citep{downes00}. With all of
the assumptions, uncertainties and survey inefficiencies detailed
above, our estimate of the NL-phase lifetime should be viewed as a
lower limit.

As we were about to submit this paper, a paper appeared by
\citet{schmidtobreick15}, who presented the results of a survey for
nova shells around 10 DNe in the 3--4\,hr period range with
low-$\dot{M}$ and 5 NLs that show low states (VY~Scl stars). They
found no shells, and used this to set a lower limit of 13\,000 years
on the nova recurrence time. This is consistent with the lifetime of
the NL phase of $\sim 2400$ yrs that we have derived, as clearly the
NL phase should be shorter than the nova recurrence time, although
their figure is arguably even more uncertain than ours.

 \subsection{Asynchronous polars}
 \label{async}

The WHT survey included two asynchronous polars, V1432 Aql and BY
Cam. The images and radial profiles of these two systems are shown in
Appendix \ref{fig2}. There are no traces of nebulosity in either the
images or the radial profiles of these systems. There are two other
asynchronous polars in the IPHAS survey, J0524$+$4244 and V1500 Cyg,
the archetypal asynchronous polar. There is no evidence of a shell
around J0524$+$4244. We did recover the previously known shell around
V1500 Cyg, originating from its nova eruption in 1975 (see
Sect.~\ref{v1500}).

Our results imply that we are unable to confirm that a recent nova
eruption is the cause of the asynchronicity in the white dwarf spin of
these systems. However, it is perhaps not surprising that we did not
find any new shells given our survey limits and the synchronisation
timescale. BY~Cam, for example, is estimated to synchronise within
$\sim1100$\,yrs (\citealt{campbell99}), an order of magnitude longer
than our 100\,yr detection window (see Sect.~{lifetime}). 

\subsection{Future surveys}
\label{future}

In hindsight, our decision to use the old nova DQ Her as a guide for
our WHT search strategy (Sect.~\ref{search_strategy}) led us to
underestimate the optimal field of view for hunting for nova shells.
This is because DQ~Her has relatively slow ejecta (350\,km\,s$^{-1}$;
\citealt{warner95a}). The angular size of the shell is determined by
the time since the nova eruption, the distance to the CV, and the
speed of the ejecta, and is given by the following scaling relation:

\begin{equation}
   R\sim 20\arcsec \,\, \frac{ t/100\,\mathrm{yr} \times
     v/1000\,\mathrm{km \,s}^{-1}}{d/\mathrm{kpc}},
\end{equation}
where $R$ is the angular radius of the shell in arcseconds, $t$ is the
time elapsed since the nova eruption, $v$ is the shell expansion
velocity, and $d$ is the distance to the CV. Hence a recent, distant
nova with slow-moving ejecta ($t=100, v=500, d=2$) would have a small
shell of radius $\sim5\arcsec$, whereas an older, nearby nova with
fast moving ejecta ($t=200, v=2000, d=0.5$) would have expanded to a
radius of $\sim 2.7\arcmin$. Hence the field of view of the Auxiliary
port on the WHT ($\sim1\arcmin$ radius) was too small to detect such
large shells. This is borne out by the size of the one shell that we
did discover around V1315 Aql, which is $\sim2.5\arcmin$ in radius,
and the two shells discovered by \citeauthor{shara07} (2007; 2012) of
radii $1.5\arcmin$ (AT~Cnc) and $15\arcmin$ (Z~Cam). Another problem
with having such a small field of view is the paucity of field stars
for the radial-profile technique (Sect.~\ref{im}).

The IPHAS survey, on the other hand, had more than enough field of
view ($\sim17\arcmin$ radius per pointing) to discover nova shells but
suffered from very short exposure times (120\,s), which we had no
control over, and from being constrained to the Galactic plane, making
it difficult to pick out nova shells from the H$\alpha$ nebulosity. A
more optimal survey for nova shells would have approximately the same
field of view as an IPHAS pointing, avoid the Galactic plane and be of
similar depth to our WHT survey.

\section{Conclusions}

We have performed an H$\alpha$-imaging survey for nova shells around
CVs. We imaged 31 CVs with the WHT, and searched the IPHAS fields
around 74 CVs. 

Our search focused on looking for shells around nova-like variables,
as the nova-induced cycle theory suggests that these systems are most
likely to have undergone a recent nova eruption. Of the 24 unique NLs
we examined we found evidence for only one shell around V1315 Aql,
which has a radius of $\sim2.5\arcmin$, indicative of a nova eruption
approximately 120 years ago.

The survey included 4 asynchronous polars, 2 observed with the WHT
(V1432 Aql and BY Cam) and 2 in IPHAS (J0524$+$4244 and V1500 Cyg) but
we found no shells around any of them, except the previously known
shell around V1500 Cyg. Hence we are unable to confirm whether the
asynchronicity of the WD spin in these systems is due to a recent nova
eruption.

We find no unambiguous detections of nova shells around other classes
of CV, but we did find tentative evidence of a faint shell around the
dwarf nova V1363 Cyg (see Sect.~\ref{v1363}). We also find evidence
for a light echo around the nova V2275 Cyg, which erupted in 2001,
indicative of an earlier nova eruption $\sim300$ years ago, thus
making V2275 Cyg a possible recurrent nova (see Sect.~\ref{v2275}).

\section*{Acknowledgements}

This paper makes use of data obtained as part of the INT Photometric
H$\alpha$ Survey of the Northern Galactic Plane (IPHAS, www.iphas.org)
carried out at the INT. The INT and WHT are operated on the island of
La Palma by the Isaac Newton Group in the Spanish Observatorio del
Roque de los Muchachos of the Instituto de Astrofisica de
Canarias. All IPHAS data are processed by the Cambridge Astronomical
Survey Unit, at the Institute of Astronomy in Cambridge. The
bandmerged DR2 catalogue was assembled at the Centre for Astrophysics
Research, University of Hertfordshire, supported by STFC grant
ST/J001333/1. We would specificlly like to thank Janet Drew for her
help in accessing the IPHAS data and for her comments on the draft
paper, and Jonathan Irwin who prepared the mosaic images. We would
also like to thank Jeanne Wilson and Joanne Caffrey for their help
with the WHT data reduction. We thank the referee, Mike Shara,
for his valuable comments on the paper.

VSD, CK \& TRM were supported under grants from the Science and
Technology Facilities Council (STFC).

\bibliographystyle{mn2e}
\bibliography{abbrev.bib,refs.bib}

\appendix

\section{WHT images (in alphabetical order of constellation)}
\label{app1}

\begin{figure}
  \vspace{10pt}
\includegraphics[width=80mm,angle=0]{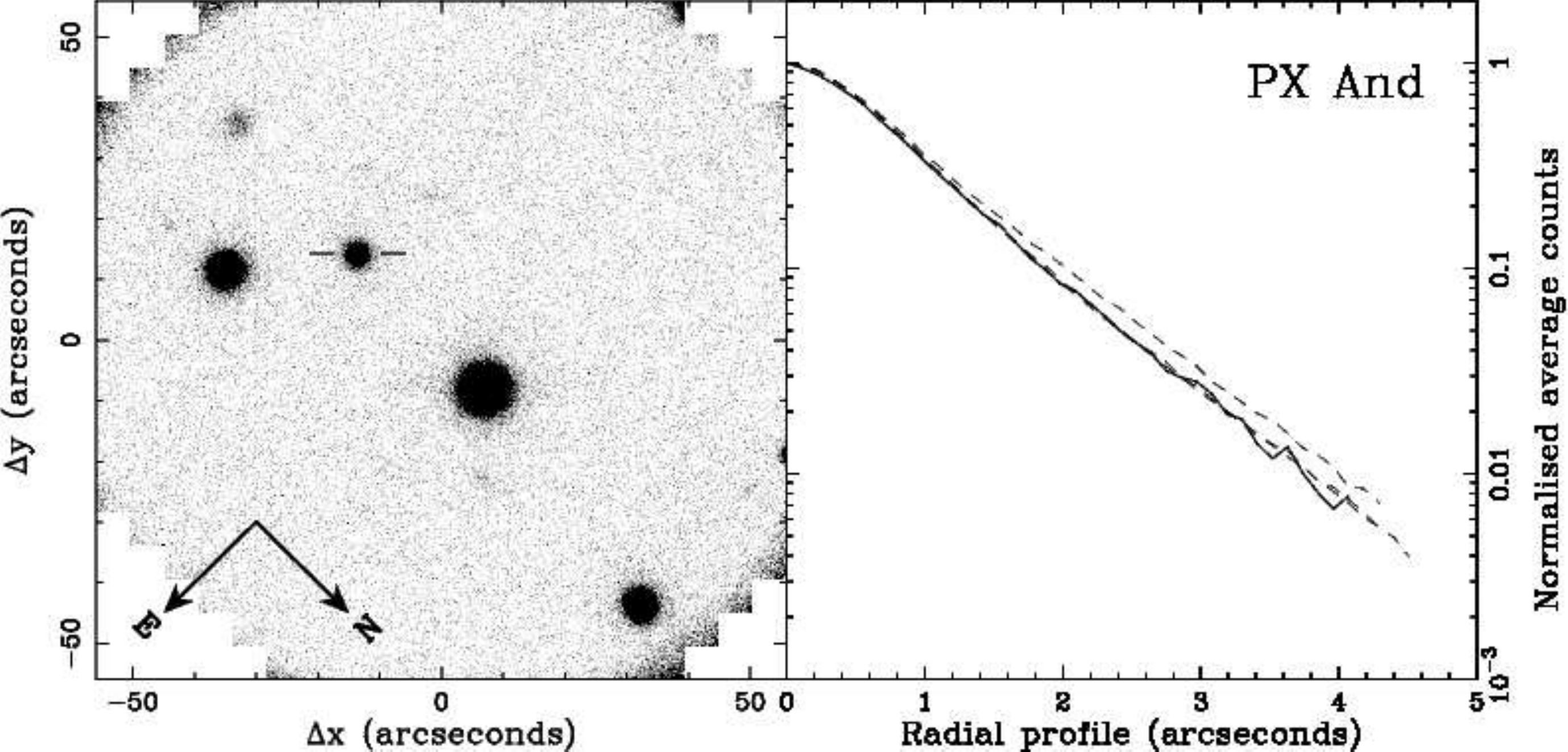}
  \vspace{1pt}
  \vspace{8pt}
\includegraphics[width=80mm,angle=0]{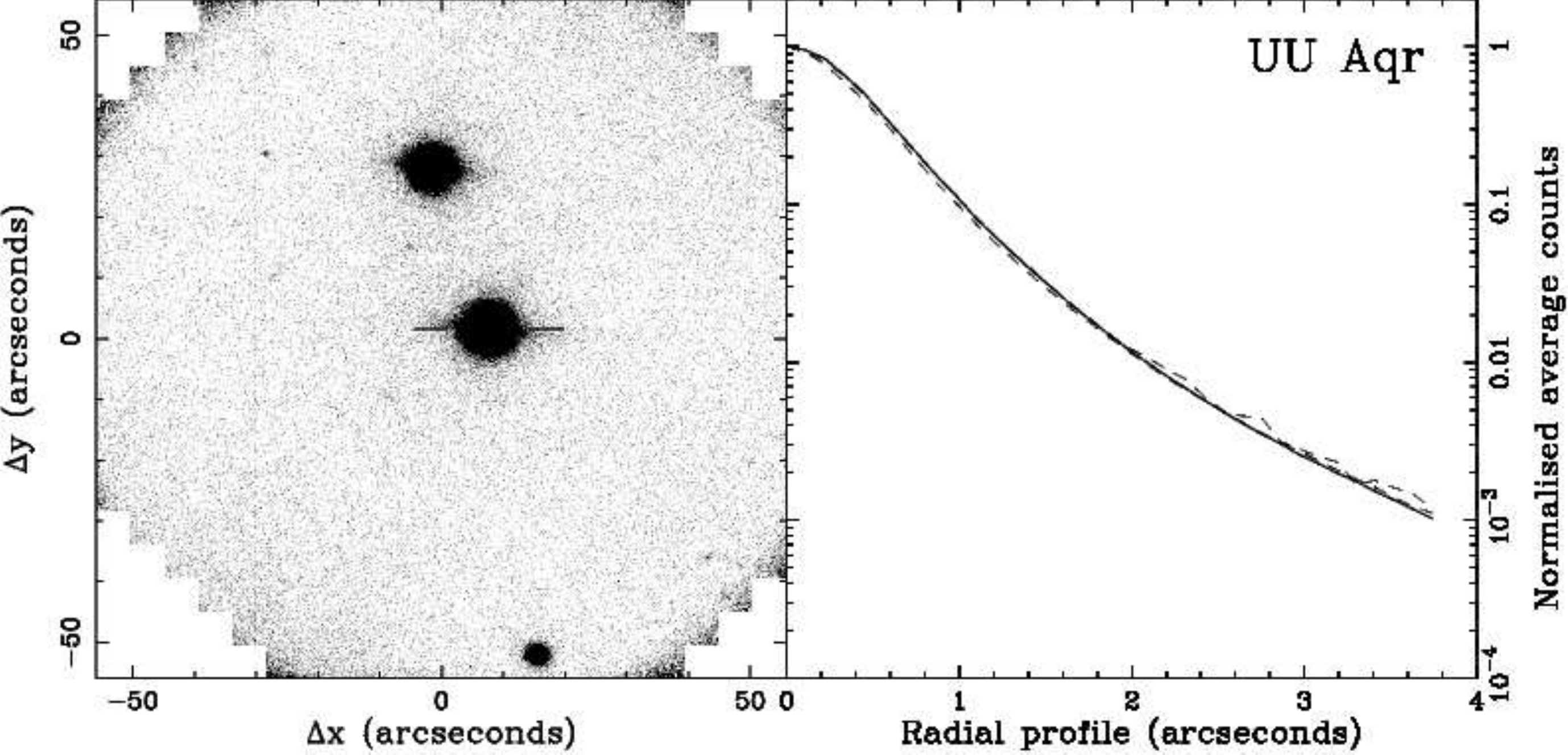}
  \vspace{1pt}
  \vspace{8pt}
\includegraphics[width=80mm,angle=0]{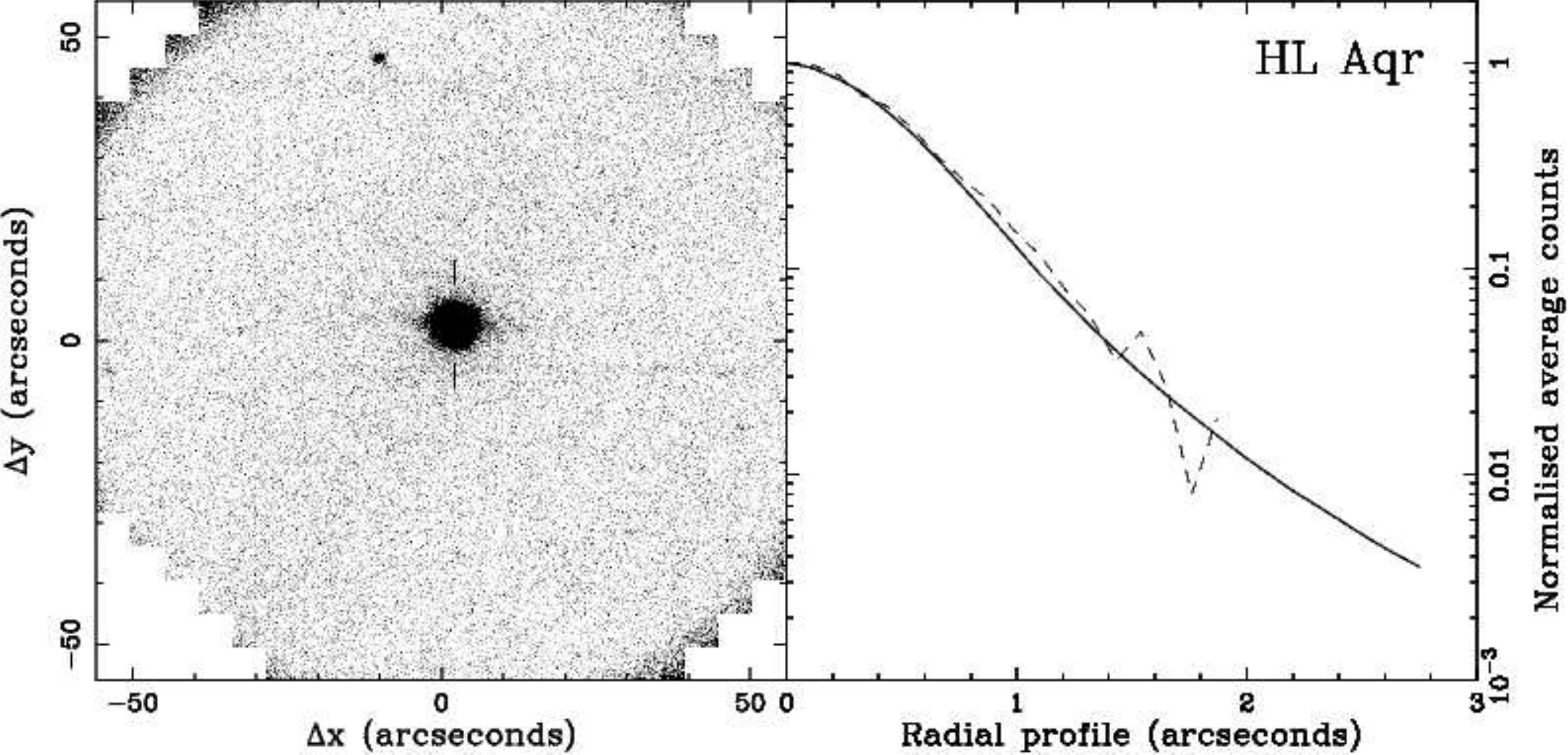}
  \vspace{1pt}
  \vspace{8pt}
\includegraphics[width=80mm,angle=0]{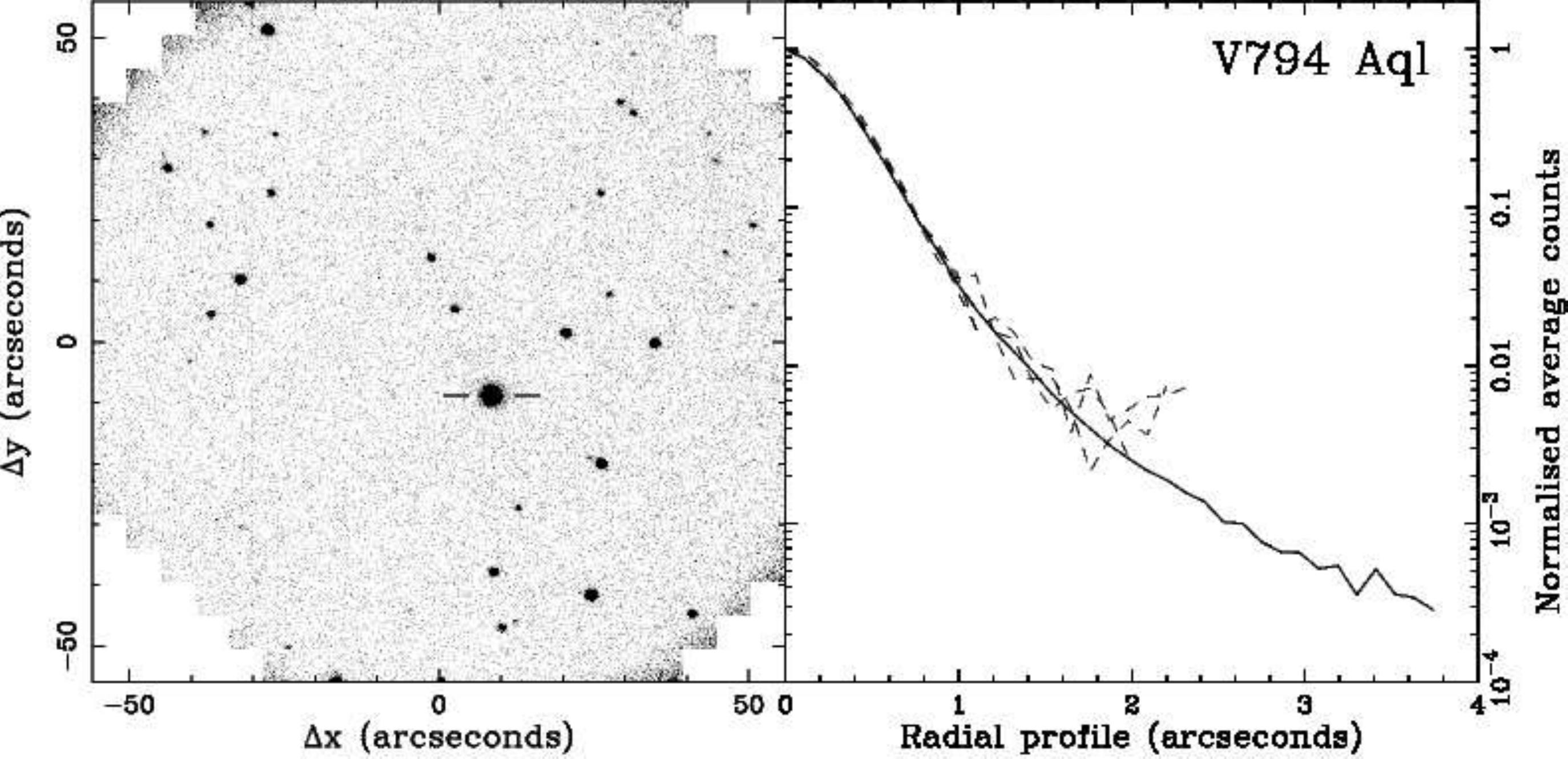}
  \vspace{1pt}
  \vspace{8pt}
\includegraphics[width=80mm,angle=0]{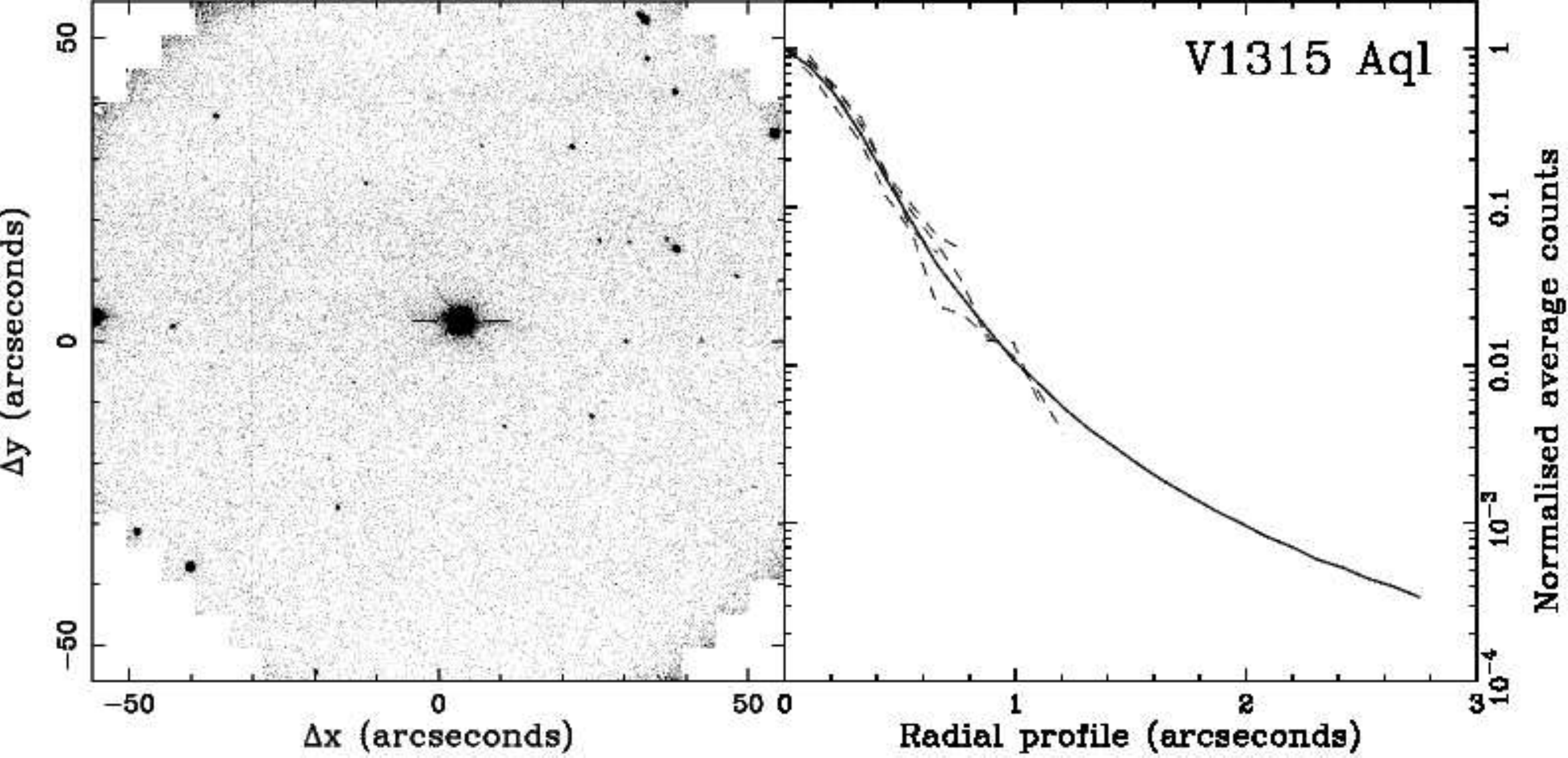}
  \vspace{1pt}
 \caption{WHT images of our target CVs (left) and the associated
   radial profiles (right); the solid line is the radial profile of
   the CV and the dashed lines are field stars. The radial profiles
   are normalised to unity and plotted until they reach $1\sigma$
   above the background level. The CVs are marked by bars and are
   located towards the centres of the images. The orientation of all
   images is the same, and is shown in the image of PX And.}
 \label{fig1}
\end{figure}

\begin{figure}
  \vspace{10pt}
\includegraphics[width=80mm,angle=0]{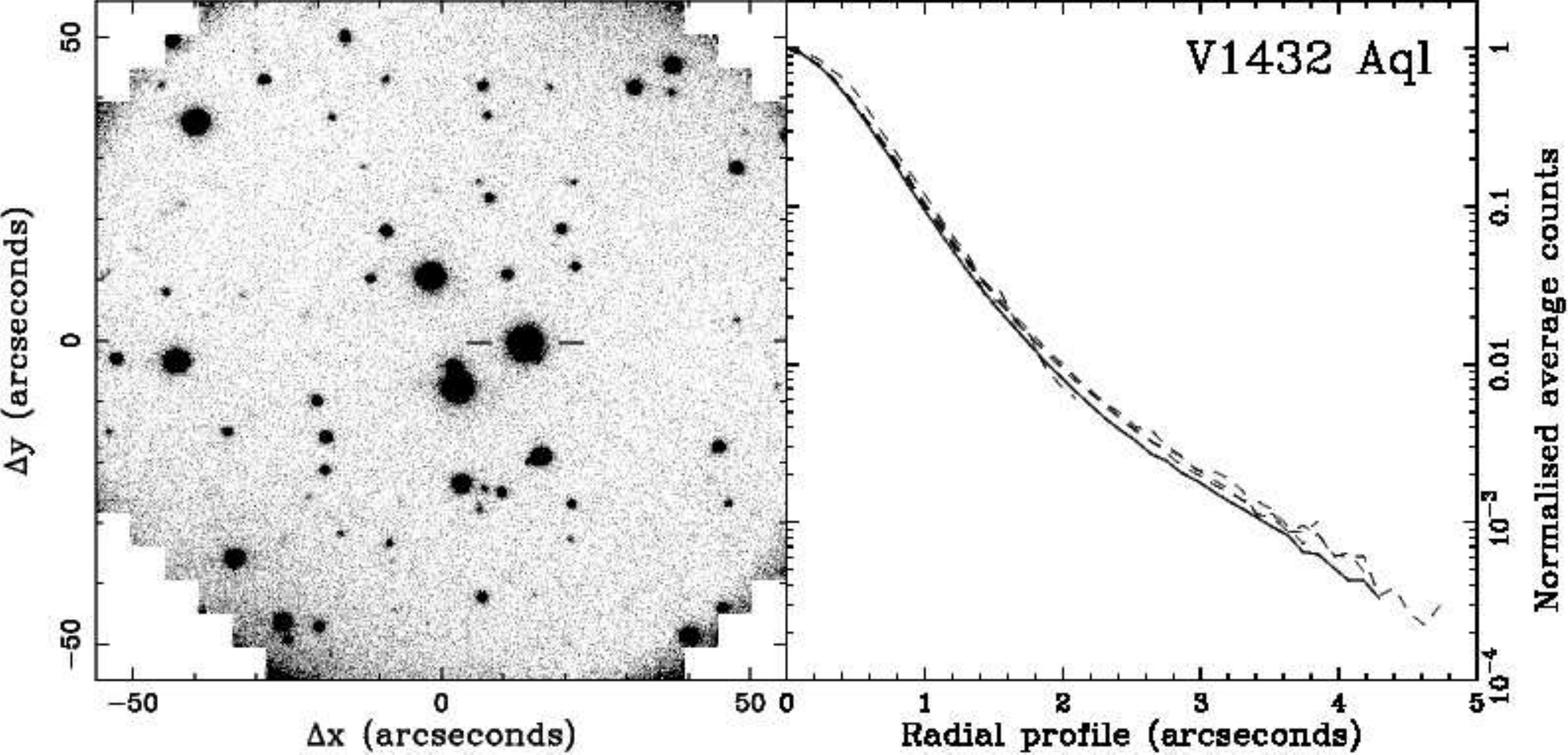}
  \vspace{1pt}
  \vspace{10pt}
\includegraphics[width=80mm,angle=0]{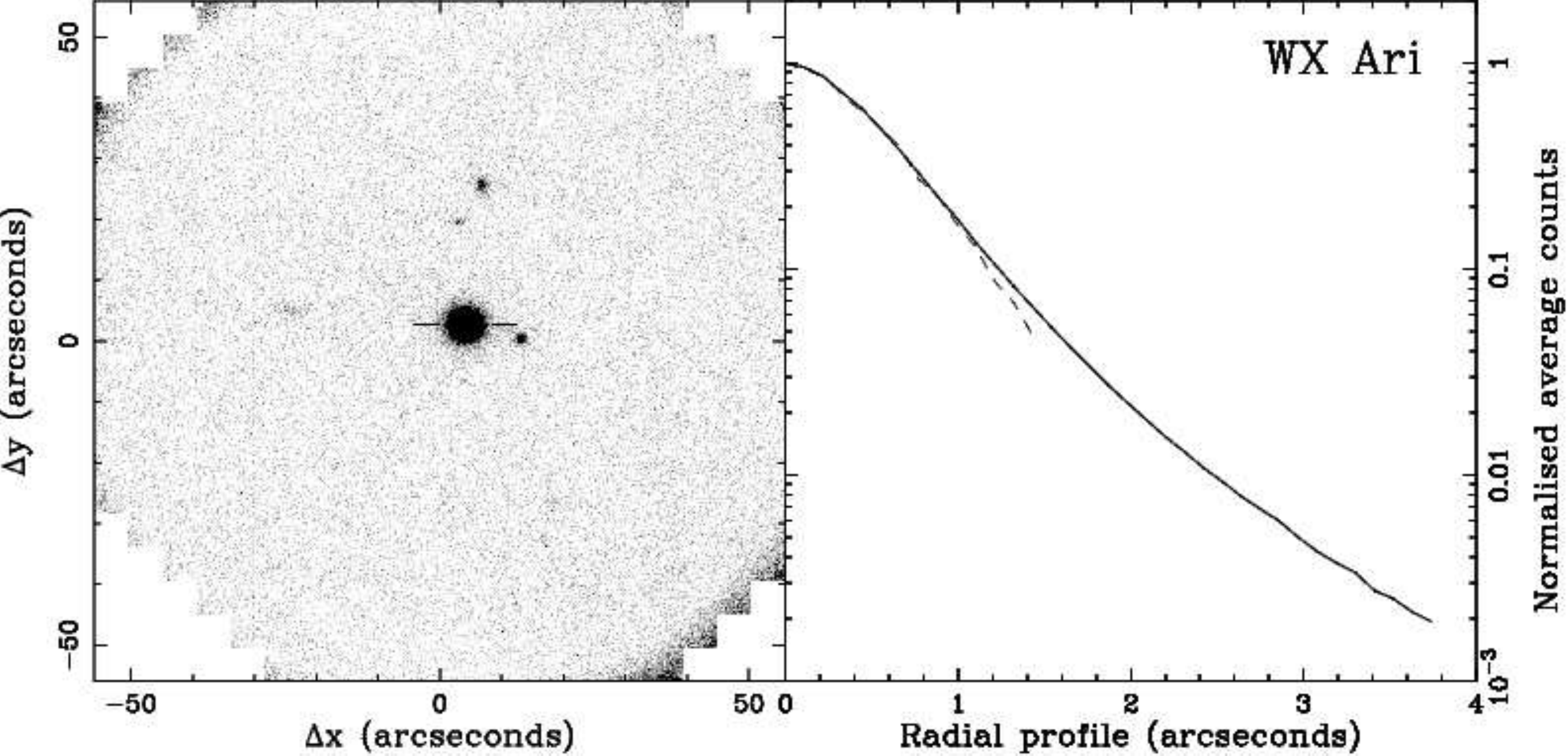}
  \vspace{1pt}
  \vspace{10pt}
\includegraphics[width=80mm,angle=0]{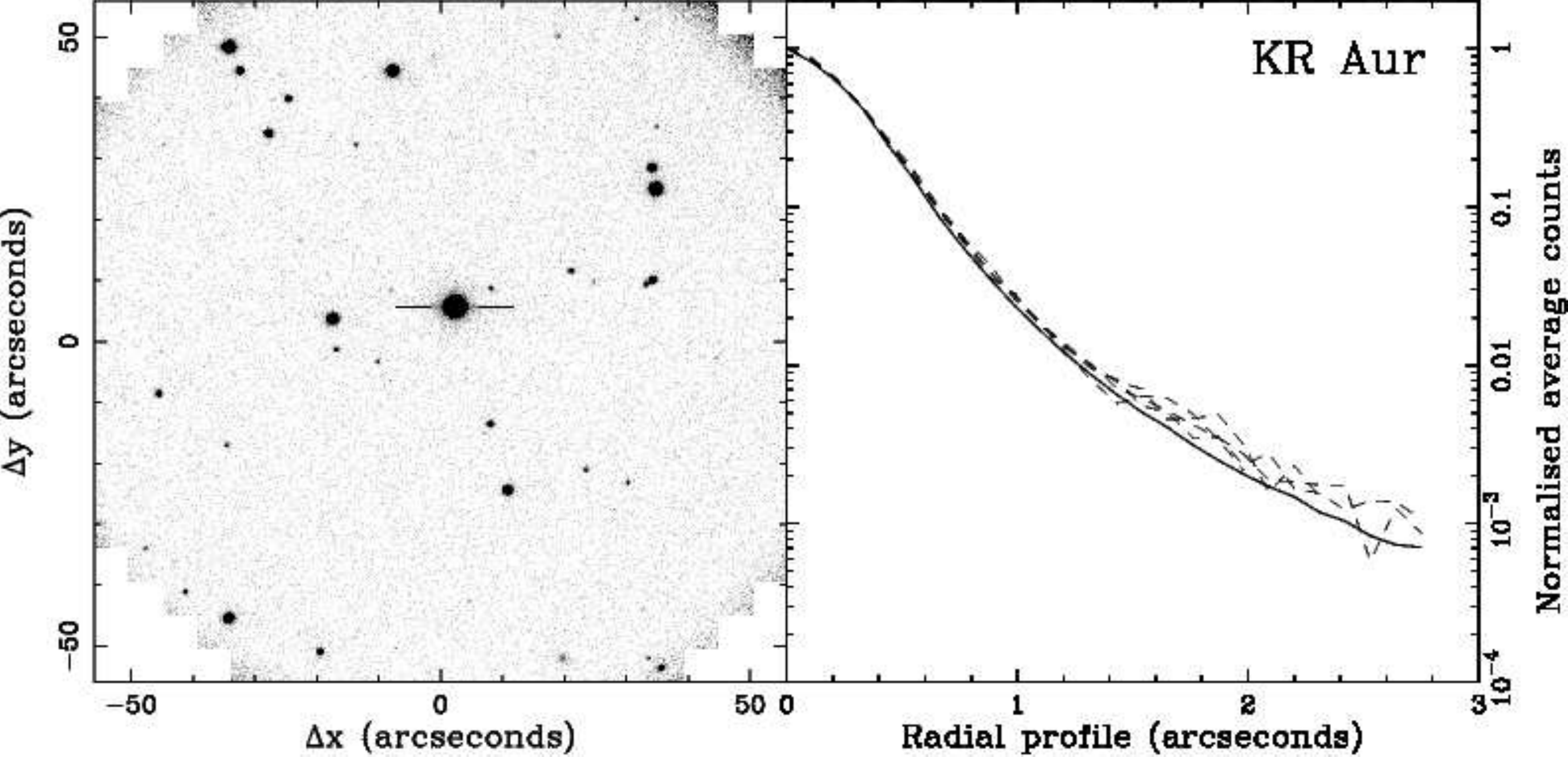}
  \vspace{1pt}
  \vspace{10pt}
\includegraphics[width=80mm,angle=0]{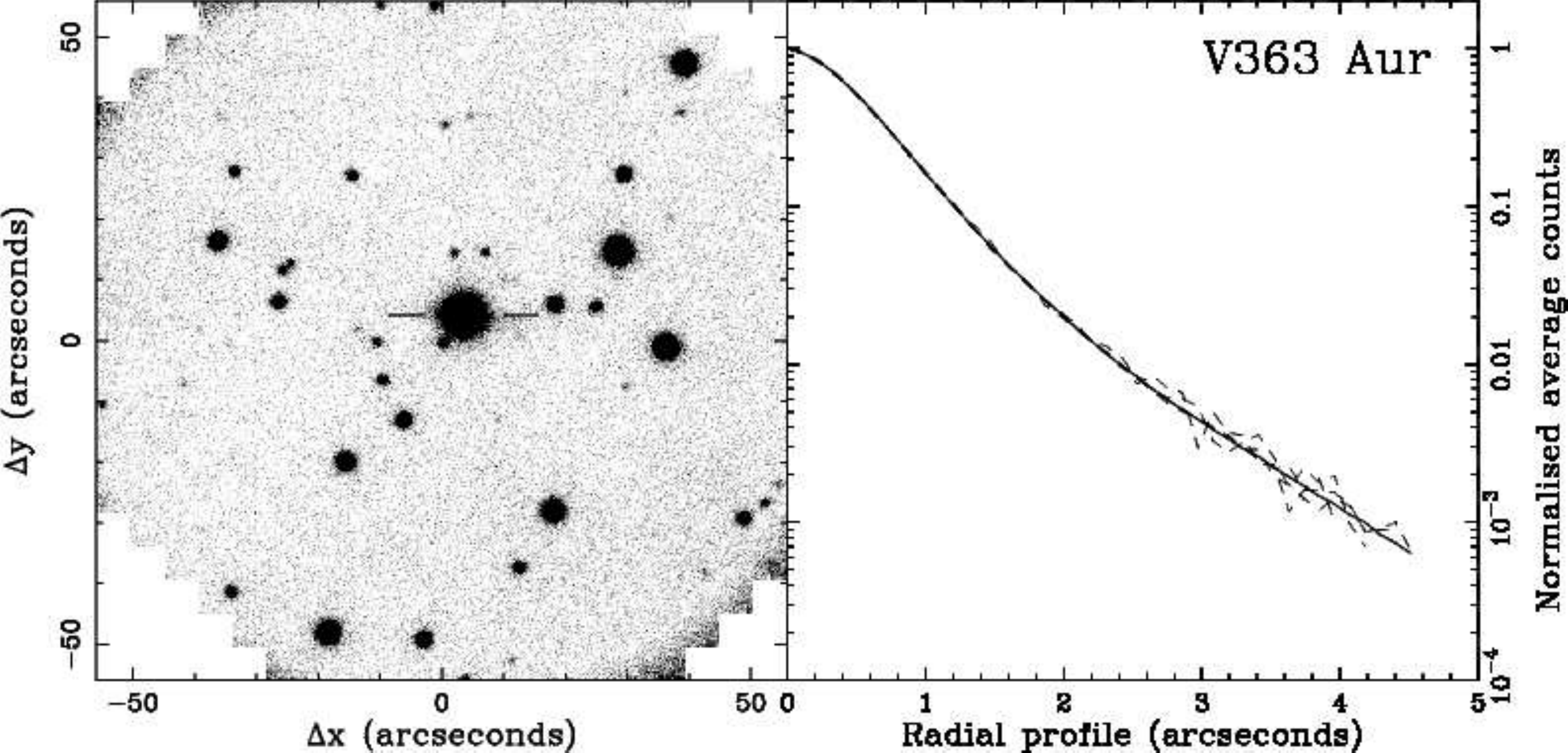}
  \vspace{1pt}
  \vspace{10pt}
\includegraphics[width=80mm,angle=0]{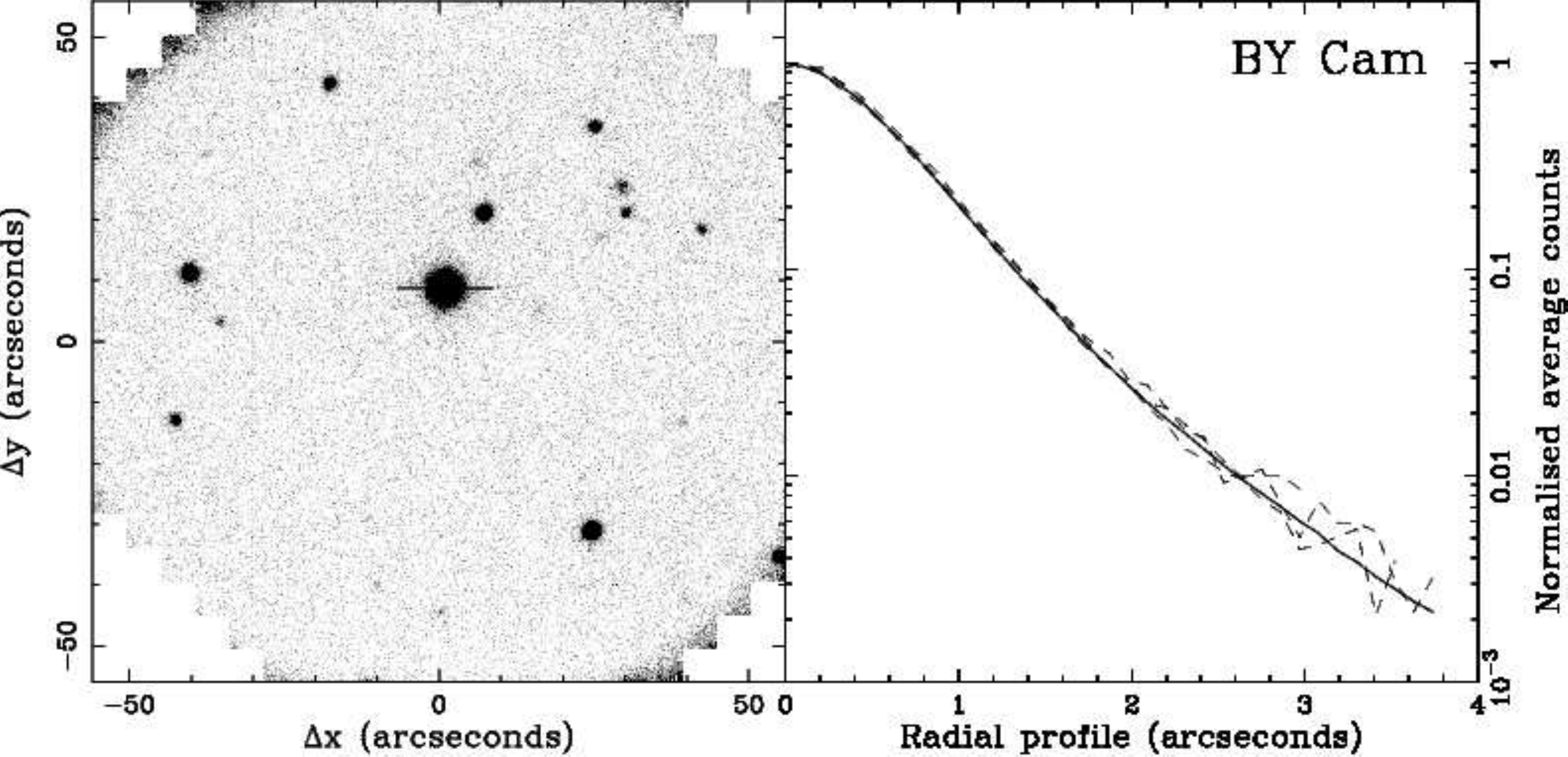}
  \vspace{1pt}
 \caption{See caption to Figure \ref{fig1} for details.}
 \label{fig2}
\end{figure}

\begin{figure}
\centering
  \vspace{10pt}
\includegraphics[width=80mm,angle=0]{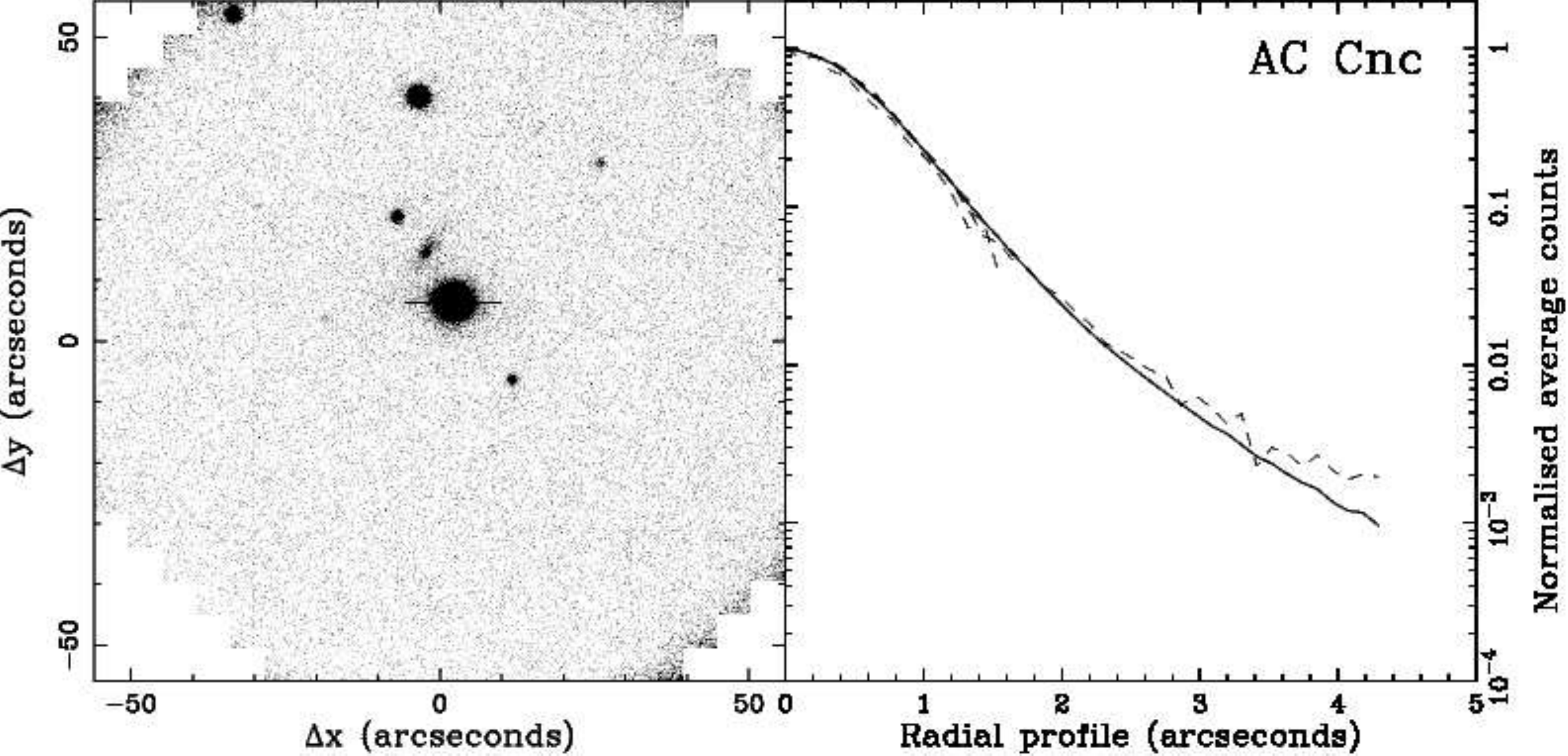}
  \vspace{1pt}
  \vspace{10pt}
\includegraphics[width=80mm,angle=0]{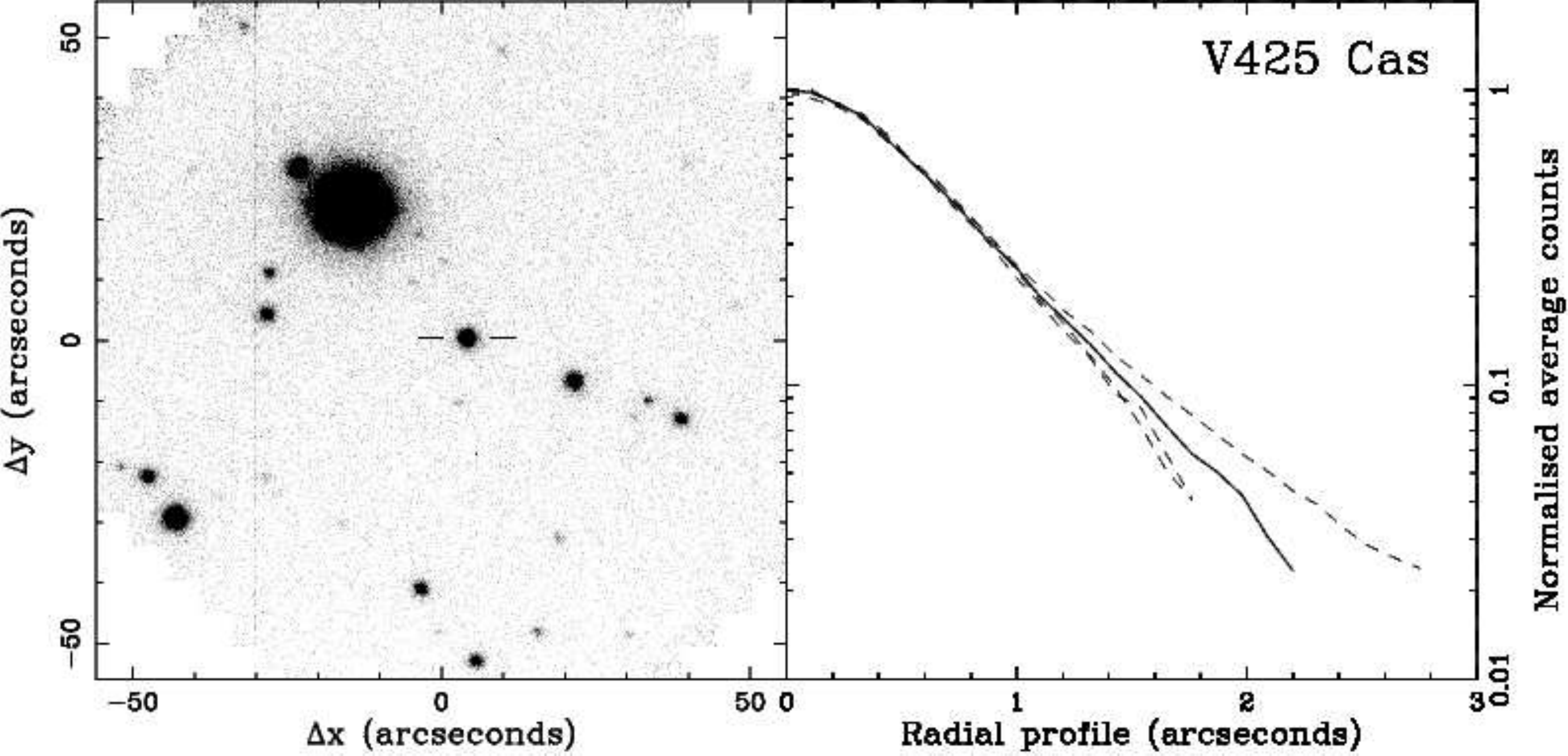}
  \vspace{1pt}
  \vspace{10pt}
\includegraphics[width=80mm,angle=0]{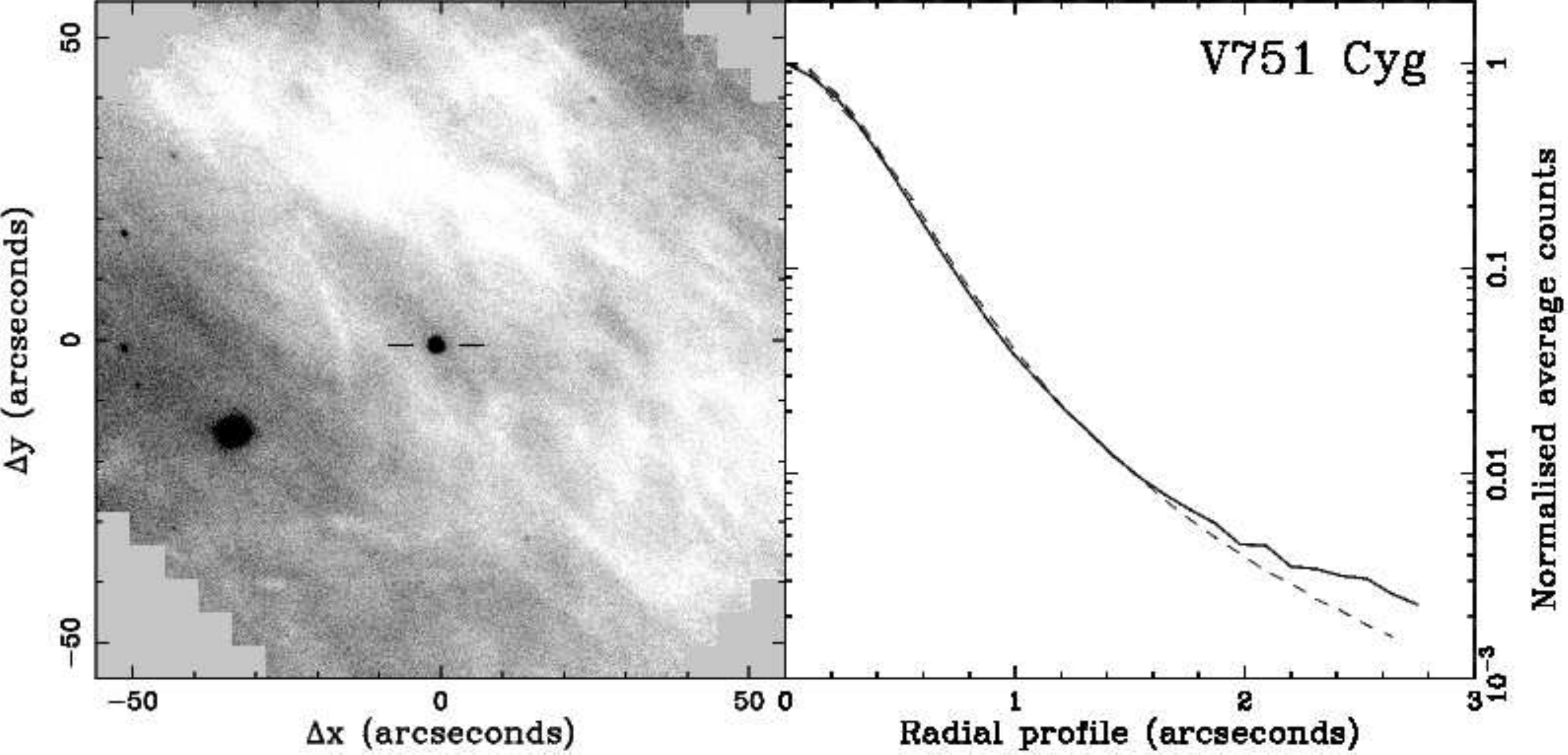}
  \vspace{1pt}
  \vspace{10pt}
\includegraphics[width=80mm,angle=0]{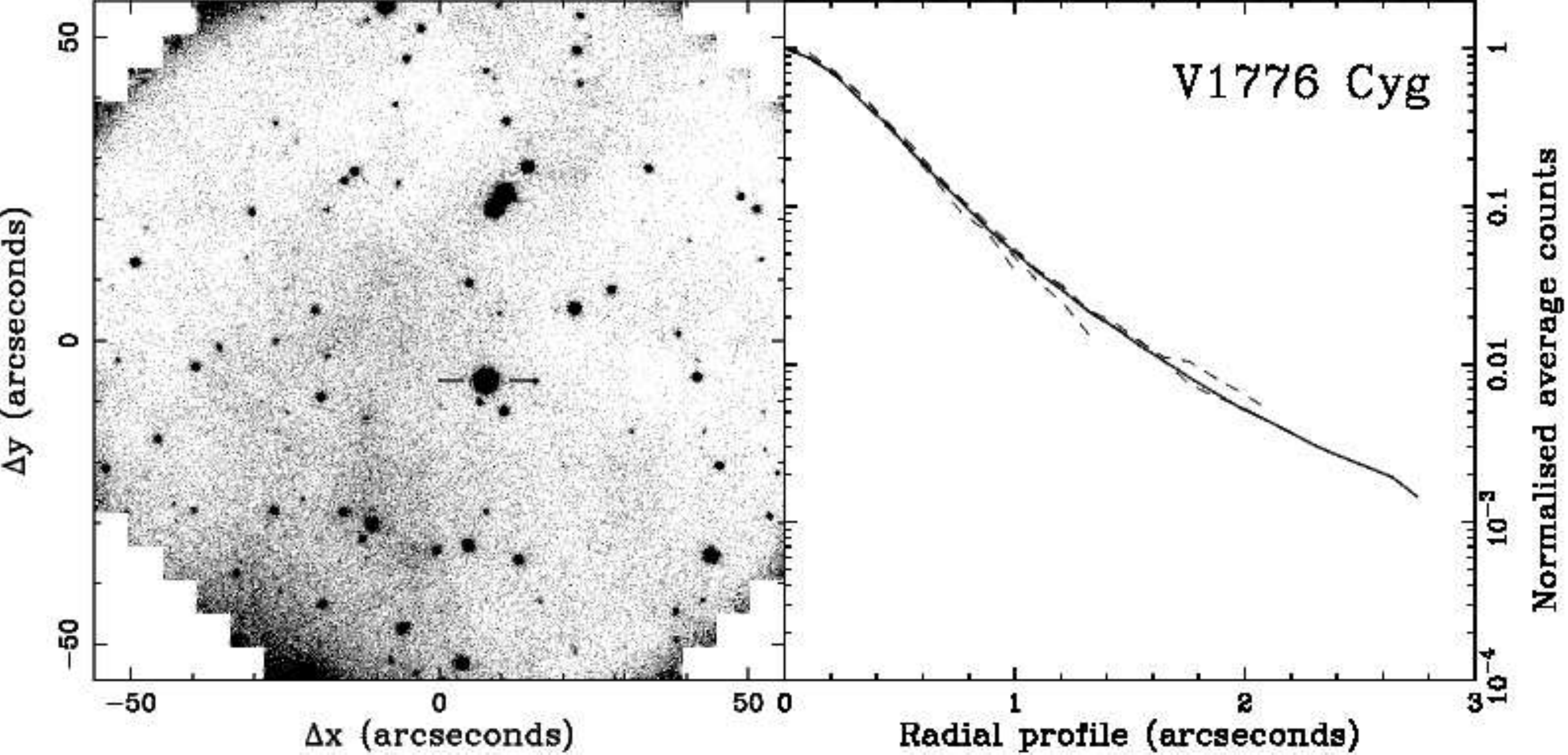}
  \vspace{1pt}
  \vspace{10pt}
\includegraphics[width=80mm,angle=0]{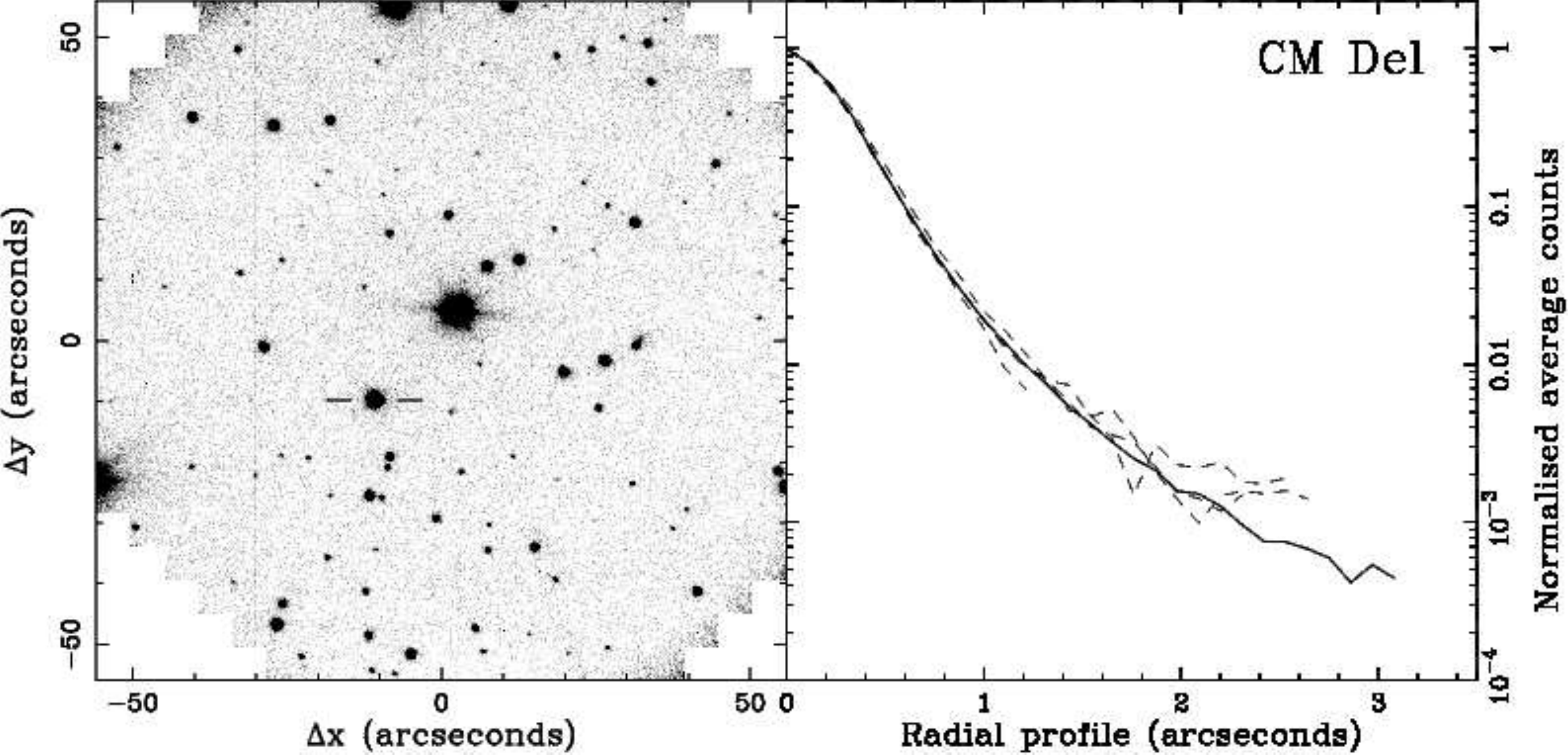}
  \vspace{1pt}
 \caption{See caption to Figure \ref{fig1} for details.}
 \label{fig3}
\end{figure}

\begin{figure}
  \vspace{10pt}
\includegraphics[width=80mm,angle=0]{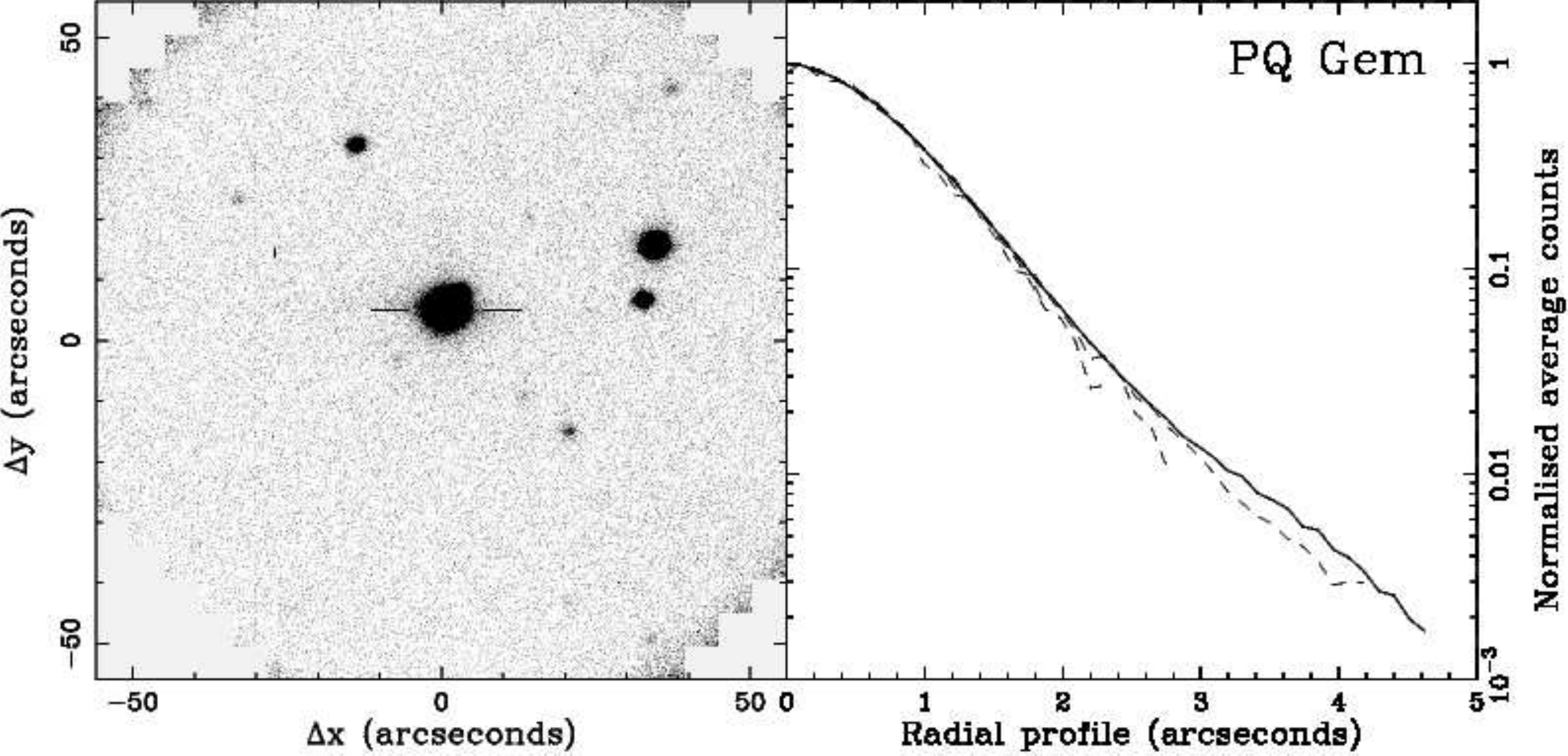}
  \vspace{1pt}
  \vspace{10pt}
\includegraphics[width=80mm,angle=0]{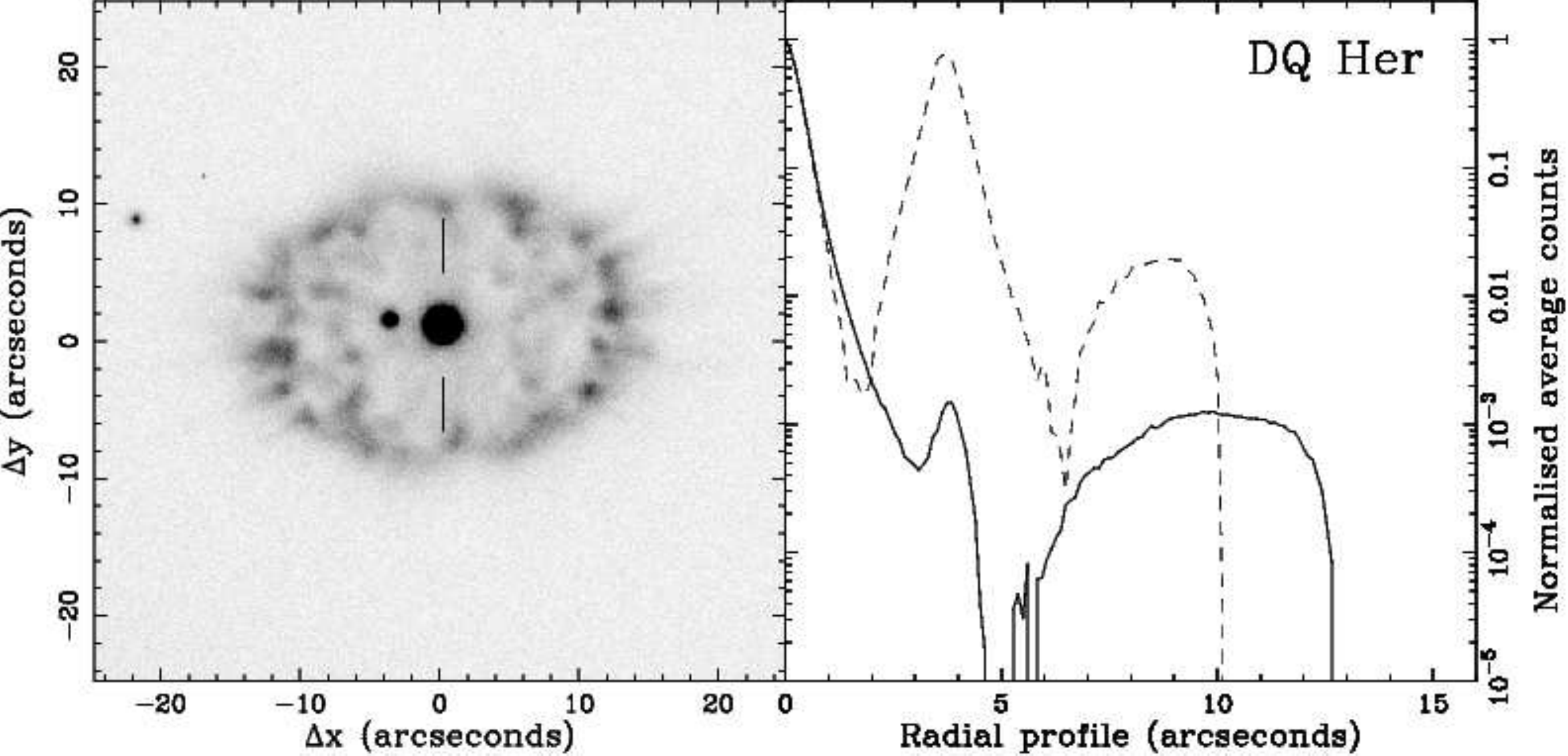}
  \vspace{1pt}
  \vspace{10pt}
\includegraphics[width=80mm,angle=0]{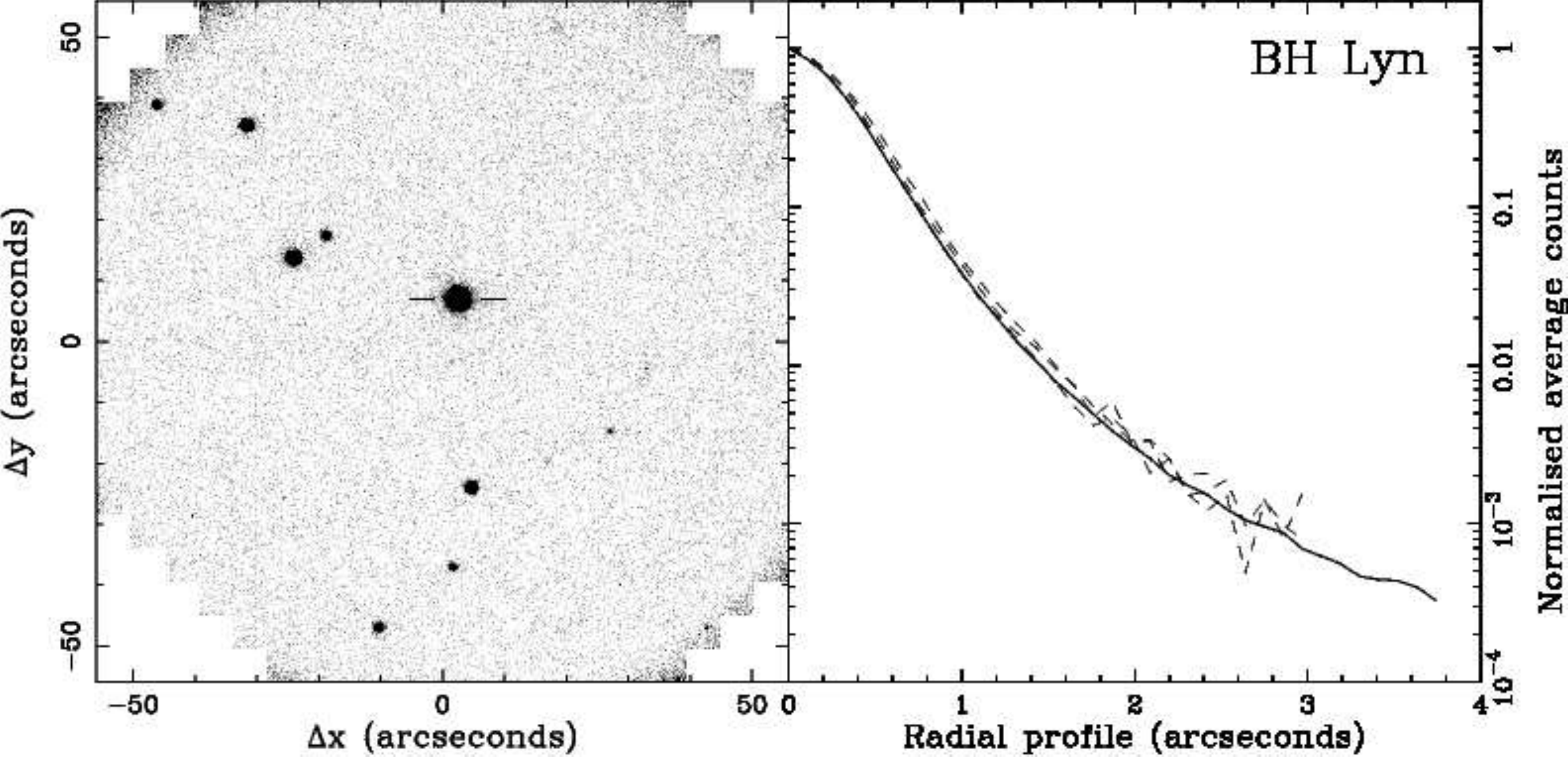}
  \vspace{1pt}
  \vspace{10pt}
\includegraphics[width=80mm,angle=0]{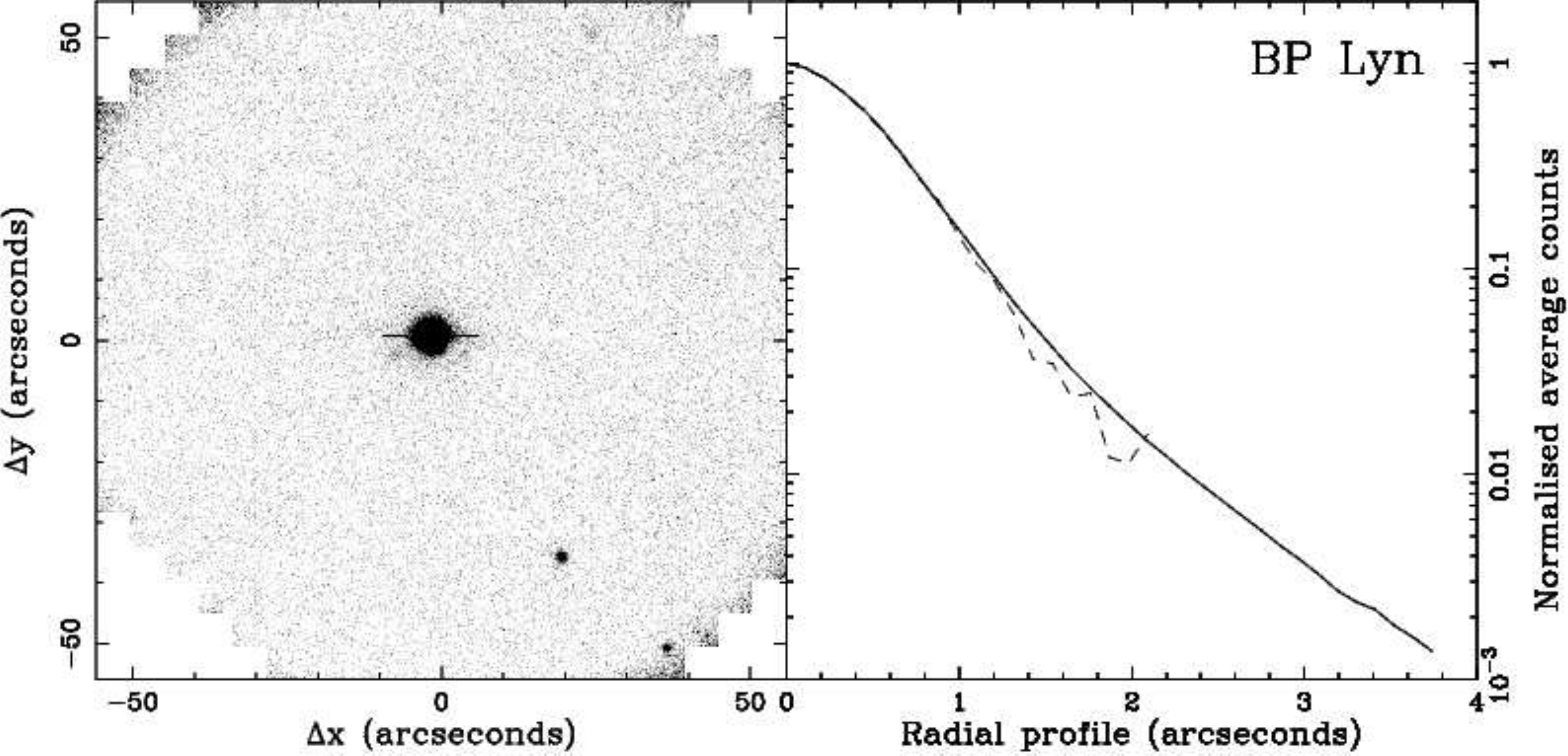}
  \vspace{1pt}
  \vspace{10pt}
\includegraphics[width=80mm,angle=0]{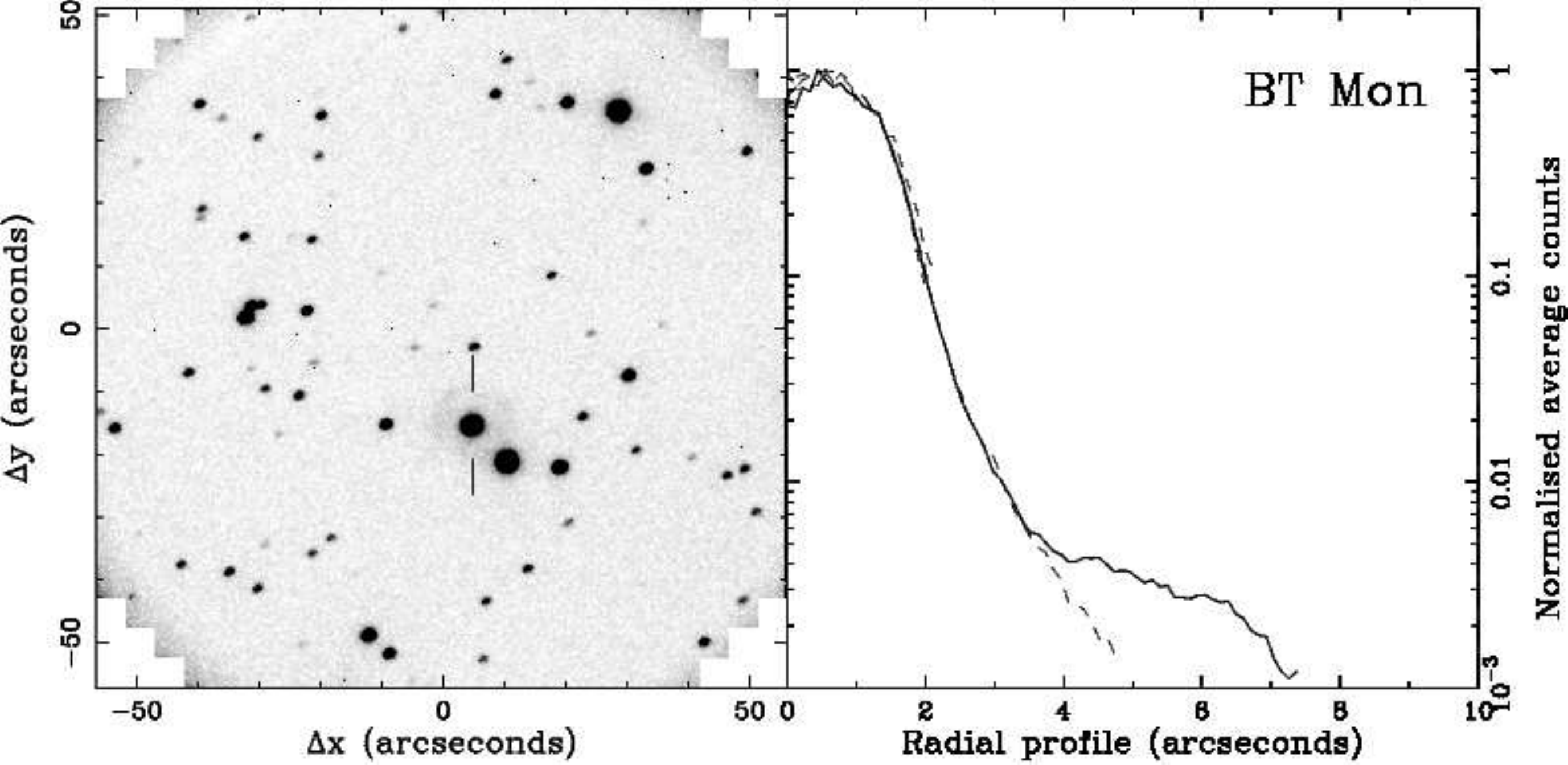}
  \vspace{1pt}
 \caption{See caption to Figure \ref{fig1} for details.}
 \label{fig4}
\end{figure}

\begin{figure}
\centering
  \vspace{10pt}
\includegraphics[width=80mm,angle=0]{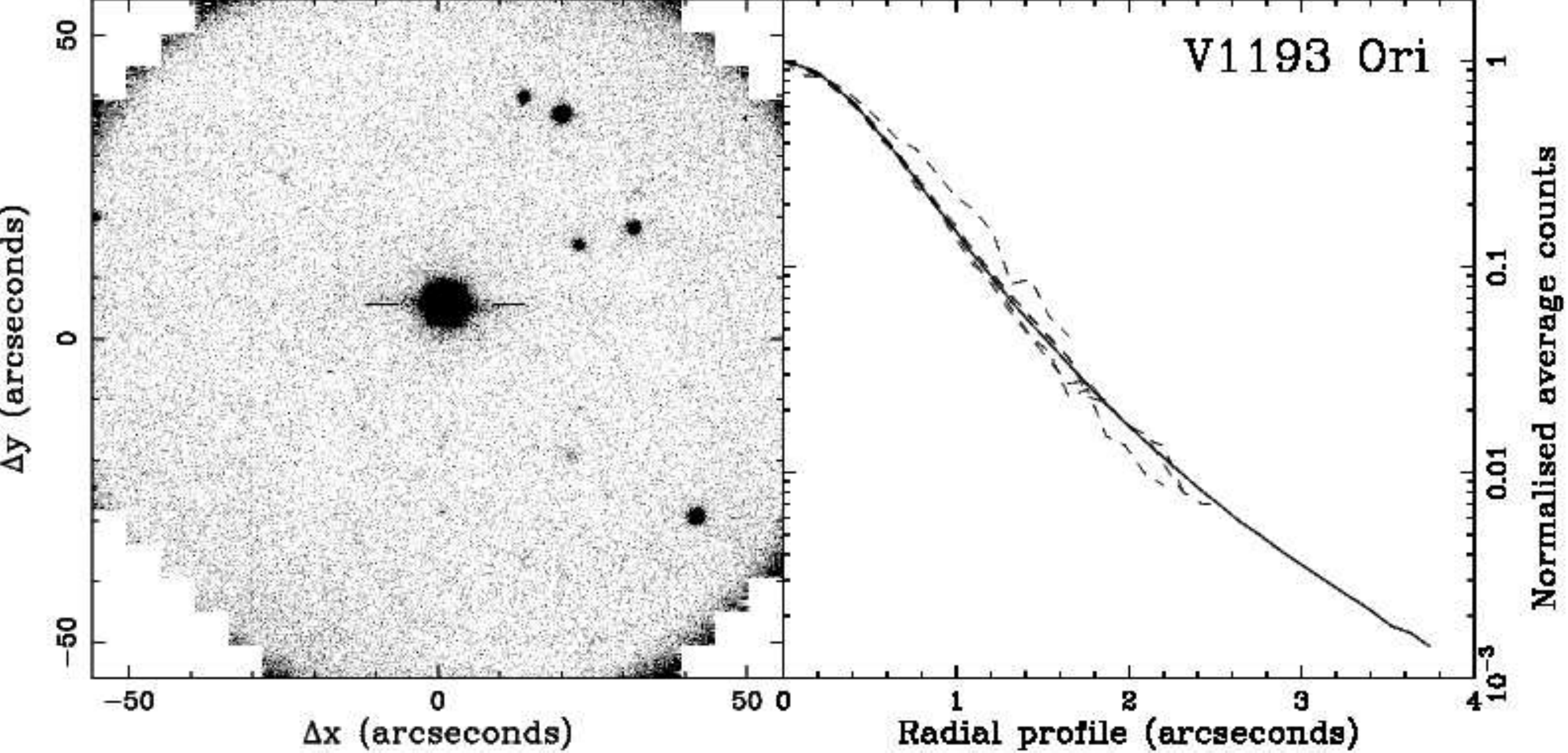}
  \vspace{1pt}
  \vspace{10pt}
\includegraphics[width=80mm,angle=0]{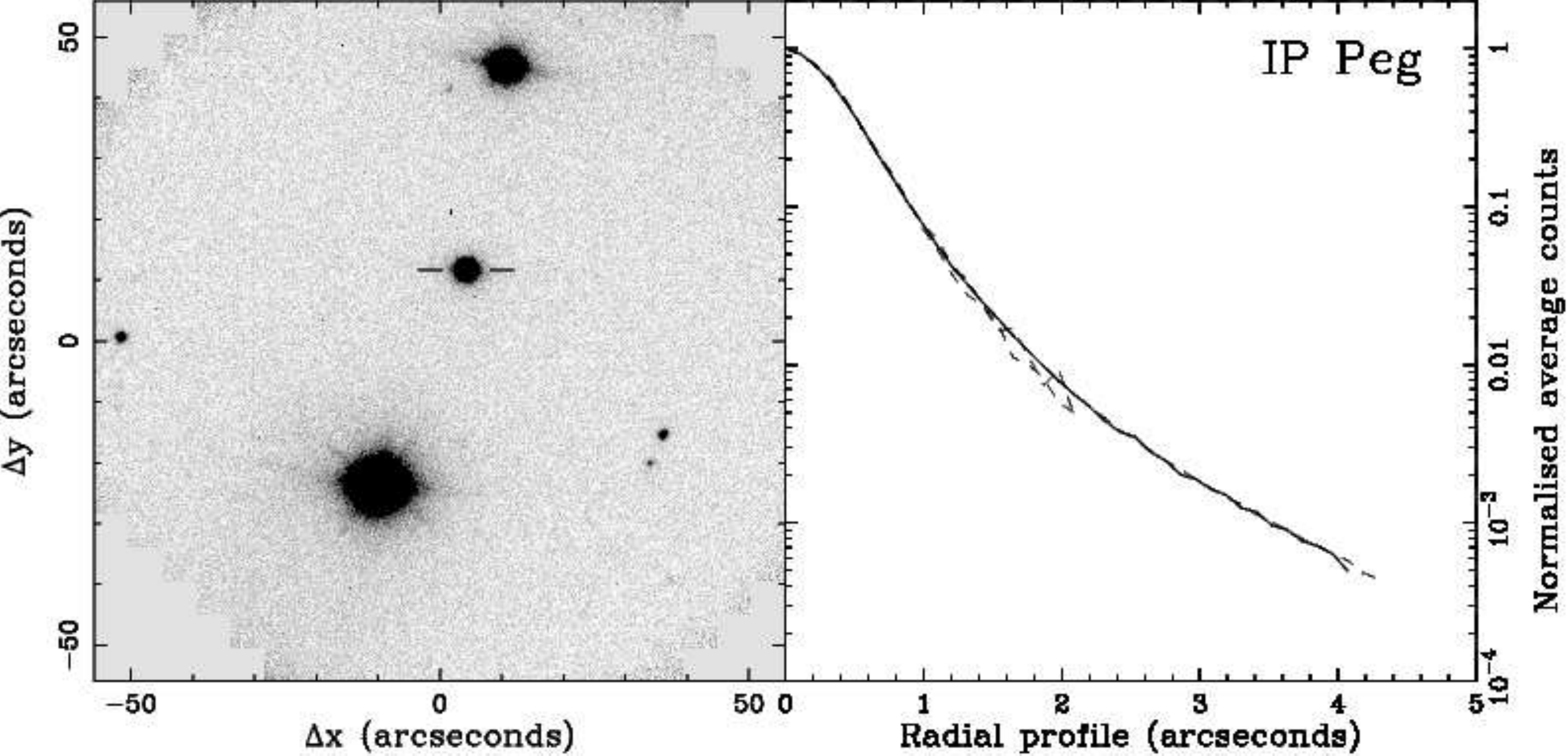}
  \vspace{1pt}
  \vspace{10pt}
\includegraphics[width=80mm,angle=0]{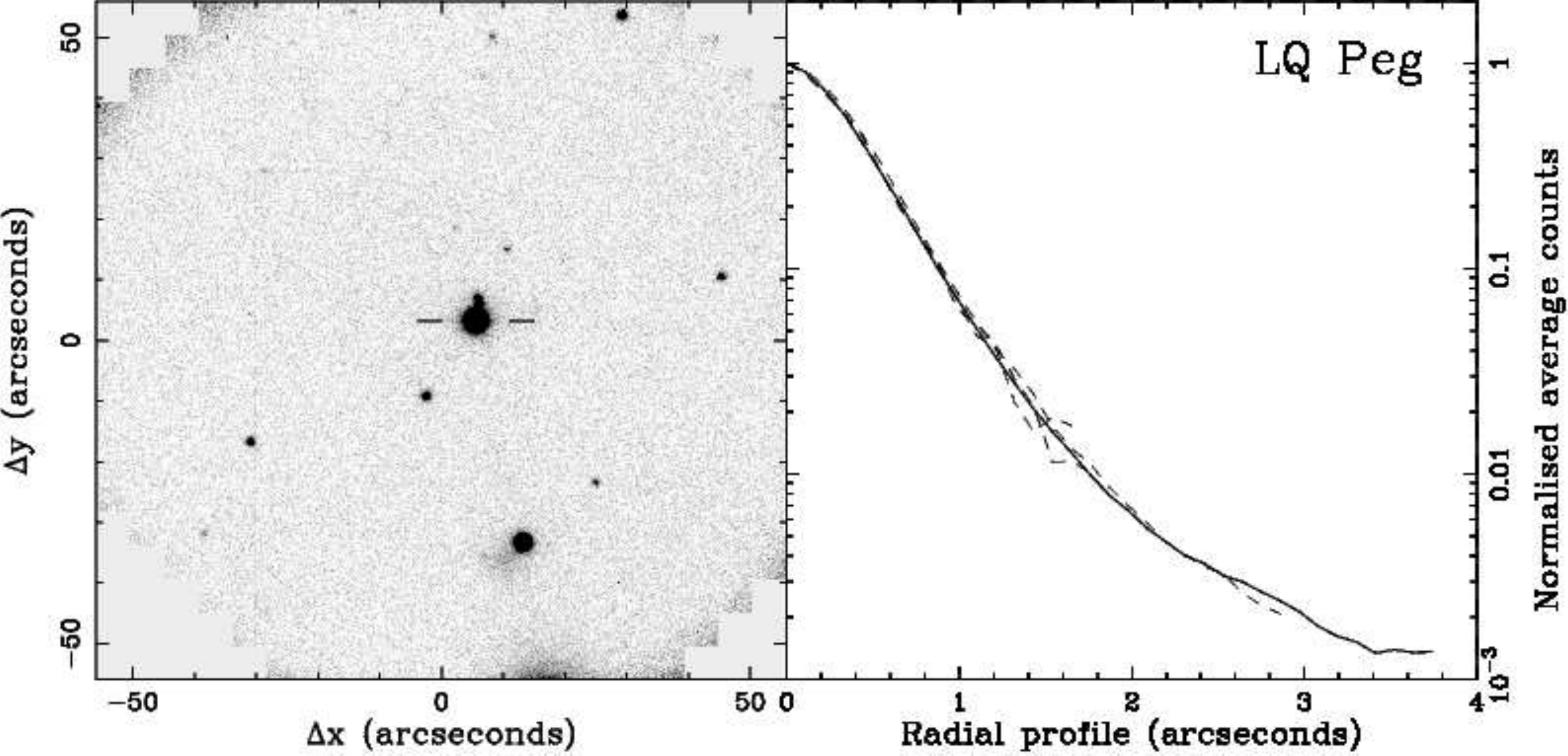}
  \vspace{1pt}
  \vspace{10pt}
\includegraphics[width=80mm,angle=0]{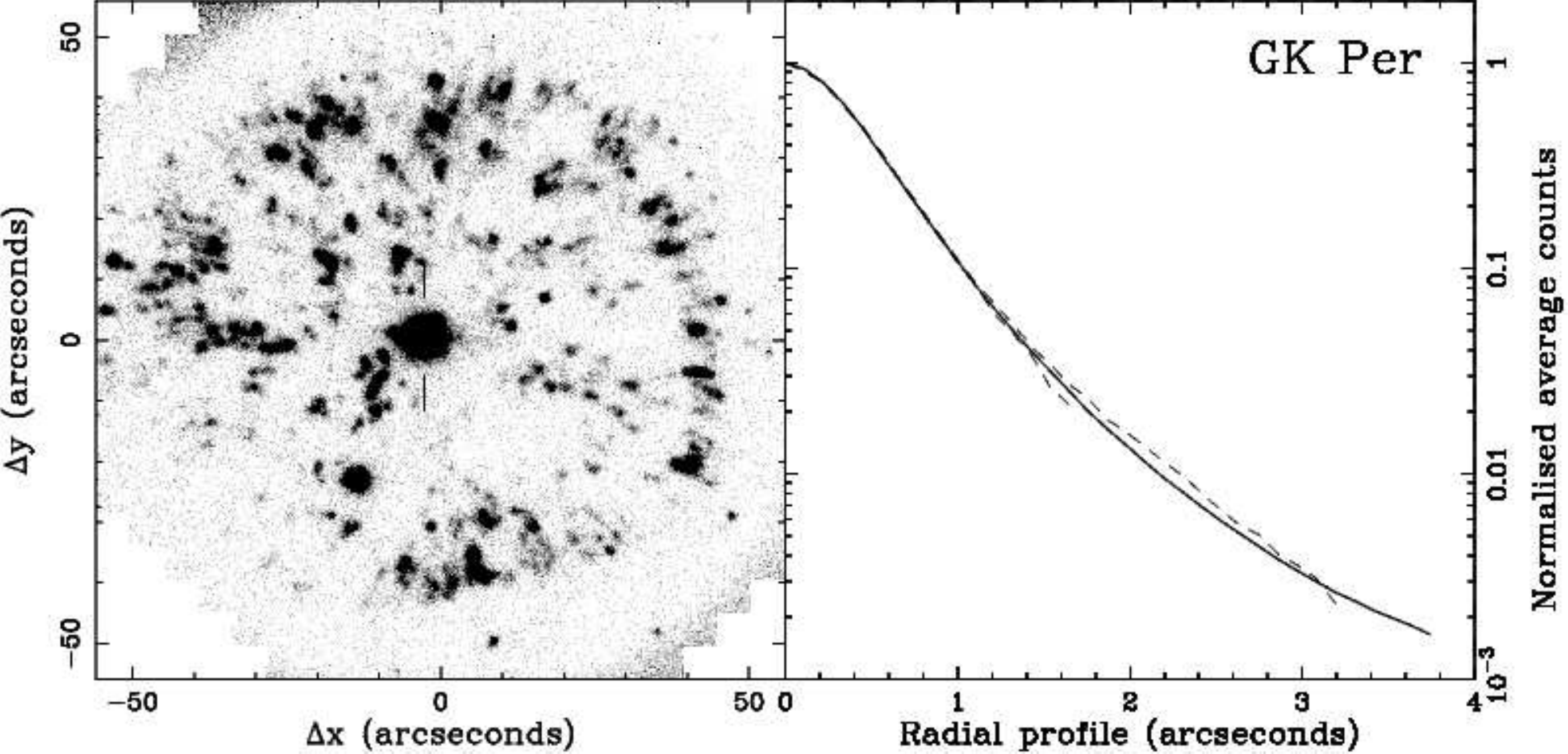}
  \vspace{1pt}
  \vspace{10pt}
\includegraphics[width=80mm,angle=0]{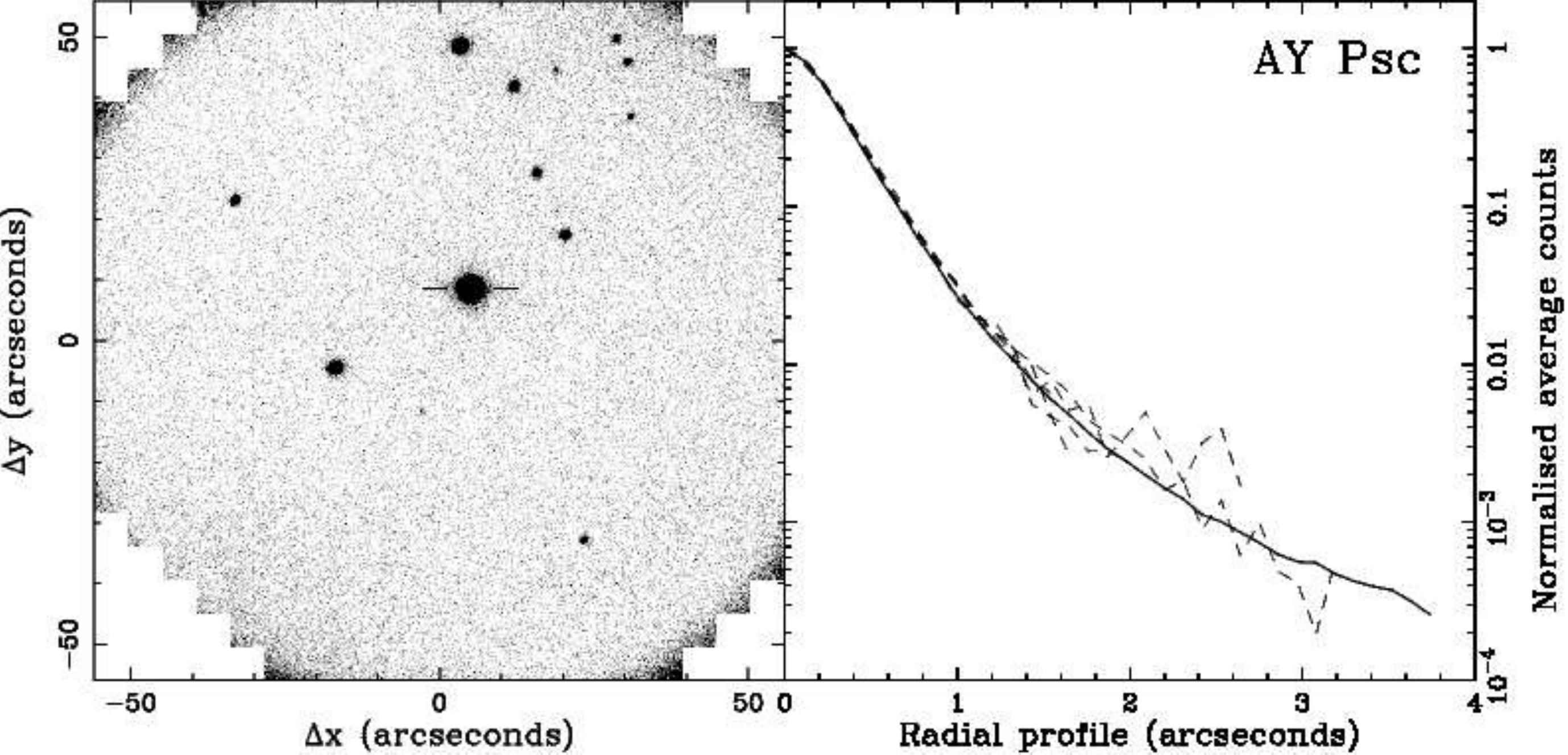}
  \vspace{1pt}
 \caption{See caption to Figure \ref{fig1} for details.}
 \label{fig5}
\end{figure}

\begin{figure}
  \vspace{10pt}
\includegraphics[width=80mm,angle=0]{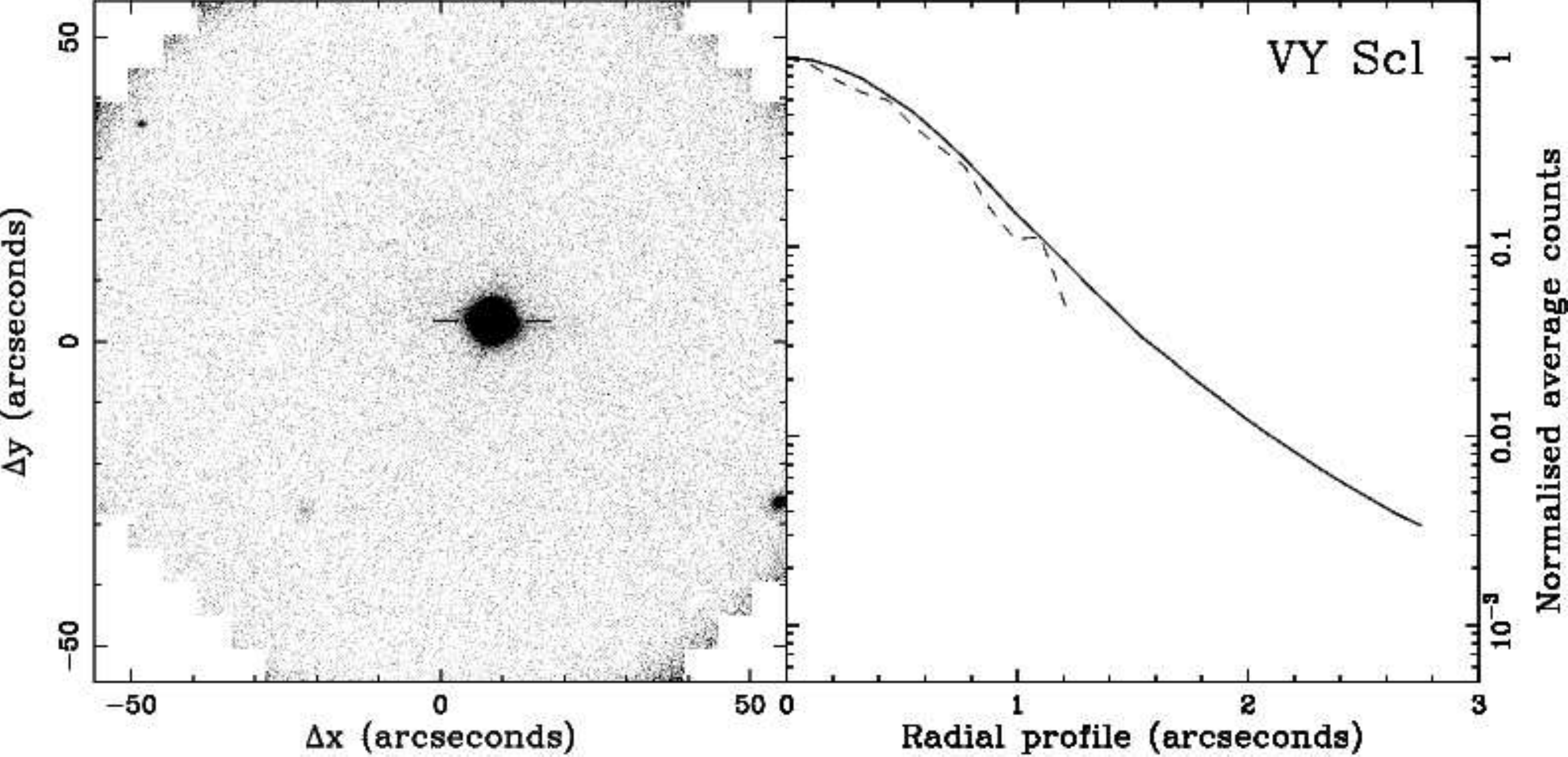}
  \vspace{1pt}
  \vspace{10pt}
\includegraphics[width=80mm,angle=0]{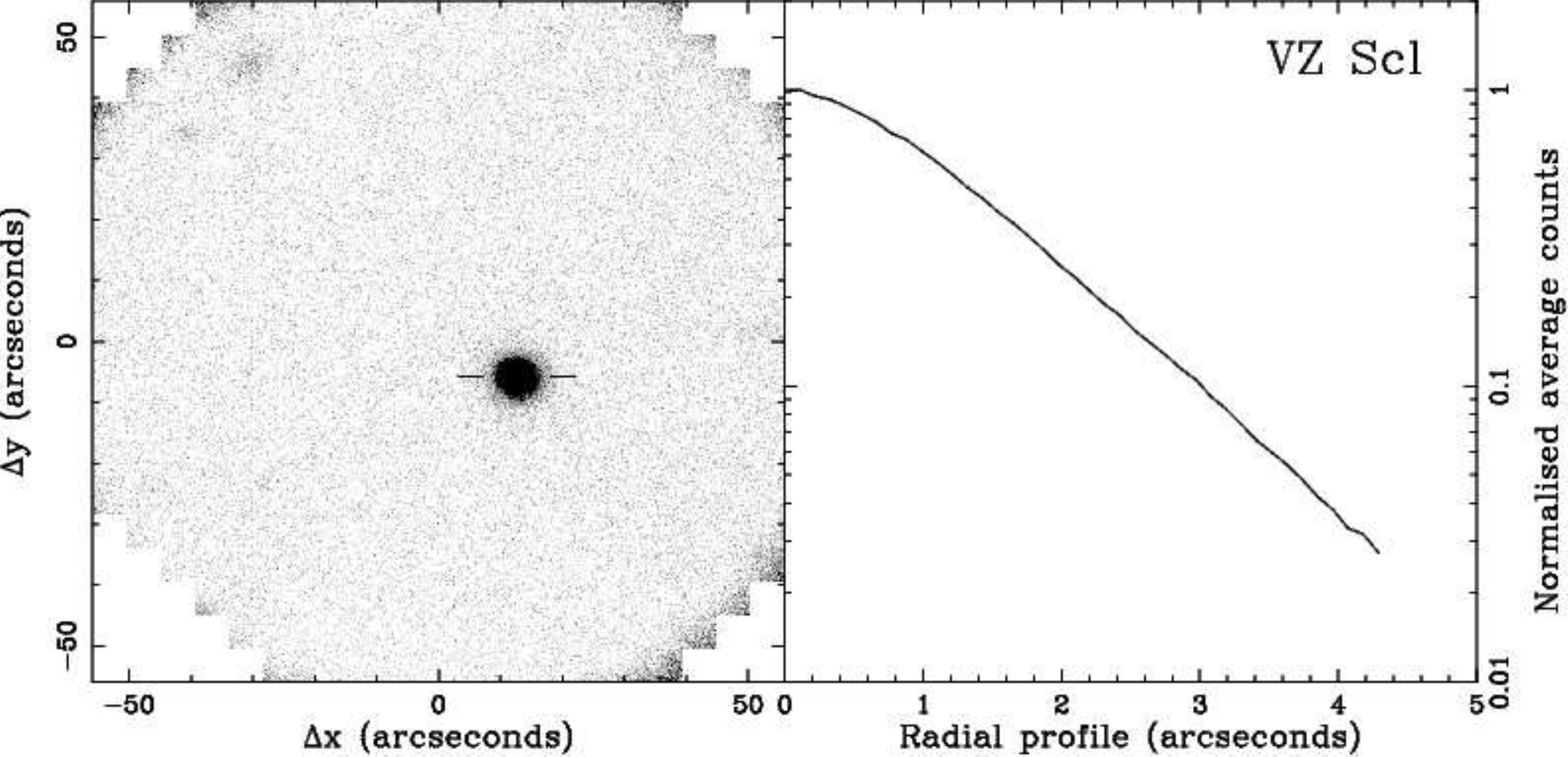}
  \vspace{1pt}
  \vspace{10pt}
\includegraphics[width=80mm,angle=0]{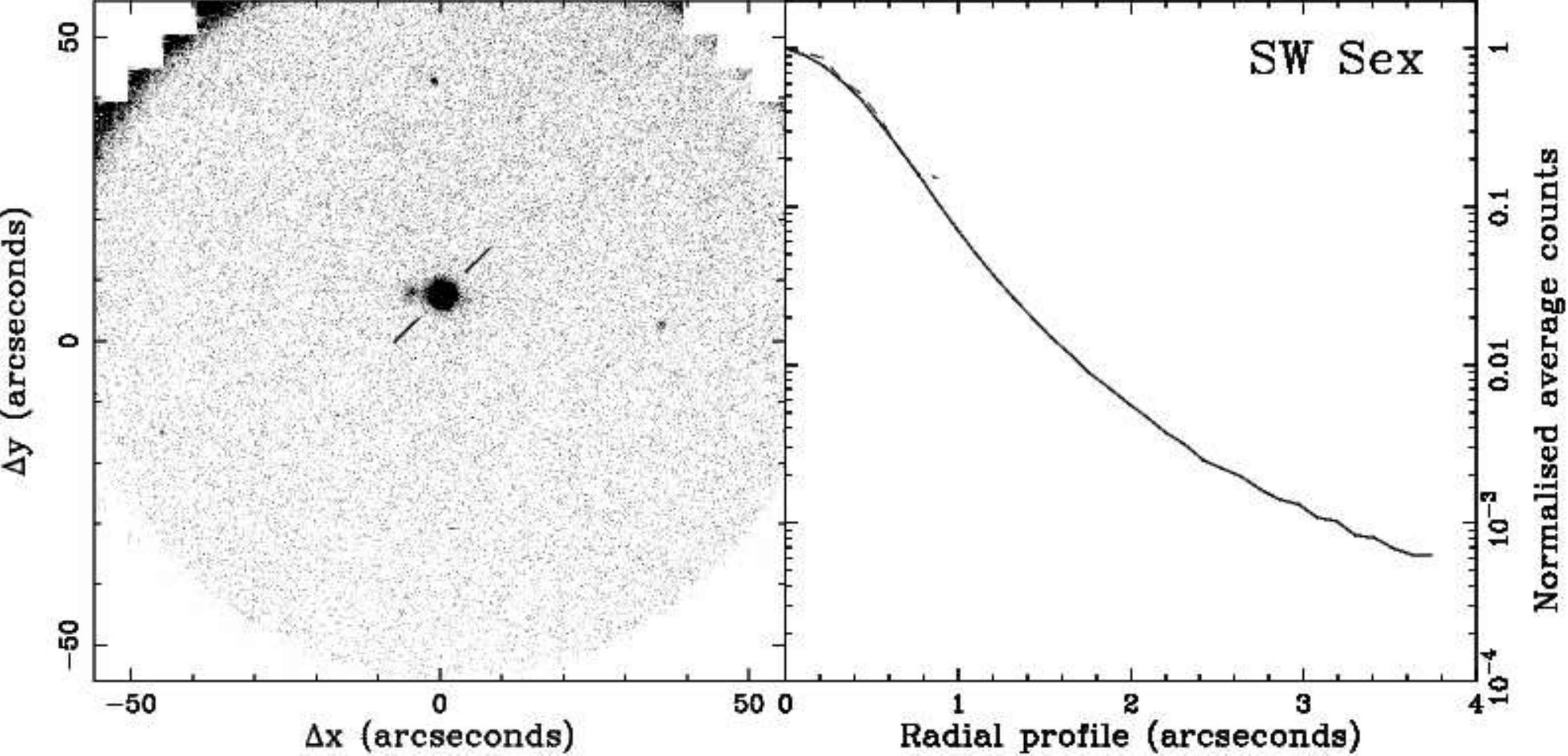}
  \vspace{1pt}
  \vspace{10pt}
\includegraphics[width=80mm,angle=0]{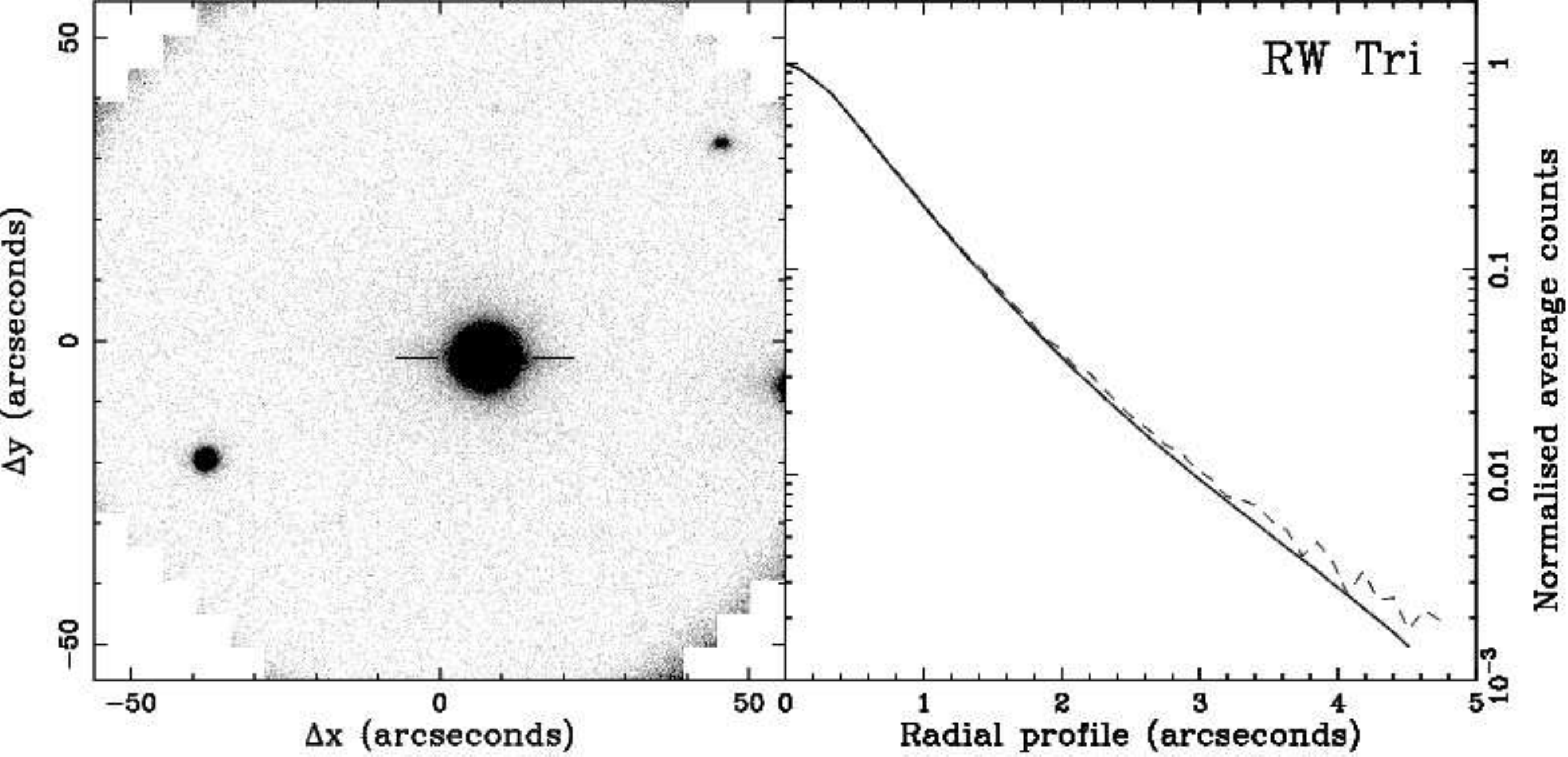}
  \vspace{1pt}
  \vspace{10pt}
\includegraphics[width=80mm,angle=0]{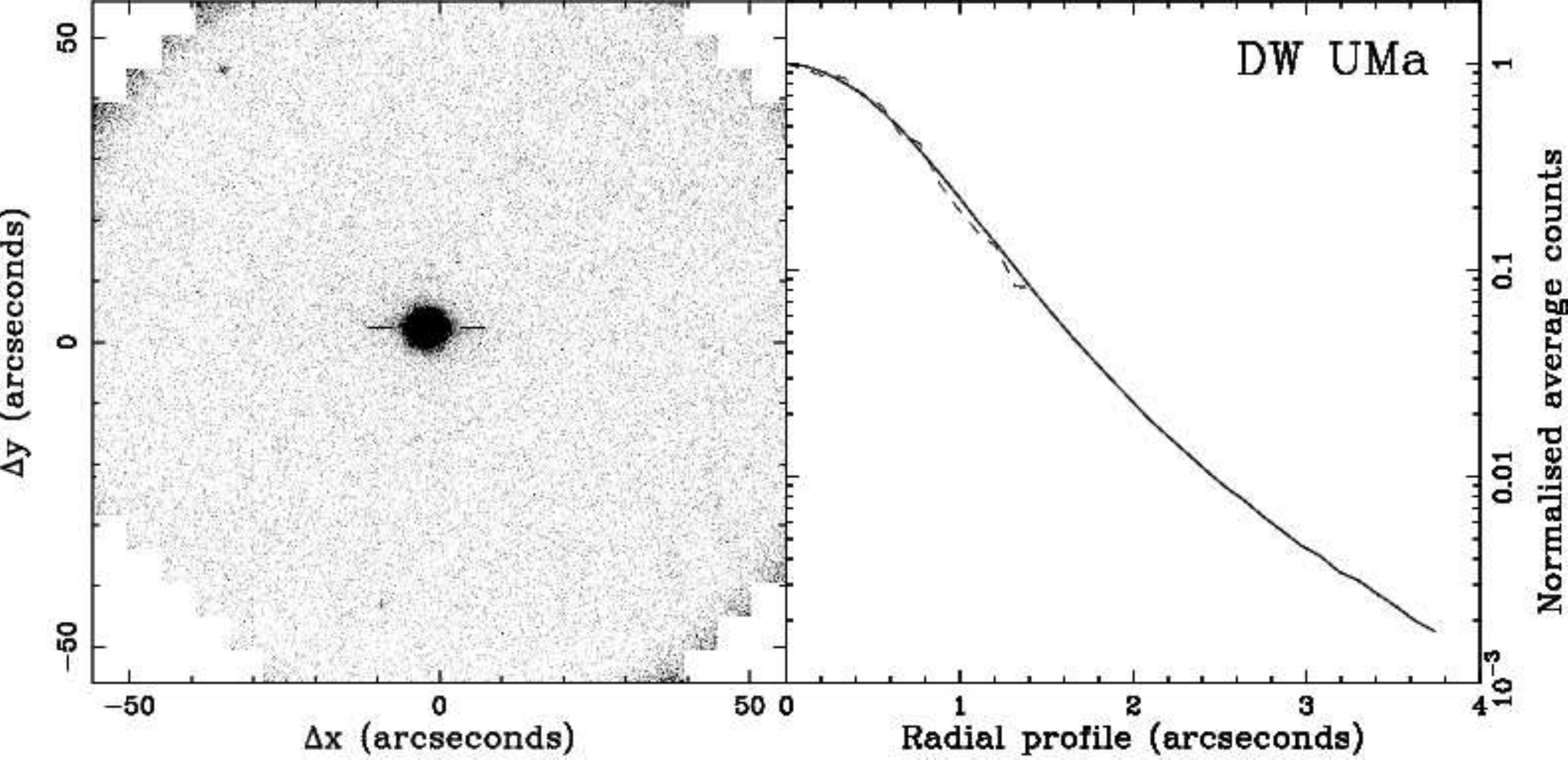}
  \vspace{1pt}
 \caption{See caption to Figure \ref{fig1} for details.}
 \label{fig6}
\end{figure}

\label{lastpage}

\end{document}